\documentclass[twocolumn,superscriptaddress,prb]{revtex4-1}

\usepackage{lipsum}
\usepackage{amsmath}
\usepackage{amsfonts}
\usepackage{color}

\usepackage{array}
\usepackage{booktabs}

\usepackage{amsthm}
\usepackage{amssymb}
\usepackage{multirow}
\usepackage{mathrsfs}
\usepackage[colorlinks,bookmarks=false,citecolor=blue,linkcolor=red,urlcolor=blue]{hyperref}
\usepackage{braket}
\setcitestyle{numbers,square}
\usepackage{overpic}

\usepackage{graphicx} 
\usepackage[table]{xcolor}

\usepackage[normalem]{ulem}

\usepackage{array}

\usepackage[caption=false]{subfig}

\usepackage[colorlinks,bookmarks=false,citecolor=blue,linkcolor=red,urlcolor=blue]{hyperref}

\newcommand{\RNum}[1]{\uppercase\expandafter{\romannumeral #1\relax}}

\definecolor{black}{rgb}{0.0,0.45,0.45}

\def \be {\begin{equation}}
\def \ee {\end{equation}}
\def \ba {\begin{array}}
\def \ea {\end{array}}
\def \bea {\begin{eqnarray}}
\def \eea {\end{eqnarray}}

\def \ble {\begin{widetext}\begin{equation}}
\def \ele {\end{equation}\end{widetext}}
\def \blea {\begin{widetext}\begin{eqnarray}}
\def \elea {\end{eqnarray}\end{widetext}}

\def \cF {{\mathcal F}}

\def \dd {\mathrm{d}}

\def \Tr {{\mathrm{Tr}}}

\def \and {{\mathrm{and}}}

\newcommand{\cFff}{\mathcal F_{\mathrm{ff}}}

\begin{document}

\title{Hidden Conformal Boundary Data in Finite-Temperature Stabilizer Entropy}

\author{Reyhaneh Khasseh}
\affiliation{Theoretical Physics III, Center for Electronic Correlations and Magnetism,\\
Institute of Physics, University of Augsburg, D-86135 Augsburg, Germany}

\author{M.~A.~Rajabpour}
\affiliation{Instituto de Física, Universidade Federal Fluminense, Av.~Gal.~Milton Tavares de Souza s/n, Gragoatá, 24210-346, Niterói, RJ, Brazil}

\begin{abstract}
We study the finite-temperature stabilizer R\'enyi entropy of the open critical
transverse-field Ising chain. At R\'enyi index one half, this observable probes
the distribution of thermal Pauli-string expectation values and can be written
as a sum over absolute values of all square minors of a finite-temperature
correlation matrix. We show that this exponentially large sum is exactly
reducible to a single Pfaffian. The Pfaffian representation reveals a block
Toeplitz--Hankel structure and allows us to extract the large-size scaling in
several thermal regimes. In the crossover window where the inverse temperature
is proportional to the system size, the stabilizer numerator factorizes into a
saturated extensive contribution and a universal finite-size scaling function.
We find that this scaling function is a level-eight eta quotient, rather than
the ordinary free-boundary Majorana thermal factor. We identify it with the normalized midpoint
disorder-field one-point function of the free-boundary Ising CFT.
The deviation is exponentially hidden at low temperature but becomes manifest in the small-\(\tau\) modular limit, where it gives a
Cardy-like asymptotic for the Pauli-string expectation-weight spectrum. These results show that finite-temperature stabilizer
entropy probes a universal Ising disorder-field amplitude that is invisible
to ordinary thermodynamic probes.
\end{abstract}

\maketitle

{\it Introduction.---}
Non-stabilizerness is the resource that allows quantum states and
operations to go beyond the classically simulable stabilizer framework
\cite{Veitch2014,HowardCampbell2017,LiuWinter2022}.  A useful way to quantify
this resource in many-body systems is through stabilizer R\'enyi entropies,
which measure how broadly a quantum state is distributed over the Pauli-string
basis \cite{Leone2022a,HaugPiroli2023,Leone2024}.  These quantities have
recently emerged as powerful diagnostics of many-body quantum complexity.  In
spin chains and tensor-network settings, stabilizer entropies have been used to
probe non-stabilizerness in ground states, criticality, Pauli-basis structure, and
matrix-product-state representations
\cite{OlivieroLeoneHamma2022,HaugPiroliMPS2023,
TarabungaDalmontePRXQ2023,Tarabunga2024,SarkarBiswasWuBiswas2026,Guglielmo2023,LamiCollura2024,
TarabungaTirritoBanulsDalmonte2024,HallamSmithPapic2026,TurkeshiDymarskySierant2025}.  They have also motivated efficient
measurement and simulation protocols, and have been applied to Clifford
disentangling, fermionic Gaussian states, and conformal tensor-network
structures
\cite{HaugLeeKimPRL2024,DingPRXQuantum2025,Collura2024,
ViscardiDalmonteHammaTirrito2025,Fan2025,Frau2025}.  Recent works have also shown that non-stabilizerness can spread differently
from entanglement under dynamics \cite{RattacasoLeoneOlivieroHamma2023,TurkeshiTirritoSierant2025}.  This
indicates that non-stabilizerness is not just another form of entanglement, but
probes a different layer of many-body quantum structure.

A particularly interesting question is what stabilizer entropies measure in
critical systems.  The connection between non-stabilizerness and conformal
field theory was already emphasized in Ref.~\cite{WhiteCaoSwingle2021,Hoshino2025,Hoshino2025b,Rajabpour2025SREShannon}.
Conventional entanglement and Shannon-type quantities are known to contain
universal information governed by conformal field theory and, in the presence
of boundaries, by boundary conformal field theory
\cite{CalabreseCardy2009,Stephan2009,Stephan2010,Stephan2014,
Alcaraz2013,Alcaraz2014,Cardy1986,Cardy1989,
AffleckLudwig1991,CardyBCFT2004}.  For quadratic fermions, stabilizer
entropies have also been related exactly to Shannon--R\'enyi entropies in
doubled free-fermion systems \cite{Rajabpour2025SREShannon}.  Recent work has
shown that stabilizer R\'enyi entropies can encode universal conformal data,
including information associated with topological defects and boundary
conditions
\cite{FrohlichFuchsRunkelSchweigert2004,Hoshino2025,Hoshino2025b}.  This
suggests that non-stabilizerness is not merely a short-distance property of a many-body wave
function, but can carry universal long-distance information
\cite{KorbanyGullansPiroli2025}.  It raises a sharper question: can the
Pauli-string expectation-value distribution reveal universal boundary or
defect-like data that are invisible to ordinary thermodynamic probes?

Most existing studies of stabilizer entropy in critical systems have focused on
pure ground states.  Much less is known about finite-temperature states, or
more generally about mixed-state settings, where the suppression of coherent
Pauli expectation values is intertwined with ordinary thermal mixing.  This suppression has two components, one coming from a genuine redistribution
of coherent Pauli weight among many strings, which is the non-stabilizer
information we want to measure, and another coming from a trivial reduction of
the overall Hilbert--Schmidt norm of the density matrix. The latter is captured by the purity, which must therefore be divided out to
isolate the stabilizer contribution. Related mixed-state perspectives have
also appeared in studies of noisy many-body systems and in purity-based
witnesses of non-stabilizerness
\cite{BallarTriguerosMarinGuzman2025,HaugTarabunga2025}.

In this Letter we study the finite-temperature stabilizer entropy of the open
critical transverse-field Ising chain.  Although the Gibbs state is Gaussian
and determined by a finite-temperature Majorana correlation matrix
\cite{LIEB1961407,Peschel2003,PeschelEisler2009}, the stabilizer entropy is not
an ordinary Gaussian thermodynamic observable.  At R\'enyi index
\(\alpha=1/2\), it involves an exponentially large sum over absolute values of
square minors of this correlation matrix.  We show that this entire sum reduces
exactly to a single Pfaffian with block Toeplitz--Hankel structure.


The main physical consequence appears in the finite-size thermal crossover
\(\beta=\tau L\). In this regime the stabilizer numerator factorizes into its
saturated zero-temperature scaling form and a universal crossover function,
which is not the ordinary free-boundary Majorana thermal factor but a
level-eight eta quotient. We identify this crossover
explicitly with the normalized free-boundary Ising cylinder one-point function
of the disorder field \(\mu\) inserted at the reflection-symmetric midpoint. This shows that finite-temperature stabilizer entropy detects defect-like
boundary information invisible to ordinary fermionic thermodynamics. We also obtain an
exact open-boundary XX--Ising doubling relation, including the purity
normalization.

{\it Model and stabilizer entropy.---} We consider the open transverse-field Ising Hamiltonian
\begin{equation}
H
=
-J\sum_{j=1}^{L-1}\sigma^x_j\sigma^x_{j+1}
-h\sum_{j=1}^{L}\sigma^z_j ,
\label{eq:H_TFI}
\end{equation}
and work at the critical point \(h=J\).  In the staggered Jordan--Wigner gauge,
the Hamiltonian is quadratic in Majorana operators, and the thermal Gibbs state
is Gaussian.  We denote the corresponding mixed Majorana correlator by
\(G(\beta)\).  At criticality, with
\(k_p=(2p-1)\pi/(2L+1)\), \(p=1,\ldots,L\), one obtains
\begin{equation}
\begin{split}
G_{j\ell}(\beta)
&=
(-1)^{j+\ell}
\frac{4}{2L+1}
\sum_{p=1}^{L}
\tanh\!\left(2\beta J\sin\frac{k_p}{2}\right)
\\
&\quad\times
\cos\!\left[\left(j-\frac12\right)k_p\right]
\sin(\ell k_p).
\end{split}
\label{eq:G_critical_main}
\end{equation}
The staggered-gauge conventions and the derivation of
Eq.~\eqref{eq:G_critical_main} are given in Sec.~S1 of the Supplemental
Material \cite{Supplement}.

For a density matrix \(\rho\), let
\(\mathcal P_L=\{I,X,Y,Z\}^{\otimes L}\) be the set of phase-free Pauli strings.
We define the Pauli-spectrum moment
\begin{equation}
\mathcal Z_\alpha(\rho)
=
2^{-L}
\sum_{P\in\mathcal P_L}
\left|\operatorname{Tr}(\rho P)\right|^{2\alpha}.
\label{eq:Aalpha_main}
\end{equation}
Moments of the form \(\mathcal Z_\alpha(\rho)\) probe the Pauli spectrum of
the state, namely the distribution of Pauli-string expectation values. Since \(\mathcal Z_1(\rho)=\operatorname{Tr}\rho^2\), the \(\alpha=1\) moment
is the purity.  The purity-normalized mixed-state stabilizer R\'enyi entropy is
therefore
\begin{equation}
M_\alpha(\rho)
=
\frac{1}{1-\alpha}
\log
\frac{\mathcal Z_\alpha(\rho)}{\mathcal Z_1(\rho)} .
\label{eq:stabilizer_renyi_main}
\end{equation}

We focus on \(\alpha=1/2\).  For the Gaussian state \(\rho_\beta\), Wick's
theorem expresses the nonzero Pauli amplitudes as minors of the correlation
matrix \(G(\beta)\).  The stabilizer numerator is therefore the absolute-minor
sum
\begin{equation}
S_L(\beta)
=
\sum_{\substack{A,B\subseteq\{1,\ldots,L\}\\ |A|=|B|}}
\left|\det G_{A,B}(\beta)\right|,
\label{eq:SL_minor_main}
\end{equation}
where the empty minor is included, and
\(\mathcal Z_{1/2}(\rho_\beta)=2^{-L}S_L(\beta)\).  Thus, with
\(\Pi_L(\beta)=\log(2^L\operatorname{Tr}\rho_\beta^2)\), one has
\begin{equation}
M_{1/2,L}(\beta)
=
2\log S_L(\beta)-2\Pi_L(\beta).
\label{eq:Mhalf_SL_Pi_main}
\end{equation}
The factor \(\Pi_L(\beta)\) is an ordinary Gaussian thermal product over
fermionic modes and is derived in Sec.~S2 of the Supplemental Material
\cite{Supplement}.  The nontrivial object is \(S_L(\beta)\): it is sensitive to
the organization of Pauli-string expectation values and is not determined by
the thermal spectrum alone. The hierarchy of leading nontrivial minors, relevant for large-\(\alpha\)
approximations, is discussed in Sec.~S9 of the Supplemental Material
\cite{Supplement}.

The key technical step is to evaluate Eq.~\eqref{eq:SL_minor_main} without
summing exponentially many minors.  We show below that the full absolute-minor
sum is exactly equal to a single finite-size Pfaffian.  This reduction turns the
finite-temperature stabilizer problem into a block Toeplitz--Hankel Pfaffian,
placing it within the broader class of determinant and Pfaffian problems that
appear in free-fermion spin chains
\cite{JinKorepin2004,Its:2008,DeiftItsKrasovsky2011,
IvanovAbanovCheianov2013,GrohaEsslerCalabrese2018}, but with a structure
specific to the absolute values of Pauli amplitudes.

{\it Exact Pfaffian representation.---}
Pfaffian structures are natural for Pauli-basis data of fermionic Gaussian
states, and have recently been used to express pure-state amplitudes in
arbitrary Pauli bases \cite{RajabpourMirjafarlouKhasseh2025}.  The mixed-state
quantity in Eq.~\eqref{eq:SL_minor_main} is different: it is an absolute-minor
sum over all Pauli sectors.  We now give the exact finite-size reduction of
this stabilizer numerator.  Although \(S_L(\beta)\) is defined as a sum over
exponentially many minors, it can be written as one Pfaffian.

For an \(L\times L\) matrix \(G\), define its antisymmetric lift
\(\mathcal A(G)\) by
\begin{equation}
\mathcal A(G)_{2i-1,2j}=G_{ij},
\qquad
\mathcal A(G)_{2j,2i-1}=-G_{ij},
\label{eq:A_lift_main}
\end{equation}
with all other entries zero.  We also introduce the universal antisymmetric
selector \(\mathcal J_{2L}\), whose entries are
\((\mathcal J_{2L})_{mn}=1\) for \(m<n\), together with antisymmetry.  In the
staggered Jordan--Wigner gauge, the selector must be conjugated by the diagonal
sign matrix
\begin{equation}
\mathcal J'_{2L}=D_s\mathcal J_{2L}D_s,
\qquad
D_s=\operatorname{diag}_{i=1}^{L}\bigl((-1)^i,(-1)^i\bigr).
\label{eq:Jprime_main}
\end{equation}
Then the absolute-minor sum is exactly
\begin{equation}
S_L(\beta)
=
(-1)^L
\operatorname{Pf}
\begin{pmatrix}
\mathcal A(G(\beta)) & I_{2L}\\
-I_{2L} & -\mathcal J'_{2L}
\end{pmatrix}
\label{eq:SL_pf_main}
\end{equation}
for every even \(L\) and every \(\beta\).

The role of \(\mathcal J'_{2L}\) is to convert a signed Pfaffian expansion into
the absolute-value sum in Eq.~\eqref{eq:SL_minor_main}.  More explicitly, the
universal selector identity expands the Pfaffian into a signed sum over all
square minors of \(G(\beta)\), while the staggered Ising kernel obeys a fixed
minor-sign rule.  The conjugation by \(D_s\) inserts precisely this sign rule,
so that each term in the Pfaffian expansion contributes
\(|\det G_{A,B}(\beta)|\).  The proof, including the selector identity and the
staggered minor-sign rule, is given in Sec.~S3 of the Supplemental Material
\cite{Supplement}.

Equation~\eqref{eq:SL_pf_main} is the main technical input of the paper.  It
turns the nonlocal Pauli-amplitude sum into the Pfaffian of a \(4L\times4L\)
antisymmetric matrix.  Using the open-chain form of \(G(\beta)\), this matrix
has an exact finite-size block Toeplitz--Hankel structure, recorded in
Sec.~S4 of the Supplemental Material \cite{Supplement}.  This structure allows
us to treat four scaling regimes in a unified way: the fixed-temperature
thermodynamic limit, the high-temperature expansion, the saturated
low-temperature Fisher--Hartwig regime, and the finite-size thermal crossover
\(\beta=\tau L\).  In the last regime the Pfaffian yields the universal
eta-quotient crossover.  The numerator and purity contributions in all four
regimes are summarized in Table~\ref{tab:scaling_regimes}.

\begin{table*}[t]
\centering
\footnotesize
\setlength{\tabcolsep}{3.5pt}
\renewcommand{\arraystretch}{1.7}

\begin{tabular*}{\textwidth}{@{\extracolsep{\fill}}llll}
\hline\hline
\rowcolor{blue!10}
\parbox[c]{0.13\textwidth}{\centering\textbf{Regime}}
&
\parbox[c]{0.17\textwidth}{\centering\textbf{Scaling window}}
&
\parbox[c]{0.39\textwidth}{\centering\textbf{Pfaffian numerator}}
&
\parbox[c]{0.25\textwidth}{\centering\textbf{Purity contribution}}
\\
\hline

\rowcolor{blue!3}
\parbox[c]{0.13\textwidth}{\centering
\textbf{I}\\
Fixed\\temperature}
&
\parbox[c]{0.17\textwidth}{\centering
\(L\to\infty\),\\
\(\beta>0\) fixed}
&
\parbox[c]{0.39\textwidth}{\centering
\[
\log S_L(\beta)
=
L f_0(\beta)+c_0(\beta)+o(1)
\]
}
&
\parbox[c]{0.25\textwidth}{\centering
\[
\Pi_L(\beta)
=
L p_0(\beta)+p_1(\beta)+o(1)
\]
}
\\[2.2em]
\hline

\parbox[c]{0.13\textwidth}{\centering
\textbf{II}\\
High\\temperature}
&
\parbox[c]{0.17\textwidth}{\centering
\(\beta\to0^+\)}
&
\parbox[c]{0.39\textwidth}{\centering
\[
\begin{gathered}
\log S_L(\beta)
=
L f_{\rm HT}(\beta)+c_{\rm HT}(\beta)+\cdots ,
\\[0.25em]
f_{\rm HT}(\beta)
=
2\beta-3\beta^2+\frac{16}{3}\beta^3+\cdots ,
\\[0.25em]
c_{\rm HT}(\beta)
=
-\beta+\frac{5}{2}\beta^2-\frac{20}{3}\beta^3+\cdots .
\end{gathered}
\]
}
&
\parbox[c]{0.25\textwidth}{\centering
\[
\begin{gathered}
\Pi_L(\beta)
=
L p_{\rm HT}(\beta)+p_{\rm bdy}(\beta)+\cdots ,
\\[0.25em]
p_{\rm HT}(\beta)
=
2\beta^2-7\beta^4
+\frac{248}{9}\beta^6+\cdots .
\end{gathered}
\]
}
\\[4.2em]
\hline

\rowcolor{blue!3}
\parbox[c]{0.13\textwidth}{\centering
\textbf{III}\\
Saturated\\low temperature}
&
\parbox[c]{0.17\textwidth}{\centering
\(L\to\infty\),\\
\(\beta/L\to\infty\)}
&
\parbox[c]{0.39\textwidth}{\centering
\[
\log S_L(\infty)
=
L f_\infty
-\frac18\log L
+c_\infty
+o(1)
\]
}
&
\parbox[c]{0.25\textwidth}{\centering
\[
\Pi_L(\infty)=L\log2
\]
}
\\[2.5em]
\hline

\parbox[c]{0.13\textwidth}{\centering
\textbf{IV}\\
Crossover}
&
\parbox[c]{0.17\textwidth}{\centering
\(\beta=\tau L\),\\
\(0<\tau<\infty\)}
&
\parbox[c]{0.39\textwidth}{\centering
\[
\begin{gathered}
\log S_L(\tau L)
=
\log S_L(\infty)
+\log\mathcal F(\tau)
+o(1),
\\[0.35em]
\mathcal F(\tau)
=
\frac{
\eta(i\tau/2)\eta(2i\tau)^{7/4}
}{
\eta(i\tau)^2\eta(4i\tau)^{1/2}
}.
\end{gathered}
\]
}
&
\parbox[c]{0.25\textwidth}{\centering
\[
\begin{gathered}
\Pi_L(\tau L)
=
L\log2+\log\mathcal P(\tau)+o(1),
\\[0.35em]
\mathcal P(\tau)
=
\frac{
\eta(i\tau/2)^2\eta(2i\tau)^4
}{
\eta(i\tau)^5\eta(4i\tau)
}.
\end{gathered}
\]
}
\\[5.0em]

\hline\hline
\end{tabular*}

\caption{
Roadmap of the four scaling regimes.  The physical mixed-state stabilizer
entropy is obtained from
\(M_{1/2,L}(\beta)=2\log S_L(\beta)-2\Pi_L(\beta)\).  Regime I is the
fixed-temperature thermodynamic limit, Regime II is its high-temperature
expansion, Regime III is the saturated low-temperature Fisher--Hartwig regime,
and Regime IV is the finite-size thermal crossover \(\beta=\tau L\).  The
derivations of the numerator and purity terms, including the Pfaffian reduction
and the crossover eta quotients, are given in the Supplemental Material
\cite{Supplement}.
}
\label{tab:scaling_regimes}
\end{table*}

{\it Boundary crossover.---}
The most revealing regime is the finite-size thermal crossover
\(\beta=\tau L\), which provides a particularly sensitive probe of the universal low-energy spectrum of the open critical chain.
In this window, the finite-energy fermionic modes, with energies of
order \(O(1)\), are already effectively frozen into their zero-temperature
form, while the low-energy finite-size fermionic modes, with energies of order
\(1/L\), remain thermally active. The
Pfaffian numerator therefore keeps the saturated zero-temperature form
\begin{equation}
\log S_L(\infty)
=
L f_\infty
-\frac18\log L
+c_\infty
+o(1),
\label{eq:S_infty_main}
\end{equation}
and acquires a universal \(O(1)\) crossover factor:
\begin{equation}
\log S_L(\tau L)
=
\log S_L(\infty)
+\log\mathcal F(\tau)
+o(1).
\label{eq:S_regimeIV_main}
\end{equation}
The saturated contribution \(\log S_L(\infty)\) is the zero-temperature
ground-state limit, whose conformal and stabilizer--Shannon aspects are
closely related to previous analyses of critical stabilizer entropy
\cite{Hoshino2025,Hoshino2025b,Rajabpour2025SREShannon}.  The new finite-temperature information is contained in the
crossover factor \(\mathcal F(\tau)\).

The natural free-fermion benchmark is the ordinary free-boundary Majorana
thermal factor
\begin{equation}
\mathcal F_{ff}(\tau)
=
\prod_{p=1}^{\infty}(1+Q^{2p-1})^{-1},
\qquad
Q=e^{-\pi\tau}.
\label{eq:Fff_main}
\end{equation}
The stabilizer Pfaffian does not give this factor alone.  Instead, the
crossover is governed by the level-eight eta quotient
\begin{equation}
\mathcal F(\tau)
=
\frac{
\eta(i\tau/2)\eta(2i\tau)^{7/4}
}{
\eta(i\tau)^2\eta(4i\tau)^{1/2}
}.
\label{eq:F_eta_main}
\end{equation}
Equivalently,
\begin{equation}
\mathcal F(\tau)
=
\mathcal F_{ff}(\tau)\,
(Q^4;Q^8)_\infty^{3/4}
(Q^8;Q^8)_\infty^{1/4},
\label{eq:F_factorized_main}
\end{equation}
%
where \((a;q)_\infty=\prod_{m=0}^{\infty}(1-aq^m)\).  The second factor is
absent from the ordinary Majorana spectrum and contributes only in the
\(Q^{4m}\) sector.  Thus the deviation from the free-boundary answer is
exponentially hidden at low temperature, but it is a universal part of the full
crossover. The full crossover can be identified explicitly with an
Ising disorder-field amplitude associated with a \(\mathbb Z_2\) disorder
line. It coincides with the normalized free-boundary Ising
cylinder one-point function of the disorder field \(\mu\) inserted at the
reflection-symmetric midpoint,
\begin{equation}
\mathcal F(\tau)
=
\frac{C_\mu(L/2;\tau)}
     {C_\mu(L/2;\infty)} ,
\label{eq:F_mu_main}
\end{equation}
where \(C_\mu\) denotes the corresponding disorder-field one-point function
\cite{KadanoffCeva1971,Watts2000}.  This identification also has a natural
physical interpretation.  The stabilizer selector is invariant under the
original Ising \(\mathbb Z_2\) spin flip, while its low-energy fermionic action
corresponds to a sign twist of the Majorana field.  The endpoint of such a
\(\mathbb Z_2\) disorder line belongs to the Ising \(h=1/16\) twist sector.  Moreover, reflection symmetry
of the open-chain construction fixes a single such insertion at the unique
reflection-invariant point \(x=L/2\). The explicit theta--eta reduction and direct
comparison are given in the Supplemental Material \cite{Supplement}.

\begin{figure}[t]
\centering
\includegraphics[width=0.48\textwidth]{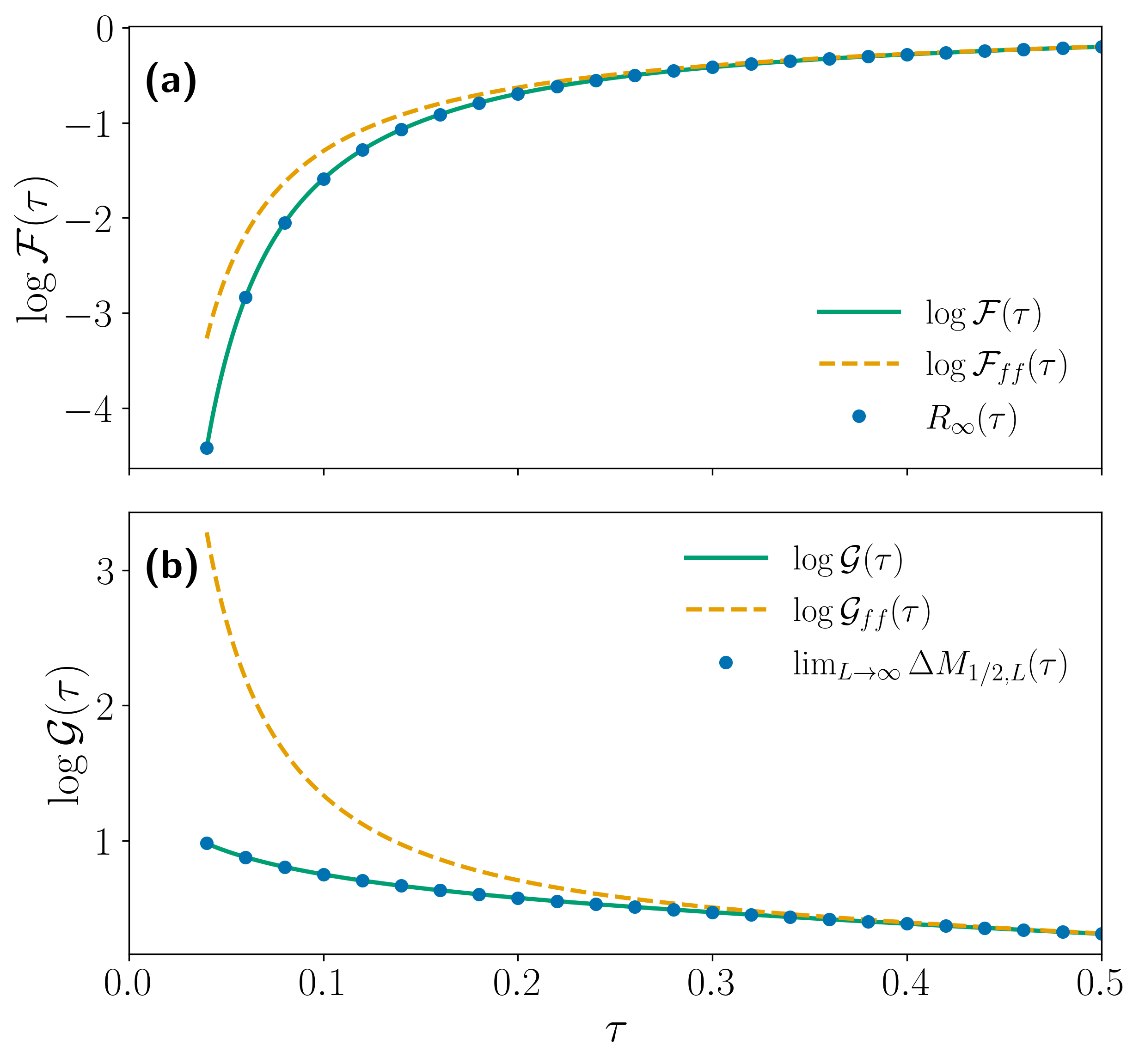}
\caption{
Regime-IV crossover in the scaling window \(\beta=\tau L\).
{(a)} Unnormalized Pauli-weight crossover.
The blue symbols are finite-size extrapolations of \(R_L(\tau)\);
the solid and dashed curves show \(\log\mathcal F(\tau)\) and the ordinary
free-boundary Majorana benchmark \(\log\mathcal F_{ff}(\tau)\), respectively.
{(b) Purity-normalized crossover.
The blue symbols are finite-size extrapolations of
\(\Delta M_{1/2,L}(\tau)\); the solid and dashed curves show
\(\log\mathcal G(\tau)\) and \(\log\mathcal G_{ff}(\tau)\), respectively.
The data follow the stabilizer eta-quotient predictions both before and after
purity normalization.}
}

\label{fig:regimeIV_main}
\end{figure}

The crossover function can be extracted directly from the exact finite-\(L\)
Pfaffian data.  We subtract the saturated bulk term, the Fisher--Hartwig
logarithm, and the saturated constant, and then extrapolate
\begin{equation}
R_\infty(\tau)
=
\lim_{L\to\infty}
\left[
\log S_L(\tau L)
-
\log S_L(\infty)
\right].
\label{eq:Rinfty_main}
\end{equation}
The prediction is \(R_\infty(\tau)=\log\mathcal F(\tau)\).

The finite-section coefficient extraction, eta-quotient reconstruction, and
residual extrapolation procedure are described in Sec.~S8 of the Supplemental
Material \cite{Supplement}.  As shown in Fig.~\ref{fig:regimeIV_main}(a), the
extrapolated data follow Eq.~\eqref{eq:F_eta_main} rather than the ordinary
Majorana factor \eqref{eq:Fff_main}.

The difference is especially transparent in the modular limit.  As
\(\tau\to0^+\), the eta quotient gives
\begin{equation}
\log\mathcal F(\tau)
=
-\frac{\pi}{16\tau}
-\frac18\log\tau
+\frac18\log2
+o(1).
\label{eq:F_small_tau_main}
\end{equation}

Through the identification
Eq.\ref{eq:F_mu_main}, these coefficients
have a direct CFT meaning: for the Ising disorder field,
\(h_\mu=\bar h_\mu=1/16\) and
\(x_\mu=h_\mu+\bar h_\mu=1/8\).  Thus the \(1/\tau\) and logarithmic terms in
Eq.~\eqref{eq:F_small_tau_main} are fixed by Ising disorder-sector
conformal data.

This singularity is stronger than the ordinary free-boundary Majorana result,
\[
\log\mathcal F_{ff}(\tau)
\sim
-\frac{\pi}{24\tau}.
\]


The sign is important: \(\mathcal F(\tau)\) is not a thermal partition
function. Since \(\mathcal F(\tau)<1\), the quantity
\(-\log\mathcal F(\tau)\) measures the thermal reduction, or depletion, of the
total absolute Pauli-string expectation weight relative to its saturated
zero-temperature value.  In particular, the \(1/\tau\) coefficient should not
be interpreted as a new bulk central charge or as a Hamiltonian
density-of-states coefficient.

If one retains the depletion notation, the leading
asymptotic corresponds to \(c_{\rm dep}^{P}=3/4\) through the conventional
parametrization
\(-\log\mathcal F(\tau)\sim\pi c_{\rm dep}^{P}/(12\tau)\).
This coefficient is only a shorthand for the leading asymptotic; its direct
CFT content is the disorder-field weight \(h_\mu=1/16\).

{\it Purity-normalized result.---}
We now return from the Pfaffian numerator to the physical mixed-state
stabilizer entropy.  The purity contribution has its own edge factor in the
same scaling window,
\[
\Pi_L(\tau L)
=
L\log2+\log\mathcal P(\tau)+o(1),
\]
so that Eq.~\eqref{eq:Mhalf_SL_Pi_main} gives
\begin{equation}
\begin{split}
M_{1/2,L}(\tau L)
={}&
2L(f_\infty-\log2)
-\frac14\log L
+2c_\infty  \\
&\quad
+\log\mathcal G(\tau)
+o(1).
\end{split}
\label{eq:M_regimeIV_main}
\end{equation}
The normalized crossover factor is
\begin{equation}
\mathcal G(\tau)
=
\frac{\mathcal F(\tau)^2}{\mathcal P(\tau)^2}
=
\frac{
\eta(i\tau)^6\,\eta(4i\tau)
}{
\eta(i\tau/2)^2\,\eta(2i\tau)^{9/2}
}.
\label{eq:G_eta_main}
\end{equation}


Thus \(\mathcal F(\tau)\) governs the unnormalized
Pauli-weight numerator, while \(\mathcal G(\tau)\) is its purity-normalized
counterpart.  In CFT terms, the midpoint disorder-field amplitude is therefore
normalized by the universal free-boundary purity factor.

For the normalized crossover shown in
Fig.~\ref{fig:regimeIV_main}(b), we define
\(\Delta M_{1/2,L}(\tau)\equiv
M_{1/2,L}(\tau L)-M_{1/2,L}(\infty)\), so that
\(\lim_{L\to\infty}\Delta M_{1/2,L}(\tau)=\log\mathcal G(\tau)\).

This distinction becomes particularly important in the
small-\(\tau\) limit:
\begin{equation}
\log\mathcal G(\tau)
=
-\frac14\log\tau
+\frac14\log2
+o(1),
\qquad
\tau\to0^+ .
\label{eq:G_small_tau_main}
\end{equation}

Hence, the \(1/\tau\) singularity of the disorder-field
amplitude \(\log\mathcal F(\tau)\) is exactly cancelled by the corresponding
purity contribution.
The edge purity factor
\(\mathcal P(\tau)\), the normalized eta quotient \(\mathcal G(\tau)\), and
their limiting forms are derived in Secs.~S2 and S8 of the Supplemental
Material \cite{Supplement}.

{\it Open XX-chain doubling.---}
As a byproduct, the open-boundary construction gives an exact relation to the
critical XX chain.  For an open XX chain on \(2L\) sites with
\(J_{\rm XX}=2J\), the finite-temperature correlation matrix is off diagonal
in the odd-even decomposition, with the off-diagonal block equal to the open
critical Ising correlator up to diagonal sign conjugations.  Hence the
absolute-minor sum factorizes as
\begin{equation}
S_{2L}^{\rm XX}(\beta)
=
\left[S_L^{\rm TFI}(\beta)\right]^2.
\label{eq:XX_TFI_S_doubling_main}
\end{equation}
The purity contribution factorizes in the same way, namely
\(\Pi_{2L}^{\rm XX}(\beta)=2\Pi_L^{\rm TFI}(\beta)\).  Therefore the
purity-normalized stabilizer entropy obeys the exact doubling relation $
M_{1/2,2L}^{\rm XX}(\beta)
=
2M_{1/2,L}^{\rm TFI}(\beta).
$ The finite-size proof is given in Sec.~S10 of the Supplemental Material
\cite{Supplement}.

{\it Conclusions.---}
We have shown that finite-temperature stabilizer entropy provides an exact and
sensitive probe of Pauli-string operator weights in the open critical Ising
chain.  At R\'enyi index one half, the exponentially large absolute-value sum
over Pauli amplitudes reduces to a single Pfaffian with block
Toeplitz--Hankel structure.  This reduction exposes a universal thermal
finite-size crossover governed by a level-eight eta quotient rather than by the
ordinary free-boundary Majorana factor.
We further identify this crossover with the normalized
free-boundary Ising cylinder one-point function of the disorder field
\(\mu\).  Its modular coefficients are therefore fixed by the Ising
disorder conformal data \(h_\mu=1/16\) and \(x_\mu=1/8\).
After purity normalization, the physical mixed-state stabilizer entropy retains
a related universal eta-quotient crossover.

These results suggest that finite-temperature stabilizer entropy defines a new
class of CFT probes.  Although our detailed analysis focused on the open
critical Ising chain, the exact XX--Ising doubling found here and the broader
stabilizer--Shannon correspondence for quadratic fermionic chains
\cite{Rajabpour2025SREShannon} indicate that analogous structures should extend
to other critical free-fermion spin chains whose stabilizer entropies reduce to
universal TFI building blocks.
Understanding how such disorder-sector amplitudes emerge
for other R\'enyi indices, boundary conditions, and critical theories is a
natural direction for future work
\cite{Hoshino2025,Hoshino2025b,Cardy1989,AffleckLudwig1991,ChoiEtAl2024,
DiatlykKhanchandaniPopovWang2024}.

{\it Acknowledgements.---}
We thank Fabian Ballar and Markus Heyl for useful comments and discussions.
We also thank CNPq and FAPERJ, grant number E-26/210.062/2023, for partial
support.

\hfill\\
\newpage
\providecommand{\href}[2]{#2}\begingroup\raggedright

\providecommand{\href}[2]{#2}
\begingroup\raggedright

\endgroup
\endgroup 

\clearpage
\onecolumngrid
\phantomsection
\begin{center}
{\LARGE\bfseries Supplemental Material: Finite-Temperature Stabilizer Entropy Reveals Hidden Boundary Defect Data\par}
\vspace{1em}
{\large Reyhaneh Khasseh\qquad M.~A.~Rajabpour\par}
\vspace{0.5em}
{\small Instituto de F\'isica, Universidade Federal Fluminense,
Av.~Gal.~Milton Tavares de Souza s/n, Gragoat\'a, 24210-346,
Niter\'oi, RJ, Brazil\par}
\end{center}
\vspace{1em}

\setcounter{equation}{0}
\setcounter{figure}{0}
\setcounter{table}{0}
\setcounter{section}{0}
\renewcommand{\theequation}{S\arabic{equation}}
\renewcommand{\thefigure}{S\arabic{figure}}
\renewcommand{\thetable}{S\Roman{table}}
\renewcommand{\theHequation}{supp.\arabic{equation}}
\renewcommand{\theHfigure}{supp.\arabic{figure}}
\renewcommand{\theHtable}{supp.\arabic{table}}
\makeatletter
\renewcommand{\bibnumfmt}[1]{[S#1]}
\renewcommand{\citenumfont}[1]{S#1}
\makeatother

\hypersetup{linkcolor=black}
\tableofcontents
\hypersetup{linkcolor=red}

\vspace{1cm}

This Supplemental Material provides the derivations supporting the main text.
We first derive the finite-temperature Majorana correlation matrix of the open
critical transverse-field Ising chain in the staggered Jordan--Wigner gauge.
We then compute the purity normalization and prove the Pfaffian representation
of the absolute-minor sum.  Finally, we analyze the resulting block
Toeplitz--Hankel Pfaffian in the four scaling regimes discussed in the main
text.

\section{Diagonalization and correlation matrix in the staggered gauge}
\label{Sec1}

In this section we derive the finite-temperature mixed Majorana correlator used in the main text.  All formulas are written directly in the staggered Jordan--Wigner gauge. The open transverse-field Ising chain is
\begin{equation}
H
=
-J\sum_{j=1}^{L-1}\sigma^x_j\sigma^x_{j+1}
-h\sum_{j=1}^{L}\sigma^z_j .
\end{equation}
We define the staggered Jordan--Wigner Majoranas
\begin{equation}
a_j
=
\left(\prod_{m<j}(-\sigma^z_m)\right)\sigma^x_j,
\qquad
b_j
=
\left(\prod_{m<j}(-\sigma^z_m)\right)\sigma^y_j .
\label{eq:app_stag_majoranas}
\end{equation}
With this convention,
\begin{equation}
\sigma^z_j=i b_j a_j,
\qquad
\sigma^x_j\sigma^x_{j+1}=i b_j a_{j+1}.
\label{eq:app_stag_spin_identities}
\end{equation}
Therefore
\begin{equation}
H
=
i\,a^TDb ,
\label{eq:app_H_D_stag}
\end{equation}
where
\begin{equation}
D_{j\ell}
=
h\delta_{j\ell}
+
J\delta_{j,\ell+1}.
\label{eq:app_D_staggered}
\end{equation}
Let \(D=U\Sigma V^T\), with
\(\Sigma=\operatorname{diag}(\sigma_1,\ldots,\sigma_L)\), be a
singular-value decomposition of \(D\).  We define
\begin{equation}
\eta_p=\sum_{j=1}^{L}U_{jp}a_j,
\qquad
\xi_p=\sum_{j=1}^{L}V_{jp}b_j .
\end{equation}
Then the Hamiltonian becomes
\begin{equation}
H
=
i\sum_{p=1}^{L}\sigma_p\,\eta_p\xi_p .
\end{equation}
For one Majorana pair we have
\begin{equation}
\left\langle \eta_p\xi_q\right\rangle_\beta
=
i\,\delta_{pq}\tanh(\beta\sigma_p).
\end{equation}
We define the mixed Majorana correlation matrix by
\begin{equation}
G_{j\ell}(\beta)
=
\frac{1}{i}\left\langle a_j b_\ell\right\rangle_\beta .
\label{eq:app_G_def_stag}
\end{equation}
Using the singular-vector expansion gives
\begin{equation}
G(\beta)
=
U\tanh(\beta\Sigma)V^T ,
\label{eq:app_G_svd_stag}
\end{equation}
or, in components,
\begin{equation}
G_{j\ell}(\beta)
=
\sum_{p=1}^{L}
\tanh(\beta\sigma_p)\,
U_{jp}V_{\ell p}.
\label{eq:app_G_modes_stag}
\end{equation}

\subsection{General open chain in the staggered gauge}

The right singular vectors solve the eigenvalue problem for \(D^TD\).  In the staggered gauge they may be written as
\begin{equation}
V_{\ell p}
=
(-1)^{\ell-1}\mathcal N_p\sin(\ell k_p),
\label{eq:app_V_general_stag}
\end{equation}
with singular values
\begin{equation}
\sigma_p
=
\sqrt{h^2+J^2-2hJ\cos k_p}.
\label{eq:app_sigma_general_stag}
\end{equation}
The allowed momenta \(k_p\in(0,\pi)\), \(p=1,\ldots,L\), are fixed by
\begin{equation}
h\sin((L+1)k_p)
=
J\sin(Lk_p).
\label{eq:app_quant_general_stag}
\end{equation}
The normalization is
\begin{equation}
\mathcal N_p
=
\left[
\sum_{r=1}^{L}\sin^2(rk_p)
\right]^{-1/2}.
\label{eq:app_N_general_stag}
\end{equation}
The left singular vectors are obtained from
\begin{equation}
Dv^{(p)}=\sigma_p u^{(p)} .
\end{equation}
Using \eqref{eq:app_V_general_stag}, one finds
\begin{equation}
U_{jp}
=
(-1)^{j-1}
\frac{\mathcal N_p}{\sigma_p}
\left[
h\sin(jk_p)
-
J\sin((j-1)k_p)
\right].
\label{eq:app_U_general_stag}
\end{equation}
Substituting \eqref{eq:app_V_general_stag} and \eqref{eq:app_U_general_stag} into \eqref{eq:app_G_modes_stag}, the finite-temperature staggered-gauge kernel is
\begin{equation}
G_{j\ell}(\beta)
=
(-1)^{j+\ell}
\sum_{p=1}^{L}
\frac{\mathcal N_p^2}{\sigma_p}
\tanh(\beta\sigma_p)
\left[
h\sin(jk_p)
-
J\sin((j-1)k_p)
\right]
\sin(\ell k_p).
\label{eq:app_G_general_stag_explicit}
\end{equation}

\subsection{Critical open chain in the staggered gauge}

At the critical point \(h=J\), the quantization condition reduces to
\(\sin((L+1)k_p)=\sin(Lk_p)\), giving
\begin{equation}
k_p
=
\frac{(2p-1)\pi}{2L+1},
\qquad
p=1,\ldots,L .
\label{eq:app_kp_stag}
\end{equation}
The singular values reduce to
\begin{equation}
\sigma_p
=
2J\sin\frac{k_p}{2}.
\label{eq:app_sigmap_stag}
\end{equation}
The normalized singular vectors are
\begin{equation}
U_{jp}
=
(-1)^{j-1}
\sqrt{\frac{4}{2L+1}}\,
\cos\!\left[\left(j-\frac12\right)k_p\right],
\qquad
V_{\ell p}
=
(-1)^{\ell-1}
\sqrt{\frac{4}{2L+1}}\,
\sin(\ell k_p).
\label{eq:app_UV_critical_stag}
\end{equation}
Therefore the critical finite-temperature staggered-gauge correlation matrix is
\begin{equation}
G_{j\ell}(\beta)
=
(-1)^{j+\ell}
\frac{4}{2L+1}
\sum_{p=1}^{L}
\tanh\!\left(2\beta J\sin\frac{k_p}{2}\right)
\cos\!\left[\left(j-\frac12\right)k_p\right]
\sin(\ell k_p),
\label{eq:app_G_critical_stag}
\end{equation}
where \(k_p\) is given in Eq.~\eqref{eq:app_kp_stag}. This is the finite-temperature open-chain Majorana kernel in the staggered Jordan--Wigner gauge.  

\section{Exact calculation of purity and its scaling limits}
\label{Sec2}

In this section we compute the purity normalization entering the mixed-state
stabilizer R\'enyi quantity.  The Pfaffian/minor expression gives the
unnormalized Pauli-amplitude numerator \(S_L(\beta)\), whereas the mixed-state
definition used in the main text is
\begin{equation}
\mathcal Z_\alpha(\rho)
=
2^{-L}
\sum_{P\in\mathcal P_L}
\left|\Tr(\rho P)\right|^{2\alpha},
\qquad
M_\alpha(\rho)
=
\frac{1}{1-\alpha}
\log
\frac{\mathcal Z_\alpha(\rho)}{\mathcal Z_1(\rho)} .
\end{equation}
For \(\alpha=1/2\), this gives
\begin{equation}
M_{1/2}(\rho)
=
2\log
\left[
\frac{
2^{-L}\sum_{P\in\mathcal P_L}|\Tr(\rho P)|
}{
\Tr(\rho^2)
}
\right].
\label{eq:M_half_main_def_supp}
\end{equation}
Since the unnormalized Pauli-amplitude numerator satisfies
\(S_L(\beta)=\sum_{P\in\mathcal P_L}|\Tr(\rho_\beta P)|\), or equivalently
\(\mathcal Z_{1/2}(\rho_\beta)=2^{-L}S_L(\beta)\), we obtain
\begin{equation}
M_{1/2,L}(\beta)
=
2\log S_L(\beta)
-
2\Pi_L(\beta).
\label{eq:M_half_S_purity_correct}
\end{equation}
where we have $\Pi_L(\beta)
=
\log\!\left(2^L\Tr\rho_\beta^2\right)$.

\subsection{Exact lattice purity product}

The purity normalization is simple because the Gibbs state factorizes over the
fermionic normal modes obtained in Sec.~\ref{Sec1}.  At criticality, the
thermal polarization of the \(p\)-th mode is
\begin{equation}
t_p(\beta)
=
\tanh\!\left(2\beta J\sin\frac{k_p}{2}\right),
\qquad
k_p=\frac{(2p-1)\pi}{2L+1},
\qquad p=1,\ldots,L .
\label{eq:t_p_def_purity}
\end{equation}
Each mode is a two-level system with eigenvalues
\((1\pm t_p)/2\).  Hence its purity is
\[
\frac12\left(1+t_p^2\right),
\]
and the full lattice purity is the exact product
\begin{equation}
\Tr\rho_\beta^2
=
\prod_{p=1}^{L}
\frac{1+t_p(\beta)^2}{2}.
\label{eq:purity_product_exact}
\end{equation}
We denote its logarithm by
\begin{equation}
\Pi_L(\beta)
:=
\log\!\left(2^L\Tr\rho_\beta^2\right)
=
\sum_{p=1}^{L}
\log\left(1+t_p(\beta)^2\right).
\label{eq:Pi_L_def}
\end{equation}
For the low-temperature and crossover regimes it is useful to separate the
saturated contribution explicitly.  Define
\begin{equation}
r_p(\beta)
:=
\exp\!\left(-4\beta J\sin\frac{k_p}{2}\right).
\label{eq:r_p_def}
\end{equation}
Since $t_p(\beta)=\big(1-r_p(\beta)\big)/\big(1+r_p(\beta)\big)$,
one has
\begin{equation}
\Pi_L(\beta)
=
L\log2
+
\sum_{p=1}^{L}
\left[
\log\!\left(1+r_p(\beta)^2\right)
-
2\log\!\left(1+r_p(\beta)\right)
\right].
\label{eq:Pi_r_form}
\end{equation}
This form makes the saturated contribution \(L\log2\) explicit.  In the
crossover window \(\beta=\tau L\), only the lowest open-chain modes give an
\(O(1)\) correction to this saturated value.

\subsection{Fixed-temperature scaling}

For fixed \(\beta>0\) and \(L\to\infty\), define
\begin{equation}
h_\beta(k)
=
\log\left[
1+\tanh^2\!\left(2\beta J\sin\frac{k}{2}\right)
\right].
\end{equation}
Then
\begin{equation}
\Pi_L(\beta)
=
\sum_{p=1}^{L}h_\beta(k_p).
\end{equation}
Using the open-chain Euler--Maclaurin formula for $k_p=(2p-1)\pi/(2L+1)$, we obtain
\begin{equation}
\Pi_L(\beta)
=
L p_0(\beta)+p_1(\beta)+O(L^{-1}),
\label{eq:Pi_fixed_beta_asymptotic}
\end{equation}
where
\begin{equation}
p_0(\beta)
=
\frac1\pi
\int_0^\pi
\log\left[
1+\tanh^2\!\left(2\beta J\sin\frac{k}{2}\right)
\right]\dd k,
\label{eq:p0_beta}
\end{equation}
and
\begin{equation}
p_1(\beta)
=
\frac12p_0(\beta)
-
\frac12
\log\left[
1+\tanh^2(2\beta J)
\right].
\label{eq:p1_beta}
\end{equation}
Thus, if the Pfaffian numerator has the fixed-temperature form
\begin{equation}
\log S_L(\beta)
=
L f_0(\beta)+c_0(\beta)+o(1),
\end{equation}
then the purity-normalized mixed-state stabilizer entropy has
\begin{equation}
M_{1/2,L}(\beta)
=
L\left(2f_0(\beta)-2p_0(\beta)\right)
+
\left(2c_0(\beta)-2p_1(\beta)\right)
+o(1).
\label{eq:M_fixed_beta_asymptotic}
\end{equation}

\subsection{High-temperature expansion}

The high-temperature expansion follows from the scalar series
\begin{equation}
\log(1+\tanh^2 x)
=
x^2-\frac76x^4+\frac{62}{45}x^6
-\frac{2159}{1260}x^8
+\frac{4526}{2025}x^{10}
-\frac{1414477}{467775}x^{12}
+O(x^{14}).
\label{eq:log_tanh_series}
\end{equation}
With $x=2\beta J\sin k/2$, this gives the expansion of \(p_0(\beta)\) and \(p_1(\beta)\).  For example, at
\(J=1\),
\begin{equation}
\begin{aligned}
p_0(\beta)
={}&
2\beta^2
-7\beta^4
+\frac{248}{9}\beta^6
-\frac{2159}{18}\beta^8
+\frac{126728}{225}\beta^{10}
-\frac{5657908}{2025}\beta^{12}
+O(\beta^{14}),
\end{aligned}
\label{eq:p0_high_temp}
\end{equation}
and
\begin{equation}
\begin{aligned}
p_1(\beta)
={}&
-\beta^2
+\frac{35}{6}\beta^4
-\frac{1364}{45}\beta^6
+\frac{66929}{420}\beta^8
-\frac{1747036}{2025}\beta^{10}  \\
&\qquad
+\frac{2243360522}{467775}\beta^{12}
+O(\beta^{14}).
\end{aligned}
\label{eq:p1_high_temp}
\end{equation}
For general \(J\), these expansions are obtained by replacing
\(\beta\) by \(J\beta\). For fixed \(L\), the exact high-temperature expansion is
\begin{equation}
\Pi_L(\beta)
=
\sum_{r\ge1}a_r(2\beta J)^{2r}
\sum_{p=1}^{L}\sin^{2r}\frac{k_p}{2},
\label{eq:Pi_fixed_L_high_temp}
\end{equation}
where
\begin{equation}
\log(1+\tanh^2 x)=\sum_{r\ge1}a_r x^{2r}.
\end{equation}
The first coefficients are
\begin{equation}
a_1=1,\qquad
a_2=-\frac76,\qquad
a_3=\frac{62}{45},\qquad
a_4=-\frac{2159}{1260}.
\end{equation}

\subsection{Saturated low-temperature limit}

In the saturated limit \(\beta\to\infty\) at fixed \(L\), all \(t_p(\beta)\to1\).
Therefore
\begin{equation}
2^L\Tr\rho_\infty^2
=
\prod_{p=1}^{L}(1+1)
=
2^L,
\end{equation}
or equivalently $\Pi_L(\infty)=L\log2.$ This is exact and is consistent with \(\Tr\rho_\infty^2=1\), since the zero-temperature Gibbs state is pure.
 If the saturated Pfaffian numerator has the form
\begin{equation}
\log S_L(\infty)
=
L f_\infty-\frac18\log L+c_\infty+o(1),
\end{equation}
then the purity-normalized zero-temperature stabilizer entropy is
\begin{equation}
M_{1/2,L}(\infty)
=
L(2f_\infty-2\log2)
-\frac14\log L
+2c_\infty
+o(1).
\label{eq:M_saturated}
\end{equation}

\subsection{Crossover scaling \texorpdfstring{\(\beta=\tau L\)}{beta = tau L}}

We now consider the crossover regime in which \(\beta=\tau L\) with fixed
\(0<\tau<\infty\). For simplicity we set \(J=1\) in the following crossover formulas.  For general
\(J\), replace \(\tau\) by \(J\tau\).  Define $Q=e^{-\pi\tau}$. For fixed \(p\),
\begin{equation}
r_p(\tau L)
=
\exp\!\left(-4\tau L\sin\frac{k_p}{2}\right)
\longrightarrow
Q^{2p-1}.
\end{equation}
Using \eqref{eq:Pi_r_form}, only the edge modes \(p=O(1)\) contribute an
\(O(1)\) correction to \(L\log2\).  Hence
\begin{equation}
\Pi_L(\tau L)
=
L\log2+\log\mathcal P(\tau)+o(1),
\label{eq:Pi_crossover}
\end{equation}
where
\begin{equation}
\mathcal P(\tau)
=
\prod_{p=1}^{\infty}
\frac{1+Q^{2(2p-1)}}{(1+Q^{2p-1})^2}.
\label{eq:P_tau_product}
\end{equation}
This is the edge purity factor. In \(q\)-Pochhammer notation,
\[
(a;q)_\infty=\prod_{m=0}^{\infty}(1-aq^m),
\]
one may write
\begin{equation}
\mathcal P(\tau)
=
\frac{
(Q^4;Q^8)_\infty (Q;Q^2)_\infty^2
}{
(Q^2;Q^4)_\infty^3
}.
\label{eq:P_tau_pochhammer}
\end{equation}

\subsection{Normalized Regime-IV crossover}

The unnormalized Pfaffian numerator crossover is
\begin{equation}
\log S_L(\tau L)
=
L f_\infty-\frac18\log L+c_\infty
+\log\mathcal F(\tau)+o(1),
\label{eq:S_regime_IV}
\end{equation}
with
\begin{equation}
\mathcal F(\tau)
=
\frac{
\eta(i\tau/2)\eta(2i\tau)^{7/4}
}{
\eta(i\tau)^2\eta(4i\tau)^{1/2}
}.
\label{eq:F_tau_eta}
\end{equation}
Combining \eqref{eq:S_regime_IV} with the purity crossover
\eqref{eq:Pi_crossover}, we find
\begin{equation}
M_{1/2,L}(\tau L)
=
L(2f_\infty-2\log2)
-\frac14\log L
+2c_\infty
+\log\mathcal G(\tau)
+o(1),
\label{eq:M_regime_IV}
\end{equation}
where the purity-normalized crossover function is
\begin{equation}
\mathcal G(\tau)
=
\frac{\mathcal F(\tau)^2}{\mathcal P(\tau)^2}.
\label{eq:G_def}
\end{equation}
Using \eqref{eq:P_tau_pochhammer} and the product representation of
\(\mathcal F(\tau)\), this becomes
\begin{equation}
\mathcal G(\tau)
=
\frac{
(Q^2;Q^4)_\infty^{4}
(Q^8;Q^8)_\infty^{1/2}
}{
(Q;Q^2)_\infty^{2}
(Q^4;Q^8)_\infty^{1/2}
},
\qquad
Q=e^{-\pi\tau}.
\label{eq:G_pochhammer}
\end{equation}
Equivalently, in eta-function form,
\begin{equation}
\mathcal G(\tau)
=
\frac{
\eta(i\tau)^6\,\eta(4i\tau)
}{
\eta(i\tau/2)^2\,\eta(2i\tau)^{9/2}
}.
\label{eq:G_eta}
\end{equation}

Thus the purity normalization changes the Regime-IV crossover from the
level-eight numerator quotient \(\mathcal F(\tau)\) to the eta quotient
\(\mathcal G(\tau)=\mathcal F(\tau)^2/\mathcal P(\tau)^2\).

\subsection{Asymptotics of the normalized crossover}

Using the eta-quotient representation of \(\mathcal G(\tau)\) in
Eq.~\eqref{eq:G_eta}, we now derive its low- and high-temperature limits. For large \(\tau\), or equivalently \(Q\ll1\), we write
\begin{equation}
\log\mathcal G(\tau)
=
\sum_{j=1}^{\infty}B_j Q^j .
\end{equation}
If \(j=2^r n\), with \(n\) odd, then the coefficients are
\begin{equation}
B_j
=
\begin{cases}
\displaystyle
\frac{2\sigma(n)}{n},
& r=0,\\[0.8em]
\displaystyle
-\frac{3\sigma(n)}{n},
& r=1,\\[0.8em]
\displaystyle
-\frac{\sigma(n)}{n},
& r\ge2,
\end{cases}
\label{eq:Bj_rule_correct}
\end{equation}
where $\sigma(n)=\sum_{d|n}d .$ Thus the first terms are
\begin{equation}
\begin{split}
\log\mathcal G(\tau)
={}&
2Q
-3Q^2
+\frac83Q^3
-Q^4
+\frac{12}{5}Q^5
-4Q^6
+\frac{16}{7}Q^7
-Q^8  \\
&\quad
+\frac{26}{9}Q^9
-\frac{18}{5}Q^{10}
+\frac{24}{11}Q^{11}
-\frac43Q^{12}
+O(Q^{13}) .
\end{split}
\label{eq:logG_large_tau_correct}
\end{equation}
Unlike the previously written level-two expression, the correctly normalized
crossover contains odd powers of \(Q\).

For the small-\(\tau\) limit, use the modular transformation
\begin{equation}
\eta(i a\tau)
=
(a\tau)^{-1/2}
\eta\!\left(\frac{i}{a\tau}\right),
\qquad a>0.
\end{equation}
Applying this to Eq.~\eqref{eq:G_eta} gives
\begin{equation}
\mathcal G(\tau)
=
2^{1/4}\tau^{-1/4}
\frac{
\eta(i/\tau)^6\,\eta(i/(4\tau))
}{
\eta(2i/\tau)^2\,\eta(i/(2\tau))^{9/2}
}.
\end{equation}
The leading exponential factors cancel.  Therefore, as \(\tau\to0^+\),
\begin{equation}
\log\mathcal G(\tau)
=
-\frac14\log\tau
+\frac14\log2
+o(1).
\label{eq:logG_small_tau_correct}
\end{equation}
Consequently,
\begin{equation}
M_{1/2,L}(\tau L)
=
L(2f_\infty-2\log2)
-\frac14\log L
+2c_\infty
-\frac14\log\tau
+\frac14\log2
+o(1).
\label{eq:M_small_tau_correct}
\end{equation}
The logarithms combine as
\[
-\frac14\log L-\frac14\log\tau
=
-\frac14\log(\tau L)
=
-\frac14\log\beta ,
\]
which is consistent with the crossover interpretation.

\section{Pfaffian representation of the absolute-minor sum}
\label{SecPfaffian}

In this section we show how the sum over absolute values of all square minors of
the correlation matrix can be written as a single Pfaffian.  The construction
has two parts.  First, for an arbitrary matrix \(G\), a universal Pfaffian
identity produces a signed sum of all square minors.  Second, for the staggered
Ising correlation matrix, the signs of the minors are fixed.  By inserting the
corresponding staggered sign into the universal selector matrix, the signed
Pfaffian expansion becomes the desired absolute-minor sum.

Let \(G=(G_{ij})_{i,j=1}^{L}\) be an \(L\times L\) matrix.  For subsets
\(A,B\subseteq\{1,\ldots,L\}\) with \(|A|=|B|\), let \(G_{A,B}\) denote the
submatrix with row set \(A\) and column set \(B\).  We define the absolute-minor
sum
\begin{equation}
\mathcal S(G)
=
\sum_{\substack{A,B\subseteq\{1,\ldots,L\}\\ |A|=|B|}}
\left|\det G_{A,B}\right|,
\label{eq:minor_sum_pf_section}
\end{equation}
where the empty minor is included and has determinant \(1\).  For the Ising
problem studied here, the stabilizer numerator is
\begin{equation}
S_L(\beta)=\mathcal S(G(\beta)).
\label{eq:SL_as_minor_sum}
\end{equation}

\subsection{Antisymmetric lift and minors}
Introduce \(2L\) labels
\[
r_1,c_1,r_2,c_2,\ldots,r_L,c_L .
\]
The antisymmetric lift \(\mathcal A(G)\) is the \(2L\times2L\) matrix defined by
\begin{equation}
\mathcal A(G)_{r_i,c_j}=G_{ij},
\qquad
\mathcal A(G)_{c_j,r_i}=-G_{ij},
\label{eq:A_lift_interleaved}
\end{equation}
with all row-row and column-column entries equal to zero.  Equivalently, in the
numerical ordering \(r_i\leftrightarrow 2i-1\), \(c_i\leftrightarrow 2i\),
\begin{equation}
\mathcal A(G)_{2i-1,2j}=G_{ij},
\qquad
\mathcal A(G)_{2j,2i-1}=-G_{ij}.
\label{eq:A_lift_numeric}
\end{equation}

For \(A=\{a_1<\cdots<a_s\}\) and \(B=\{b_1<\cdots<b_s\}\), define
\begin{equation}
I(A,B):=\{r_a:a\in A\}\cup\{c_b:b\in B\}.
\end{equation}
We also define the inversion statistic
\begin{equation}
\nu(A,B)
:=
\#\{(a,b)\in A\times B:\ a>b\}.
\label{eq:nu_AB_pf_section}
\end{equation}
It counts the inversions between the selected row and column indices, and
therefore records the sign generated when passing from interleaved to grouped
row-column order. The principal Pfaffian selected by \(I(A,B)\) is equal to the corresponding
minor, up to a known ordering sign:
\begin{equation}
\operatorname{Pf}\!\left(\mathcal A(G)_{I(A,B),I(A,B)}\right)
=
(-1)^{\nu(A,B)+\binom{s}{2}}
\det G_{A,B},
\qquad s=|A|=|B|.
\label{eq:minor_as_principal_pf}
\end{equation}
This follows from the standard bipartite identity
\begin{equation}
\operatorname{Pf}
\begin{pmatrix}
0 & M\\
-M^T & 0
\end{pmatrix}
=
(-1)^{\binom{s}{2}}\det M,
\label{eq:bipartite_pf_det}
\end{equation}
together with the extra factor \((-1)^{\nu(A,B)}\), which reorders the selected
interleaved labels into grouped row-column order.

\subsection{Universal selector identity}

The second ingredient is a universal selector matrix which sums principal
Pfaffians.  Let \(\mathcal J_{2L}\) be the antisymmetric matrix
\begin{equation}
(\mathcal J_{2L})_{mn}
=
\begin{cases}
1, & m<n,\\
-1, & m>n,\\
0, & m=n.
\end{cases}
\label{eq:J_selector_pf_section}
\end{equation}
Every even principal submatrix of \(\mathcal J_{2L}\) has Pfaffian equal to one:
\begin{equation}
\operatorname{Pf}\!\left((\mathcal J_{2L})_{J,J}\right)=1,
\qquad |J|\ \text{even}.
\label{eq:selector_pf_one}
\end{equation}
Consequently, for any \(2L\times2L\) antisymmetric matrix \(X\),
\begin{equation}
(-1)^L
\operatorname{Pf}
\begin{pmatrix}
X & I_{2L}\\
-I_{2L} & -\mathcal J_{2L}
\end{pmatrix}
=
\sum_{\substack{J\subseteq\{1,\ldots,2L\}\\ |J|\ {\rm even}}}
\operatorname{Pf}(X_{J,J}).
\label{eq:universal_pf_sum}
\end{equation}
This identity is purely algebraic.  In the Pfaffian expansion, the identity
block pairs each unselected top-layer index with its copy in the auxiliary
layer, while the selector matrix contributes one for the selected indices.

Applying this identity to \(X=\mathcal A(G)\), only subsets containing the same
number of \(r\)-labels and \(c\)-labels contribute, because \(\mathcal A(G)\) is
bipartite.  Therefore the sum reduces to square minors of \(G\).  Combining
Eq.~\eqref{eq:universal_pf_sum} with Eq.~\eqref{eq:minor_as_principal_pf} gives
\begin{equation}
(-1)^L
\operatorname{Pf}
\begin{pmatrix}
\mathcal A(G) & I_{2L}\\
-I_{2L} & -\mathcal J_{2L}
\end{pmatrix}
=
\sum_{\substack{A,B\subseteq\{1,\ldots,L\}\\ |A|=|B|}}
(-1)^{\nu(A,B)+\binom{|A|}{2}}
\det G_{A,B}.
\label{eq:signed_minor_pf_formula}
\end{equation}
This formula holds for every \(L\times L\) matrix \(G\).  It is a signed
minor-sum identity.

\subsection{Staggered sign rule and absolute values}

We now specialize to the staggered Ising correlation matrix \(G(\beta)\).  The
kernel derived in Sec.~\ref{Sec1} contains an alternating factor
\((-1)^{i+j}\).  It is useful to encode this factor by defining
\begin{equation}
\epsilon(A):=\prod_{a\in A}(-1)^a
=
(-1)^{\sum_{a\in A}a}.
\label{eq:epsilon_A_def}
\end{equation}
For the staggered Ising kernel, the required minor sign rule is
\begin{equation}
\operatorname{sgn}\det G_{A,B}(\beta)
=
(-1)^{\nu(A,B)+\binom{|A|}{2}}
\epsilon(A)\epsilon(B).
\label{eq:minor_sign_rule_pf_section}
\end{equation}
Equivalently,
\begin{equation}
\left|\det G_{A,B}(\beta)\right|
=
(-1)^{\nu(A,B)+\binom{|A|}{2}}
\epsilon(A)\epsilon(B)
\det G_{A,B}(\beta).
\label{eq:absolute_minor_sign_rewrite}
\end{equation}
Thus the sign already produced by the interleaved Pfaffian representation must
be supplemented by the staggered factor \(\epsilon(A)\epsilon(B)\).

This additional factor is implemented by conjugating the selector matrix.  Define
\begin{equation}
\mathcal J'_{2L}
=
D_s\,\mathcal J_{2L}\,D_s,
\label{eq:Jprime_def_pf}
\end{equation}
where
\begin{equation}
D_s
=
\operatorname{diag}
\bigl(
(-1)^1,(-1)^1,
(-1)^2,(-1)^2,
\ldots,
(-1)^L,(-1)^L
\bigr).
\label{eq:Ds_def_pf}
\end{equation}
The two entries \((-1)^i,(-1)^i\) correspond to the labels \(r_i\) and \(c_i\).
In the Pfaffian formula we keep the correlation-matrix lift
\(\mathcal A(G)\) fixed and apply the conjugation only to the selector matrix.
For a selected set \(I(A,B)\), this changes the selector contribution by the
factor \(\epsilon(A)\epsilon(B)\). Therefore the signed identity \eqref{eq:signed_minor_pf_formula} becomes an
absolute-minor identity for the Ising kernel:
\begin{equation}
S_L(\beta)
=
\mathcal S(G(\beta))
=
(-1)^L
\operatorname{Pf}K_L(\beta),
\label{eq:absolute_minor_pf_formula}
\end{equation}
with
\begin{equation}
K_L(\beta)
=
\begin{pmatrix}
 \mathcal A(G(\beta)) & I_{2L}\\
-I_{2L} & -\mathcal J'_{2L}
\end{pmatrix}.
\label{eq:K_stag_def_TH}
\end{equation}
Every term in the Pfaffian expansion of \(K_L(\beta)\) is then equal to
\(|\det G_{A,B}(\beta)|\) for some pair of subsets \(A,B\) with
\(|A|=|B|\).  Hence the exponentially large stabilizer numerator is reduced to
the Pfaffian of a \(4L\times4L\) antisymmetric matrix.

\section{Block Toeplitz--Hankel structure of the Pfaffian matrix}
\label{Sec3}

In this section we record the finite block Toeplitz--Hankel structure of the
Pfaffian matrix entering the stabilizer numerator.  The purpose is to make the
finite-size structure explicit before taking scaling limits.  The only inputs
are the critical open-chain kernel \(G(\beta)\) and the Pfaffian matrix
\(K_L(\beta)\) defined in the previous section.

\subsection{Toeplitz--Hankel form of the staggered kernel}

We start from the critical staggered-gauge kernel derived in
Eq.~\eqref{eq:app_G_critical_stag}.  For compactness we write
\[
t_p(\beta)
=
\tanh\!\left(2\beta J\sin\frac{k_p}{2}\right),
\qquad
k_p=\frac{(2p-1)\pi}{2L+1}.
\]
The open boundary condition leads to a decomposition into a Toeplitz part,
depending on \(k-j\), and a Hankel part, depending on \(k+j-1\). Using
\[
2\cos x\sin y=\sin(y+x)+\sin(y-x),
\]
we define the scalar coefficients
\begin{equation}
q_m^{(L,\beta)}
=
(-1)^m
\frac{2}{2L+1}
\sum_{p=1}^{L}
t_p(\beta)
\sin\!\left[\left(m+\frac12\right)k_p\right],
\qquad m\in\mathbb Z .
\label{eq:q_m_stag_TH}
\end{equation}
Then
\begin{equation}
G_{jk}(\beta)
=
q_{k-j}^{(L,\beta)}
-
q_{k+j-1}^{(L,\beta)}.
\label{eq:G_TH_stag_correct}
\end{equation}
With the staggered factor \((-1)^{j+k}\) absorbed into the definition of
\(q_m^{(L,\beta)}\), the two trigonometric terms combine with the relative
minus sign in Eq.~\eqref{eq:G_TH_stag_correct}.

\subsection{Site ordering}

To display the Toeplitz--Hankel structure, we reorder the variables site by
site as
\[
r_1,c_1,\bar r_1,\bar c_1,\,
r_2,c_2,\bar r_2,\bar c_2,\ldots,
r_L,c_L,\bar r_L,\bar c_L .
\]
Here barred labels refer to the auxiliary sector introduced in the Pfaffian
representation of Sec.~\ref{SecPfaffian}. In this ordering, \(K_L(\beta)\) becomes an \(L\times L\) matrix of \(4\times4\)
blocks.  The dependence of the blocks on \(j-i\) and \(i+j-1\) then gives the
Toeplitz and Hankel parts, respectively.

\subsection{Block Toeplitz--Hankel coefficients}

Let
\begin{equation}
\Omega
=
\begin{pmatrix}
0&1\\
-1&0
\end{pmatrix},
\qquad
\mathsf U
=
\begin{pmatrix}
1&1\\
1&1
\end{pmatrix}.
\end{equation}
The staggered selector contributes a \(2\times2\) barred-sector block
\begin{equation}
B_d
=
(-1)^d
\begin{cases}
-\mathsf U, & d>0,\\
-\Omega, & d=0,\\
\mathsf U, & d<0,
\end{cases}
\qquad d=j-i.
\label{eq:Bd_stag_TH}
\end{equation}
The sign \((-1)^d\) comes from the conjugated selector
\(\mathcal J'_{2L}\), while the three cases distinguish whether the column site
lies to the right of, equal to, or to the left of the row site.
The block Toeplitz coefficients are
\begin{equation}
\Phi_d
=
\begin{pmatrix}
C_d & \delta_{d0}I_2\\
-\delta_{d0}I_2 & B_d
\end{pmatrix},
\qquad d=-(L-1),\ldots,L-1,
\label{eq:Phi_d_stag_TH}
\end{equation}
where
\begin{equation}
C_d
=
\begin{pmatrix}
0 & q_d\\
-q_{-d} & 0
\end{pmatrix}.
\label{eq:Cd_stag_TH}
\end{equation}
Here \(C_d\) is the Toeplitz part of the antisymmetric lift
\(\mathcal A(G)\), while the off-diagonal identity blocks come from the
identity matrices in \(K_L(\beta)\).  The remaining contribution from
\(\mathcal A(G)\) is the Hankel part.  Its coefficients are
\begin{equation}
\Psi_m
=
\begin{pmatrix}
-q_m\,\Omega & 0\\
0 & 0
\end{pmatrix},
\qquad m=1,\ldots,2L-1.
\label{eq:Psi_m_stag_TH}
\end{equation}
The minus sign in \eqref{eq:Psi_m_stag_TH} is the sign coming from the
staggered Hankel part in \eqref{eq:G_TH_stag_correct}. Therefore, in site order,
\begin{equation}
K_L(\beta)
=
T_L(\Phi_L)+H_L(\Psi_L),
\label{eq:K_TH_stag_final}
\end{equation}
where
\begin{equation}
\bigl(T_L(\Phi_L)\bigr)_{ij}
=
\Phi_{j-i},
\qquad
\bigl(H_L(\Psi_L)\bigr)_{ij}
=
\Psi_{i+j-1}.
\end{equation}

\subsection{Symbol notation}

Equivalently, introduce
\begin{equation}
 Q_L(z)
=
\sum_{d=-(L-1)}^{L-1}
 q_d\,z^d,
\end{equation}
and
\begin{equation}
R_L(z)
=
\sum_{m=1}^{2L-1}
q_m\,z^m.
\end{equation}
These finite scalar functions encode the Toeplitz and Hankel coefficients,
respectively. The Toeplitz part has the \(4\times4\)
matrix-valued symbol
\begin{equation}
\Phi_L(z)
=
\begin{pmatrix}
0 & Q_L(z) & 1 & 0\\
-Q_L(z^{-1}) & 0 & 0 & 1\\
-1 & 0 & B_L^{11}(z) & B_L^{12}(z)\\
0 & -1 & B_L^{21}(z) & B_L^{22}(z)
\end{pmatrix},
\label{eq:Phi_symbol_stag_TH}
\end{equation}
where
\begin{equation}
B_L(z)
=
\sum_{d=-(L-1)}^{L-1}
B_d z^d.
\end{equation}
The Hankel part is generated by
\begin{equation}
\Psi_L(z)
=
-R_L(z)
\begin{pmatrix}
\Omega & 0\\
0 & 0
\end{pmatrix}.
\label{eq:Psi_symbol_stag_TH}
\end{equation}
Thus, in the staggered gauge, the stabilizer numerator has the exact finite-size
block Toeplitz--Hankel representation
\begin{equation}
S_L(\beta)
=
(-1)^L
\operatorname{Pf}\!\left(T_L(\Phi_L)+H_L(\Psi_L)\right).
\label{eq:SL_TH_stag_final}
\end{equation}
This is an exact rewriting of the finite-\(L\) Pfaffian matrix, not an
asymptotic approximation.

\section{Asymptotic scaling regimes for the Pfaffian formula: Regime I}
\label{Sec4}

In this section we study the fixed-temperature large-\(L\) asymptotics of the
Pfaffian numerator \(S_L(\beta)\).  Throughout this section \(\beta>0\) is kept
fixed as \(L\to\infty\).  We call this fixed-temperature limit Regime I.

The starting point is the Pfaffian representation derived in
Sec.~\ref{SecPfaffian} and the block Toeplitz--Hankel form recorded in
Sec.~\ref{Sec3}..  The purpose of
the present section is to extract the leading extensive term and to define the
constant term in a way that is compatible with the staggered Jordan--Wigner gauge
used throughout the Supplemental Material.  We first reduce the Pfaffian to a
determinant, then identify the limiting determinant symbol, and finally separate
the bulk contribution from the finite-size constant.

\subsection{Schur-reduced determinant}

The \(4L\times4L\) Pfaffian matrix \(K_L(\beta)\) in
Eq.~\eqref{eq:K_stag_def_TH} has a natural two-sector structure: the physical
sector coming from \(\mathcal A(G)\), and the barred auxiliary sector coming
from the selector matrix.  Taking the Schur complement of the barred-sector
selector gives a reduced \(2L\times2L\) antisymmetric matrix.  We define
\begin{equation}
\mathcal D_L(\beta)
:=
\mathcal A(G(\beta))
-
\left(\mathcal J^\prime_{2L}\right)^{-1}.
\label{eq:DL_schur_regimeI}
\end{equation}
Since \(\det(-\mathcal J^\prime_{2L})=1\), the determinant of \(K_L(\beta)\)
reduces to
\begin{equation}
\det K_L(\beta)
=
\det \mathcal D_L(\beta).
\end{equation}
Since \(S_L(\beta)=(-1)^L\operatorname{Pf}K_L(\beta)>0\), we may write
\begin{equation}
\log S_L(\beta)
=
\frac12 \log \det \mathcal D_L(\beta).
\label{eq:logS_detDL_regimeI}
\end{equation}
Thus the asymptotics of the Pfaffian numerator are reduced to the asymptotics of
a \(2L\times2L\) determinant.

The matrix \(\mathcal D_L(\beta)\) is again a \(2\times2\) block
Toeplitz--Hankel matrix.  Its Toeplitz part contains two contributions: the
Toeplitz component of the staggered Majorana kernel and the inverse selector
\((\mathcal J^\prime_{2L})^{-1}\).  Its Hankel part contains the reflected
component of the kernel.  We write this schematically as
\begin{equation}
\mathcal D_L(\beta)
=
T_L(a_L)+H_L(b_L),
\label{eq:DL_TH_regimeI}
\end{equation}
where \(a_L\) and \(b_L\) are \(2\times2\) matrix-valued finite symbols.

\subsection{Limiting reduced symbol}

In the fixed-temperature limit, the finite symbols \(a_L,b_L\) converge to
limiting symbols \(a_\beta,b_\beta\).  The Toeplitz symbol may be represented as
\begin{equation}
a_\beta(e^{i\theta})
=
\begin{pmatrix}
-C(\theta)&1+Q_\beta(e^{i\theta})+C(\theta)\\
-1-Q_\beta(e^{-i\theta})+C(\theta)&-C(\theta)
\end{pmatrix},
\label{eq:a_beta_regimeI}
\end{equation}
where
\begin{equation}
C(\theta)=i\cot\frac{\theta}{2}.
\label{eq:C_theta_regimeI}
\end{equation}
The cotangent term comes from the inverse of the selector, while
\(Q_\beta\) comes from the Toeplitz part of the staggered kernel.  With the
normalization of Eq.~\eqref{eq:app_G_critical_stag}, its boundary value is
\begin{equation}
Q_\beta(e^{i\theta})
=
i e^{-i\theta/2}
\tanh\!\left(2\beta J\sin\frac{\theta}{2}\right).
\label{eq:Q_beta_regimeI}
\end{equation}
The corresponding Hankel symbol \(b_\beta\) is determined by the reflected part
of the same kernel.  Its explicit form is not needed for the leading extensive
term, but it does enter the constant term.

Although the matrix representative \eqref{eq:a_beta_regimeI} contains the
cotangent term \(C(\theta)\), the singularity cancels in the determinant of the
symbol.  This cancellation is useful because the leading bulk term depends only
on the scalar determinant of the Toeplitz symbol.  Define
\begin{equation}
s(\theta)=\sin\frac{\theta}{2},
\qquad
t_\beta(\theta)
=
\tanh\!\left(2\beta J s(\theta)\right).
\end{equation}
Then
\begin{equation}
\Delta_\beta(\theta)
:=
\det a_\beta(e^{i\theta})
=
1
+
\frac{2t_\beta(\theta)}{s(\theta)}
+
t_\beta(\theta)^2 .
\label{eq:Delta_beta_regimeI}
\end{equation}
The apparent singularity at \(\theta=0\) is removable because
\[
t_\beta(\theta)
\sim
2\beta J s(\theta),
\qquad
\theta\to0.
\]
Therefore
\begin{equation}
\Delta_\beta(0)=1+4\beta J.
\label{eq:Delta_zero_regimeI}
\end{equation}
At the other endpoint,
\begin{equation}
\Delta_\beta(\pi)
=
1+2\tanh(2\beta J)+\tanh^2(2\beta J).
\label{eq:Delta_pi_regimeI}
\end{equation}
For every fixed \(\beta>0\), \(\Delta_\beta(\theta)\) is positive on
\([0,\pi]\).

\subsection{Bulk coefficient}

The leading extensive contribution to \(\log\det \mathcal D_L(\beta)\) is
determined by the determinant of the limiting Toeplitz symbol.  Since
\(\log S_L(\beta)=\frac12\log\det\mathcal D_L(\beta)\), the Regime-I bulk
coefficient is
\begin{equation}
f_0(\beta)
=
\frac{1}{2\pi}
\int_0^\pi
\log \Delta_\beta(\theta)\,\dd\theta=
\frac{1}{4\pi}
\int_0^{2\pi}
\log \Delta_\beta(\theta)\,\dd\theta  .
\label{eq:f0_regimeI}
\end{equation}
where in the second equality we used the even \(2\pi\)-periodic extension of \(\Delta_\beta\).

The cancellation in \eqref{eq:Delta_beta_regimeI} is important: at fixed
\(\beta>0\), the determinant-level symbol has no nonintegrable endpoint
singularity.  Thus the expected Regime-I expansion contains no
Fisher--Hartwig logarithmic term.  In this regime the leading contribution is
therefore purely extensive, with possible corrections starting at the constant
level.

\begin{figure}[t]
    \centering
    \includegraphics[width=0.5\textwidth]{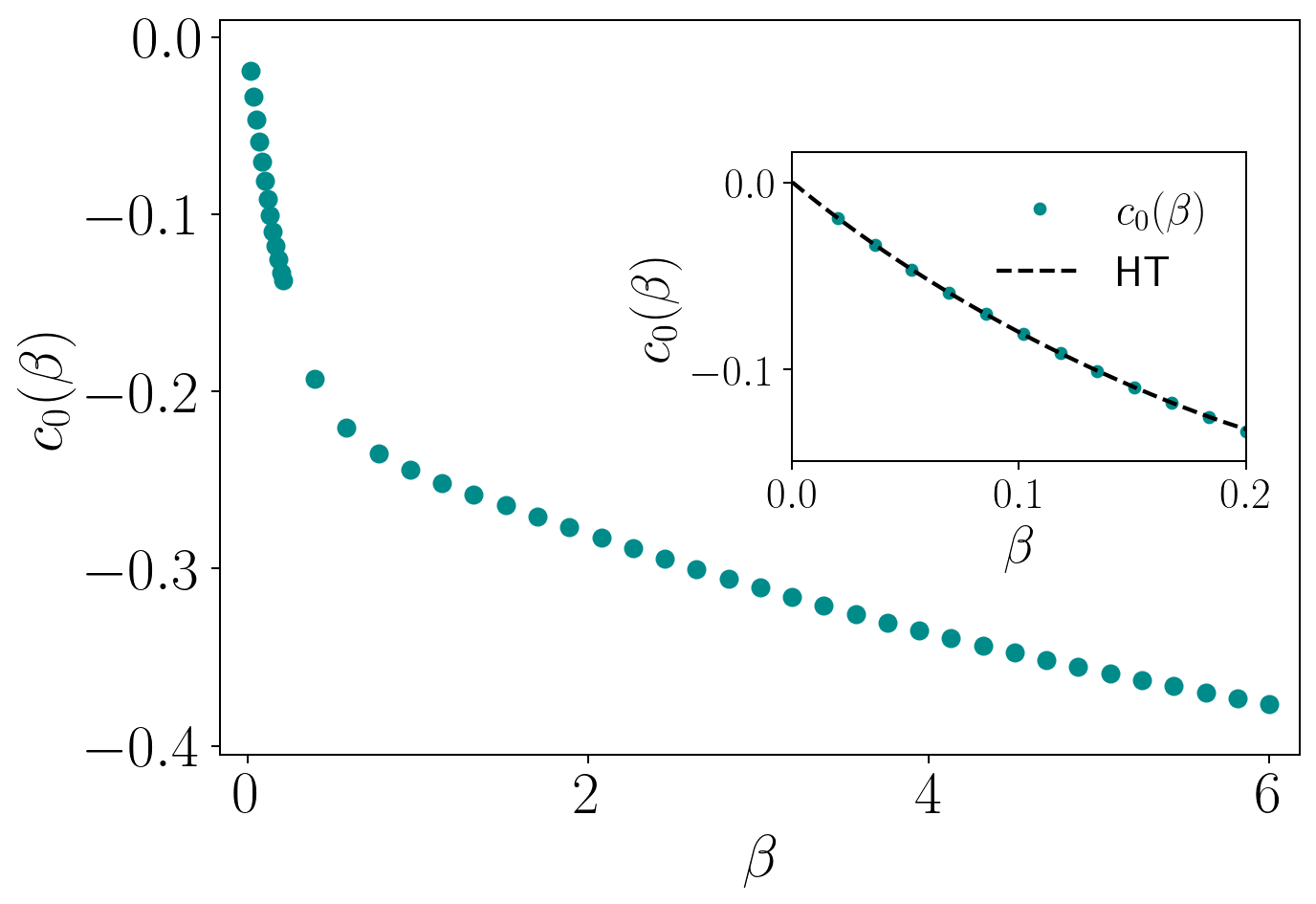}
    \caption{
    Regime-I finite part \(c_0(\beta)\) as a function of the inverse temperature
    \(\beta\).  For each value of \(\beta\), the finite-size quantity
    \(c_L(\beta)=\log S_L(\beta)-L f_0(\beta)\) is computed from the exact
    Schur-reduced determinant representation of the Pfaffian numerator and then
    extrapolated to \(L\to\infty\) by fitting in powers of \(1/L\). 
    }
    \label{fig:c0_beta_regimeI}
\end{figure}

\subsection{Definition of the constant term}

We define the finite-\(L\) renormalized constant by subtracting the bulk term
from the exact determinant expression:
\begin{equation}
c_L(\beta)
:=
\frac12\log\det\mathcal D_L(\beta)
-
L f_0(\beta).
\label{eq:cL_regimeI}
\end{equation}
The Regime-I constant is then
\begin{equation}
c_0(\beta)
:=
\lim_{L\to\infty}c_L(\beta),
\label{eq:c0_regimeI}
\end{equation}
whenever the limit exists.  Equivalently,
\begin{equation}
c_0(\beta)
=
\frac12
\lim_{L\to\infty}
\left[
\log\det\mathcal D_L(\beta)
-
\frac{L}{\pi}
\int_0^\pi
\log\Delta_\beta(\theta)\,\dd\theta
\right].
\label{eq:c0_renormalized_regimeI}
\end{equation}
This definition keeps the full block Toeplitz--Hankel structure at finite \(L\)
and subtracts only the universal bulk term determined by \(\Delta_\beta\). The behavior of \(c_0(\beta)\) is shown in Fig.~\ref{fig:c0_beta_regimeI}.

\begin{figure}[t]
    \centering

    \includegraphics[width=0.4\textwidth]{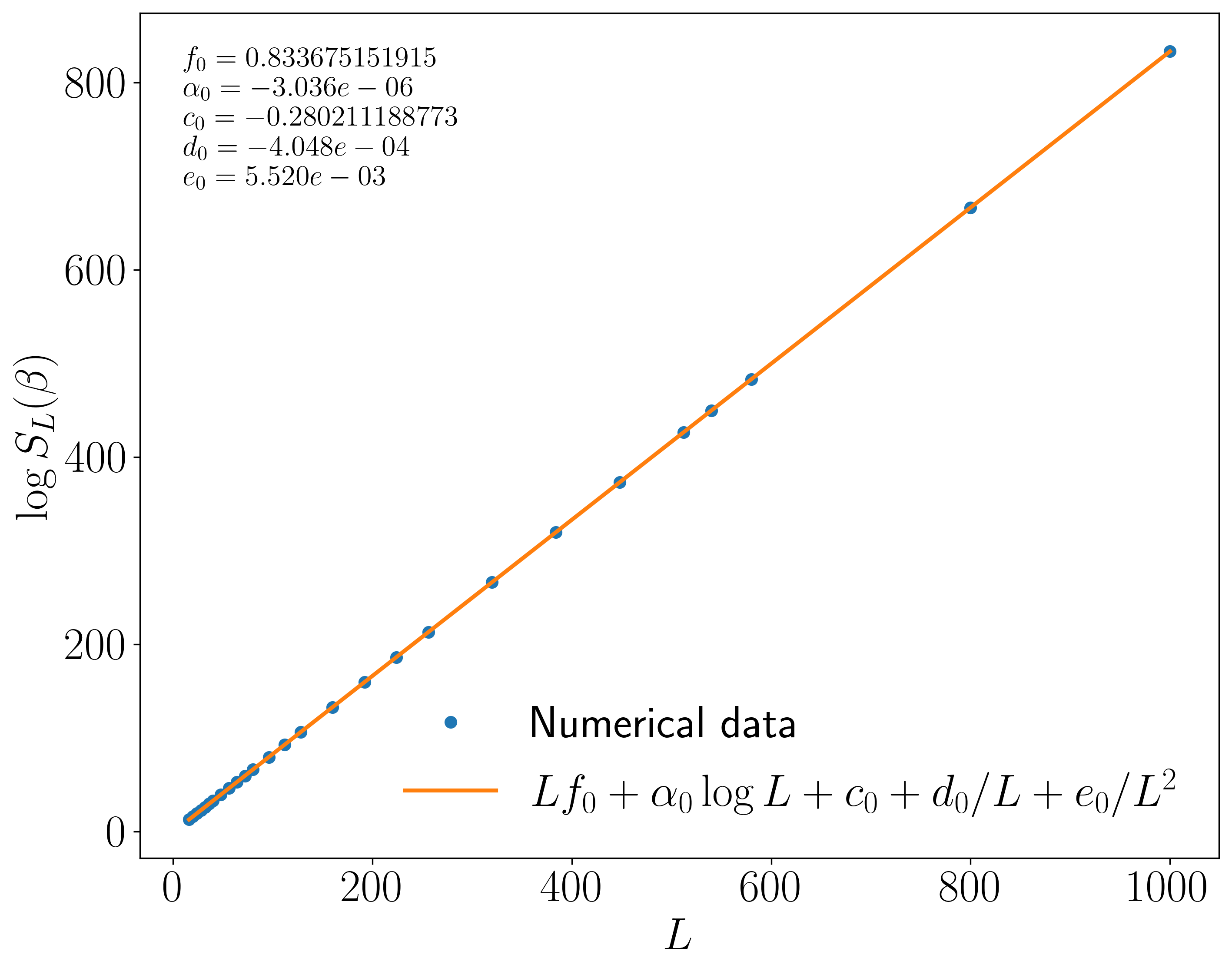}
    \includegraphics[width=0.4\textwidth]{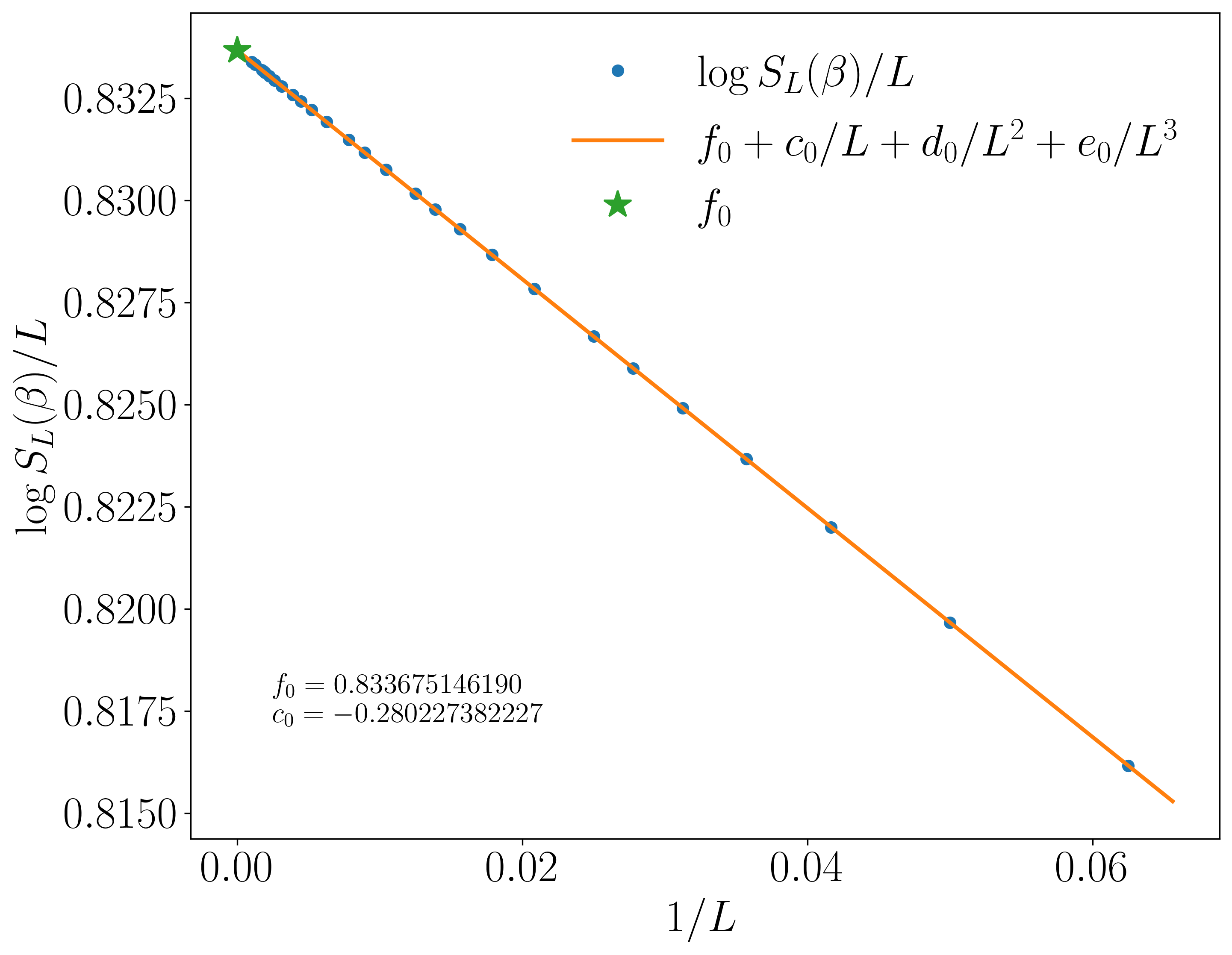}

  \caption{
Numerical confirmation of the Regime-I fixed-temperature scaling form for the
staggered-gauge Pfaffian numerator at \(\beta=2\) and \(J=1\).
Left: finite-size scaling of \(\log S_L(\beta)\) using the Regime-I fitting
form.  The fitted logarithmic coefficient is consistent with
\(\alpha_0(\beta)=0\).  Right: thermodynamic-limit extrapolation of
\(\log S_L(\beta)/L\) against \(1/L\).  The intercept gives the bulk coefficient
\(f_0(\beta)\), in agreement with the analytic symbol prediction.
}
    \label{fig:regimeI_scaling_checks}
\end{figure}

The expected fixed-temperature expansion is therefore
\begin{equation}
\log S_L(\beta)
=
L f_0(\beta)
+
c_0(\beta)
+
\frac{d_0(\beta)}{L}
+
o(L^{-1}).
\label{eq:S_regimeI_expansion}
\end{equation}
More generally, one may fit the finite-size data using
\begin{equation}
\log S_L(\beta)
=
L f_0(\beta)
+
\alpha_0(\beta)\log L
+
c_0(\beta)
+
\frac{d_0(\beta)}{L}
+
\frac{e_0(\beta)}{L^2}
+\cdots .
\label{eq:S_regimeI_fit}
\end{equation}
The regular-symbol prediction is $\alpha_0(\beta)=0$. Figure~\ref{fig:regimeI_scaling_checks} illustrates this Regime-I finite-size
scaling at \(\beta=2\), showing both the direct fit of \(\log S_L(\beta)\) and
the extrapolation of \(\log S_L(\beta)/L\) to the bulk coefficient \(f_0(\beta)\).

\subsection{Contribution to the normalized stabilizer quantity}

The Pfaffian numerator gives the unnormalized contribution.  The mixed-state
stabilizer quantity also contains the purity normalization discussed earlier.
Combining \eqref{eq:S_regimeI_expansion} with the purity expansion derived in
Sec.~\ref{Sec2}, the normalized mixed-state quantity has the fixed-temperature
form
\begin{equation}
M_{1/2,L}(\beta)
=
L\left(2f_0(\beta)-2p_0(\beta)\right)
+
\left(2c_0(\beta)-2p_1(\beta)\right)
+
o(1),
\label{eq:M_regimeI_from_S}
\end{equation}
where \(p_0(\beta)\) and \(p_1(\beta)\) are given in
Eqs.~\eqref{eq:p0_beta} and \eqref{eq:p1_beta}.  Thus the Regime-I extensive
coefficient of the normalized quantity is
\begin{equation}
m_0(\beta)
=
2f_0(\beta)-2p_0(\beta),
\label{eq:m0_regimeI}
\end{equation}
and the Regime-I constant is
\begin{equation}
m_1(\beta)
=
2c_0(\beta)-2p_1(\beta).
\label{eq:m1_regimeI}
\end{equation}

\subsection{Differential identity for the constant}

It is useful to introduce the determinant-level constant
\begin{equation}
C_D(\beta):=2c_0(\beta).
\end{equation}
This removes the factor \(1/2\) coming from the Pfaffian-to-determinant
relation.  From \eqref{eq:c0_renormalized_regimeI},
\begin{equation}
C_D(\beta)
=
\lim_{L\to\infty}
\left[
\log\det\mathcal D_L(\beta)
-
L\ell_\beta
\right],
\label{eq:CD_regimeI}
\end{equation}
where
\begin{equation}
\ell_\beta:=2f_0(\beta)
=
\frac1\pi
\int_0^\pi
\log\Delta_\beta(\theta)\,\dd\theta .
\label{eq:ell_beta_regimeI}
\end{equation}
Differentiating formally gives
\begin{equation}
C_D'(\beta)
=
\lim_{L\to\infty}
\left[
\Tr\left(
\mathcal D_L(\beta)^{-1}
\partial_\beta \mathcal D_L(\beta)
\right)
-
L\ell_\beta'
\right].
\label{eq:CD_prime_regimeI}
\end{equation}
Only the correlation matrix \(G(\beta)\) depends on \(\beta\), so
\begin{equation}
\partial_\beta \mathcal D_L(\beta)
=
\mathcal A\!\left(\partial_\beta G(\beta)\right).
\label{eq:dDL_regimeI}
\end{equation}

The bulk derivative is explicit.  From
\eqref{eq:Delta_beta_regimeI},
\begin{equation}
\partial_\beta \Delta_\beta(\theta)
=
4J\operatorname{sech}^2\!\left(2\beta J\sin\frac{\theta}{2}\right)
\left[
1+
\sin\frac{\theta}{2}\,
t_\beta(\theta)
\right].
\label{eq:dDelta_regimeI}
\end{equation}
Therefore
\begin{equation}
\ell_\beta'
=
\frac1\pi
\int_0^\pi
\frac{
4J\operatorname{sech}^2\!\left(2\beta J\sin\frac{\theta}{2}\right)
\left[
1+
\sin\frac{\theta}{2}\,
t_\beta(\theta)
\right]
}{
\Delta_\beta(\theta)
}
\,\dd\theta .
\label{eq:ell_prime_regimeI}
\end{equation}
At \(\beta=0\), one has \(\det\mathcal D_L(0)=1\) and
\(\Delta_0(\theta)=1\), so both the determinant term and the bulk term in
Eq.~\eqref{eq:c0_renormalized_regimeI} vanish.  Thus \(c_0(0)=0\), and
\begin{equation}
c_0(\beta)
=
\frac12
\int_0^\beta
C_D'(u)\,\dd u .
\label{eq:c0_integral_regimeI}
\end{equation}

\subsection{Remarks on the constant term}

The scalar function \(\Delta_\beta(\theta)=\det a_\beta(e^{i\theta})\)
determines the leading coefficient \(f_0(\beta)\), but it does not determine the
constant \(c_0(\beta)\).  The constant depends on the full block
Toeplitz--Hankel structure, namely on both limiting symbols \(a_\beta\) and
\(b_\beta\), not only on the scalar determinant of \(a_\beta\).

Equivalently, a scalar strong-Szegő expression built only from the Fourier
coefficients of \(\log\Delta_\beta\) gives the correct bulk term but not, in
general, the correct constant.  This is the standard matrix-symbol phenomenon:
the constant term depends on the noncommutative factorization data of the full
matrix symbol, and in the present problem it also receives the Hankel reflection
contribution.

For practical purposes, the safest numerical definition of \(c_0(\beta)\) is
the finite-section limit \eqref{eq:c0_regimeI}.  Assuming the leading correction
is \(O(L^{-1})\), one may use the Richardson estimate
\begin{equation}
c_0(\beta)
\approx
2c_{2L}(\beta)-c_L(\beta).
\label{eq:richardson_two_size_regimeI}
\end{equation}
If the next correction is included and data at \(L,2L,4L\) are available, then
\begin{equation}
c_0(\beta)
\approx
\frac13 c_L(\beta)
-
2c_{2L}(\beta)
+
\frac83 c_{4L}(\beta).
\label{eq:richardson_three_size_regimeI}
\end{equation}

\section{Asymptotic scaling regimes for the Pfaffian formula: Regime II(high temperature)}\label{Sec5}

In this section we study the high-temperature expansion of the Pfaffian
numerator \(S_L(\beta)\).  In contrast to Regime I, where \(\beta>0\) is fixed
and \(L\to\infty\), Regime II treats \(\beta\to0^+\) perturbatively.  The
expansion is performed at finite \(L\), and only afterwards the coefficients are
split into bulk and boundary contributions.

\subsection{Small-\texorpdfstring{\(\beta\)}{beta} expansion of the staggered kernel}

The expansion
\begin{equation}
\tanh x
=
x-\frac{x^3}{3}
+\frac{2x^5}{15}
-\frac{17x^7}{315}
+\frac{62x^9}{2835}
-\frac{1382x^{11}}{155925}
+O(x^{13})
\end{equation}
implies that only odd powers of \(\beta\) appear in \(G(\beta)\).  We write
\begin{equation}
G(\beta)
=
\sum_{r\ge0}
\beta^{2r+1}G_{2r+1}.
\label{eq:G_regimeII_odd}
\end{equation}

It is useful to introduce the lower bidiagonal matrix
\begin{equation}
\mathcal M_L
=
\begin{pmatrix}
1&0&0&\cdots&0\\
1&1&0&\cdots&0\\
0&1&1&\cdots&0\\
\vdots&&\ddots&\ddots&\vdots\\
0&\cdots&0&1&1
\end{pmatrix}.
\label{eq:Mplus_regimeII}
\end{equation}
The positive subdiagonal is a direct consequence of the staggered factor
\((-1)^{j+k}\) in \(G_{jk}\).  Using
\[
2\sin\frac{k}{2}\cos\!\left[\left(j-\frac12\right)k\right]
=
\sin(jk)-\sin((j-1)k)
\]
together with the staggered factor, one obtains $G_1=\mathcal M_L.$ More generally, if
\begin{equation}
\tanh x=\sum_{r\ge0}\tau_r x^{2r+1},
\end{equation}
then
\begin{equation}
G_{2r+1}
=
\tau_r\,
\mathcal M_L
\left(\mathcal M_L^T\mathcal M_L\right)^r.
\label{eq:G_coeff_regimeII_stag}
\end{equation}
In particular,
\begin{align}
G_1&=\mathcal M_L,\\
G_3&=-\frac13\mathcal M_L(\mathcal M_L^T\mathcal M_L),\\
G_5&=\frac{2}{15}\mathcal M_L(\mathcal M_L^T\mathcal M_L)^2,\\
G_7&=-\frac{17}{315}\mathcal M_L(\mathcal M_L^T\mathcal M_L)^3,\\
G_9&=\frac{62}{2835}\mathcal M_L(\mathcal M_L^T\mathcal M_L)^4,\\
G_{11}&=-\frac{1382}{155925}\mathcal M_L(\mathcal M_L^T\mathcal M_L)^5.
\end{align}
For general \(J\), each coefficient of order \(\beta^n\) is multiplied by
\(J^n\).  In the coefficient tables below we set \(J=1\).

\subsection{Leading path model}

At leading order,
\begin{equation}
G(\beta)
=
\beta\mathcal M_L
+
O(\beta^3).
\end{equation}
The nonzero pattern of \(\mathcal M_L\) is again the bipartite path
\[
r_1-c_1-r_2-c_2-\cdots-r_L-c_L.
\]
Therefore the leading path-counting problem is unchanged combinatorially: the
number of nonzero \(s\times s\) minors of \(\mathcal M_L\) is the number of
size-\(s\) matchings in a path with \(2L\) vertices,
\begin{equation}
N_{L,s}
=
\binom{2L-s}{s}.
\end{equation}
The difference from the older normalization is the weight of each matched edge:
with the present convention the leading weight is \(\beta\), not \(\beta/2\).
Thus the linearized numerator is
\begin{equation}
S_L^{\rm lin}(\beta)
=
\sum_{s=0}^{L}
\binom{2L-s}{s}\beta^s.
\label{eq:Slin_regimeII}
\end{equation}
Equivalently,
\begin{equation}
S_L^{\rm lin}(\beta)=P_{2L}(\beta),
\qquad
P_n(x)=P_{n-1}(x)+xP_{n-2}(x),
\qquad
P_0(x)=P_1(x)=1.
\end{equation}
Writing
\begin{equation}
r_\pm
=
\frac{1\pm\sqrt{1+4\beta}}{2},
\end{equation}
one obtains
\begin{equation}
S_L^{\rm lin}(\beta)
=
\frac{r_+^{2L+1}-r_-^{2L+1}}{\sqrt{1+4\beta}}.
\label{eq:Slin_closed_regimeII}
\end{equation}
Therefore
\begin{align}
f_{\rm lin}(\beta)
&=
2\log\left(\frac{1+\sqrt{1+4\beta}}{2}\right),\\
c_{\rm lin}(\beta)
&=
\log\left(\frac{1+\sqrt{1+4\beta}}{2}\right)
-\frac12\log(1+4\beta).
\end{align}
Their small-\(\beta\) expansions begin as
\begin{align}
f_{\rm lin}(\beta)
&=
2\beta-3\beta^2+\frac{20}{3}\beta^3
-\frac{35}{2}\beta^4+O(\beta^5),\\
c_{\rm lin}(\beta)
&=
-\beta+\frac52\beta^2-\frac{22}{3}\beta^3
+\frac{93}{4}\beta^4+O(\beta^5).
\end{align}
The full kernel differs from this path approximation starting at order
\(\beta^3\).

\subsection{Schur-complement trace expansion}

The direct minor expansion becomes inefficient beyond the first few orders.  The
systematic expansion follows from the Schur-complement determinant
\eqref{eq:logS_detDL_regimeI}.  Since
\[
\mathcal D_L(0)=-(\mathcal J'_{2L})^{-1},
\]
we may write
\begin{equation}
\mathcal D_L(\beta)
=
-(\mathcal J'_{2L})^{-1}
\left[
I-\mathcal J'_{2L}\mathcal A(G(\beta))
\right].
\end{equation}
Because \(\det[-(\mathcal J'_{2L})^{-1}]=1\), this gives
\begin{equation}
\log S_L(\beta)
=
\frac12
\operatorname{Tr}
\log\left[
I-\mathcal J'_{2L}\mathcal A(G(\beta))
\right].
\label{eq:trace_log_regimeII}
\end{equation}
For each odd \(j=2r+1\), define
\begin{equation}
A_j:=\mathcal A(G_j),
\qquad
X_j:=-\mathcal J'_{2L}A_j.
\label{eq:Xj_regimeII_stag}
\end{equation}
Then
\begin{equation}
-\mathcal J'_{2L}\mathcal A(G(\beta))
=
\sum_{r\ge0}
\beta^{2r+1}X_{2r+1}.
\end{equation}
Hence
\begin{equation}
\log S_L(\beta)
=
\frac12
\operatorname{Tr}
\log\left[
I+\sum_{r\ge0}\beta^{2r+1}X_{2r+1}
\right].
\label{eq:trace_log_series_regimeII}
\end{equation}

Writing
\begin{equation}
\log S_L(\beta)=\sum_{n\ge1}\ell_n(L)\beta^n,
\end{equation}
the finite-\(L\) coefficients are
\begin{equation}
\ell_n(L)
=
\frac12
\sum_{m=1}^{n}
\frac{(-1)^{m+1}}{m}
\sum_{\substack{j_1+\cdots+j_m=n\\ j_a\ge1\ {\rm odd}}}
\operatorname{Tr}
\left(
X_{j_1}X_{j_2}\cdots X_{j_m}
\right).
\label{eq:elln_regimeII}
\end{equation}
This expression is exact for every finite \(L\).  For example,
\begin{align}
\ell_1&=\frac12\operatorname{Tr}X_1,\\
\ell_2&=-\frac14\operatorname{Tr}X_1^2,\\
\ell_3&=\frac12\operatorname{Tr}X_3+\frac16\operatorname{Tr}X_1^3,\\
\ell_4&=-\frac12\operatorname{Tr}(X_1X_3)-\frac18\operatorname{Tr}X_1^4,\\
\ell_5&=\frac12\operatorname{Tr}X_5
+\frac12\operatorname{Tr}(X_1^2X_3)
+\frac1{10}\operatorname{Tr}X_1^5,\\
\ell_6&=-\frac12\operatorname{Tr}(X_1X_5)
-\frac14\operatorname{Tr}X_3^2
-\frac12\operatorname{Tr}(X_1^3X_3)
-\frac1{12}\operatorname{Tr}X_1^6.
\end{align}

\subsection{Bulk-boundary split and coefficients through order twelve}

For fixed perturbative order \(n\), the coefficient \(\ell_n(L)\) becomes
linear in \(L\) once \(L\) is large compared with \(n\).  This follows from the
finite propagation range of the banded matrices \(G_{2r+1}\).  We therefore
write
\begin{equation}
\ell_n(L)
=
L f_n+c_n
\qquad
(L\ge L_n).
\label{eq:ell_split_regimeII}
\end{equation}
Equivalently,
\begin{equation}
f_{\rm HT}(\beta)=\sum_{n\ge1}f_n\beta^n,
\qquad
c_{\rm HT}(\beta)=\sum_{n\ge1}c_n\beta^n.
\end{equation}
Through order \(\beta^{12}\), the staggered-gauge coefficients are as follows:
\begin{equation}
\begin{array}{c|c|c}
n & f_n & c_n\\
\hline
1 & 2 & -1\\
2 & -3 & \frac52\\
3 & \frac{16}{3} & -\frac{20}{3}\\
4 & -\frac{85}{6} & \frac{259}{12}\\
5 & \frac{128}{3} & -\frac{1108}{15}\\
6 & -\frac{5908}{45} & \frac{11462}{45}\\
7 & \frac{131584}{315} & -\frac{31264}{35}\\
8 & -\frac{192261}{140} & \frac{892539}{280}\\
9 & \frac{13101056}{2835} & -\frac{32656196}{2835}\\
10 & -\frac{224339588}{14175} & \frac{119145142}{2835}\\
11 & \frac{8574681088}{155925} & -\frac{24100725664}{155925}\\
12 & -\frac{30157148242}{155925} & \frac{89240450263}{155925}
\end{array}
\label{eq:regimeII_coeff_table}
\end{equation}
Thus
\begin{align}
f_{\rm HT}(\beta)
={}&
2\beta
-3\beta^2
+\frac{16}{3}\beta^3
-\frac{85}{6}\beta^4
+\frac{128}{3}\beta^5
-\frac{5908}{45}\beta^6
+\frac{131584}{315}\beta^7
\notag\\
&-\frac{192261}{140}\beta^8
+\frac{13101056}{2835}\beta^9
-\frac{224339588}{14175}\beta^{10}
+\frac{8574681088}{155925}\beta^{11}
\notag\\
&-\frac{30157148242}{155925}\beta^{12}
+O(\beta^{13}),
\label{eq:fHT_regimeII_stag}
\end{align}
and
\begin{align}
c_{\rm HT}(\beta)
={}&
-\beta
+\frac52\beta^2
-\frac{20}{3}\beta^3
+\frac{259}{12}\beta^4
-\frac{1108}{15}\beta^5
+\frac{11462}{45}\beta^6
-\frac{31264}{35}\beta^7
\notag\\
&+\frac{892539}{280}\beta^8
-\frac{32656196}{2835}\beta^9
+\frac{119145142}{2835}\beta^{10}
-\frac{24100725664}{155925}\beta^{11}
\notag\\
&+\frac{89240450263}{155925}\beta^{12}
+O(\beta^{13}).
\label{eq:cHT_regimeII_stag}
\end{align}

\subsection{Comparison with Regime I}

The Regime-II bulk series must agree with the small-\(\beta\) expansion of the
Regime-I bulk coefficient.  With the present normalization, the Regime-I
determinant-level symbol gives
\begin{equation}
\Delta_\beta(\theta)
=
1
+
\frac{2t_\beta(\theta)}{\sin(\theta/2)}
+
t_\beta(\theta)^2,
\qquad
t_\beta(\theta)
=
\tanh\!\left(2\beta J\sin\frac{\theta}{2}\right).
\label{eq:Delta_regimeII_consistent}
\end{equation}
Therefore
\begin{equation}
f_0(\beta)
=
\frac{1}{2\pi}
\int_0^\pi
\log\Delta_\beta(\theta)\,d\theta.
\label{eq:f0_regimeII_compare}
\end{equation}
Expanding \eqref{eq:f0_regimeII_compare} at small \(\beta\) reproduces
\(f_{\rm HT}(\beta)\) above.  In particular,
\begin{equation}
f_0(\beta)
=
f_{\rm HT}(\beta)
\end{equation}
as a formal high-temperature expansion.

Similarly, the boundary series \(c_{\rm HT}(\beta)\) is the high-temperature
expansion of the Regime-I constant \(c_0(\beta)\):
\begin{equation}
c_0(\beta)
=
c_{\rm HT}(\beta)
\end{equation}
as a formal small-\(\beta\) series.  This gives a perturbative expansion of the
Toeplitz--Hankel constant which is otherwise defined nonperturbatively by the
finite-section limit in Regime I.

\section{Asymptotic scaling regimes for the Pfaffian formula: Regime III(the saturated low-temperature regime)}\label{Sec6}

In this section we study the saturated low-temperature regime of the Pfaffian
numerator \(S_L(\beta)\).  This regime is obtained when all single-particle
modes are saturated.  Since the smallest open-chain momentum is of order
\(L^{-1}\), the thermodynamic saturated regime is not simply fixed large
\(\beta\).  Rather, the relevant scaling condition is
\begin{equation}
L\to\infty,
\qquad
\frac{\beta}{L}\to\infty .
\label{eq:regimeIII_condition}
\end{equation}
Equivalently, the thermal corrections vanish on the scale
\(\exp(-\pi J\beta/L)\).

Throughout this section we keep the staggered-gauge convention used in the
previous sections.  Thus \(G(\beta)\) includes the factor \((-1)^{j+k}\), and
the Pfaffian matrix is built using the staggered selector
\(\mathcal J'_{2L}\).

\subsection{Saturated staggered kernel}

We start from the critical staggered-gauge kernel derived in
Eq.~\eqref{eq:app_G_critical_stag}. For fixed \(L\), the saturated kernel is obtained by sending
\(\beta\to\infty\):
\begin{equation}
G_{jk}^{(\infty)}
=
(-1)^{j+k}
\frac{4}{2L+1}
\sum_{p=1}^{L}
\cos\!\left[\left(j-\frac12\right)k_p\right]
\sin(k k_p).
\label{eq:G_regimeIII_infty}
\end{equation}
The saturated numerator is then
\begin{equation}
S_L(\infty)
=
\sum_{\substack{A,B\subseteq\{1,\ldots,L\}\\ |A|=|B|}}
\left|\det G^{(\infty)}_{A,B}\right|.
\label{eq:S_regimeIII_infty}
\end{equation}
Assuming the same minor sign rule as before, this is represented by the
Pfaffian formula
\begin{equation}
S_L(\infty)
=
(-1)^L\operatorname{Pf}K_L(\infty),
\end{equation}
where
\begin{equation}
K_L(\infty)
=
\begin{pmatrix}
\mathcal A(G^{(\infty)}) & I_{2L}\\
-I_{2L} & -\mathcal J'_{2L}
\end{pmatrix}.
\label{eq:K_regimeIII_infty}
\end{equation}
Equivalently, after the Schur complement,
\begin{equation}
\log S_L(\infty)
=
\frac12
\log\det \mathcal D_L(\infty),
\qquad
\mathcal D_L(\infty)
=
\mathcal A(G^{(\infty)})
-
(\mathcal J'_{2L})^{-1}.
\label{eq:DL_regimeIII_infty}
\end{equation}

\subsection{Toeplitz--Hankel form at saturation}

Using
\[
2\cos x\sin y=\sin(y+x)+\sin(y-x),
\]
the saturated kernel has a Toeplitz--Hankel form.  Define
\begin{equation}
q_m^{(\infty,L)}
=
(-1)^m
\frac{2}{2L+1}
\sum_{p=1}^{L}
\sin\!\left[\left(m+\frac12\right)k_p\right],
\qquad m\in\mathbb Z .
\label{eq:q_regimeIII_finite}
\end{equation}
Then
\begin{equation}
G_{jk}^{(\infty)}
=
q_{k-j}^{(\infty,L)}
-
q_{k+j-1}^{(\infty,L)}.
\label{eq:G_regimeIII_TH}
\end{equation}
The minus sign is the same staggered Hankel sign already present at finite
temperature.

The finite sum in \eqref{eq:q_regimeIII_finite} can be evaluated explicitly.
Using
\[
\sum_{p=1}^{L}\sin((2p-1)x)
=
\frac{\sin^2(Lx)}{\sin x},
\]
with $x=(m+1/2)\pi/(2L+1),$
we obtain
\begin{equation}
q_m^{(\infty,L)}
=
(-1)^m
\frac{2}{2L+1}
\frac{
\sin^2\!\left[
\dfrac{L(m+\frac12)\pi}{2L+1}
\right]
}{
\sin\!\left[
\dfrac{(m+\frac12)\pi}{2L+1}
\right]
}.
\label{eq:q_regimeIII_explicit}
\end{equation}
For fixed \(m\) and \(L\to\infty\),
\begin{equation}
q_m^{(\infty,L)}
\longrightarrow
q_m^{(\infty)}
=
\frac{(-1)^m}{\pi(m+\frac12)}.
\label{eq:q_regimeIII_limit}
\end{equation}
The slow \(1/(m+1/2)\) decay is the origin of the endpoint singularity in
the saturated symbol.

Equivalently, in the shifted Toeplitz variable used in the reduced symbol, the
Abel-summed generating function is
\begin{equation}
Q_\infty(e^{i\theta})
:=
\sum_{m\in\mathbb Z}
q_m^{(\infty)}(-e^{i\theta})^m
=
i e^{-i\theta/2},
\qquad
0<\theta<2\pi .
\label{eq:Q_regimeIII_infty}
\end{equation}
This is the \(\beta\to\infty\) limit of the Regime-I symbol
\[
Q_\beta(e^{i\theta})
=
i e^{-i\theta/2}
\tanh\!\left(2\beta J\sin\frac{\theta}{2}\right).
\]

\subsection{Fixed-\(L\) approach to saturation}

Let
\begin{equation}
s_p=\sin\frac{k_p}{2},
\qquad
\mu_p=4J s_p
=
4J\sin\frac{(2p-1)\pi}{4L+2}.
\label{eq:mu_regimeIII}
\end{equation}
Then
\begin{equation}
\tanh(2\beta J s_p)
=
1-2e^{-\mu_p\beta}
+2e^{-2\mu_p\beta}
-2e^{-3\mu_p\beta}
+\cdots .
\label{eq:tanh_regimeIII_lowT}
\end{equation}
The smallest exponent is
\begin{equation}
\mu_1
=
4J\sin\frac{\pi}{4L+2}
=
\frac{\pi J}{L}
+
O(L^{-2}).
\label{eq:mu1_regimeIII}
\end{equation}

Define the staggered rank-one mode matrices
\begin{equation}
(B_p^{(L)})_{jk}
=
(-1)^{j+k}
\frac{4}{2L+1}
\cos\!\left[\left(j-\frac12\right)k_p\right]
\sin(k k_p).
\label{eq:Bp_regimeIII}
\end{equation}
Then
\begin{equation}
G(\beta)
=
\sum_{p=1}^{L}
\tanh(2\beta J s_p)\,B_p^{(L)},
\qquad
G^{(\infty)}
=
\sum_{p=1}^{L}
B_p^{(L)}.
\end{equation}
Therefore
\begin{equation}
G(\beta)
=
G^{(\infty)}
+
2\sum_{n=1}^{\infty}
(-1)^n
\sum_{p=1}^{L}
e^{-n\mu_p\beta}
B_p^{(L)}.
\label{eq:G_regimeIII_lowT}
\end{equation}
The first few thermal scales are
\begin{equation}
G(\beta)-G^{(\infty)}
=
-2e^{-\mu_1\beta}B_1^{(L)}
+
2e^{-2\mu_1\beta}B_1^{(L)}
-
2e^{-\mu_2\beta}B_2^{(L)}
+\cdots .
\label{eq:G_regimeIII_first_scales}
\end{equation}
For \(L\ge2\), $2\mu_1<\mu_2<3\mu_1.$
Let
\begin{equation}
K_\infty^{(L)}:=K_L(\infty),
\qquad
R_\infty^{(L)}:=(K_\infty^{(L)})^{-1}.
\end{equation}
For an \(L\times L\) matrix \(B\), define the lifted perturbation
\begin{equation}
\widehat B
=
\begin{pmatrix}
\mathcal A(B)&0\\
0&0
\end{pmatrix}.
\end{equation}
Then
\begin{equation}
K_L(\beta)
=
K_\infty^{(L)}
+
2\sum_{n=1}^{\infty}
(-1)^n
\sum_{p=1}^{L}
e^{-n\mu_p\beta}
\widehat B_p^{(L)}.
\label{eq:K_regimeIII_lowT}
\end{equation}
Set
\begin{equation}
X_p^{(L)}
=
R_\infty^{(L)}\widehat B_p^{(L)}.
\label{eq:Xp_regimeIII}
\end{equation}
Expanding
\[
\log\operatorname{Pf}(K+\Delta K)
=
\log\operatorname{Pf}K
+
\frac12\operatorname{Tr}\log(I+K^{-1}\Delta K)
\]
gives, at fixed \(L\),
\begin{equation}
\log S_L(\beta)
=
\log S_L(\infty)
+
a_1(L)e^{-\mu_1\beta}
+
a_2(L)e^{-2\mu_1\beta}
+
a_3(L)e^{-\mu_2\beta}
+
O_L(e^{-3\mu_1\beta}),
\label{eq:fixedL_regimeIII_thermal}
\end{equation}
where
\begin{align}
a_1(L)&=-\operatorname{Tr}X_1^{(L)},\\
a_2(L)&=\operatorname{Tr}X_1^{(L)}
-\operatorname{Tr}\!\left[(X_1^{(L)})^2\right],\\
a_3(L)&=-\operatorname{Tr}X_2^{(L)}.
\end{align}
This fixed-\(L\) expansion explains the thermodynamic saturation condition
\(\beta/L\to\infty\).

\subsection{Saturated reduced symbol and bulk coefficient}

The limiting reduced Toeplitz symbol is obtained from the Regime-I symbol by
setting \(t_\beta(\theta)=1\).  Thus
\begin{equation}
a_\infty(e^{i\theta})
=
\begin{pmatrix}
-C(\theta)&1+Q_\infty(e^{i\theta})+C(\theta)\\
-1-Q_\infty(e^{-i\theta})+C(\theta)&-C(\theta)
\end{pmatrix},
\label{eq:a_regimeIII_infty}
\end{equation}
where
\begin{equation}
C(\theta)=i\cot\frac{\theta}{2},
\qquad
Q_\infty(e^{i\theta})=i e^{-i\theta/2},
\qquad
Q_\infty(e^{-i\theta})=-i e^{i\theta/2}.
\end{equation}
\begin{figure}[t]
    \centering

    \includegraphics[width=0.4\textwidth]{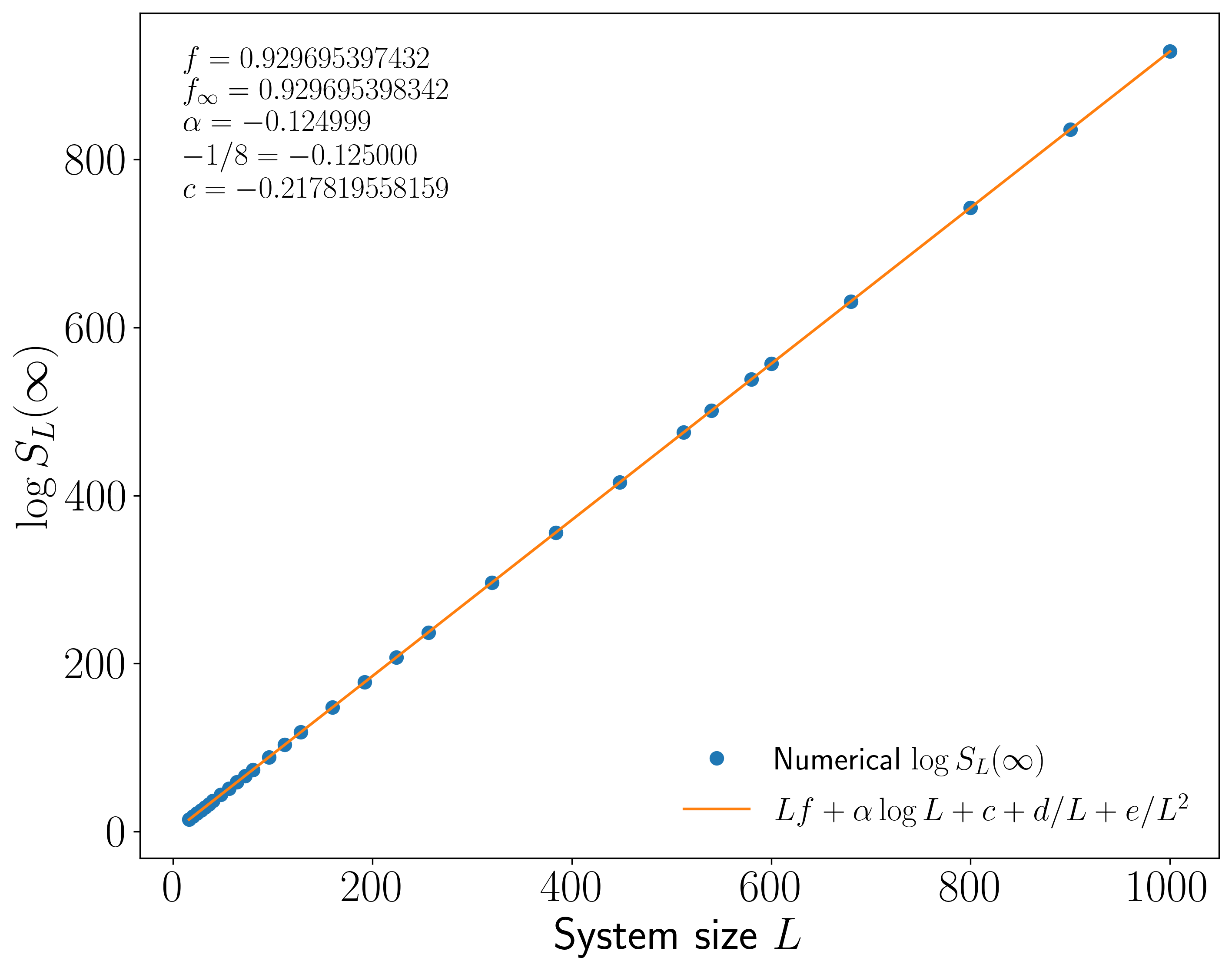}
    \includegraphics[width=0.4\textwidth]{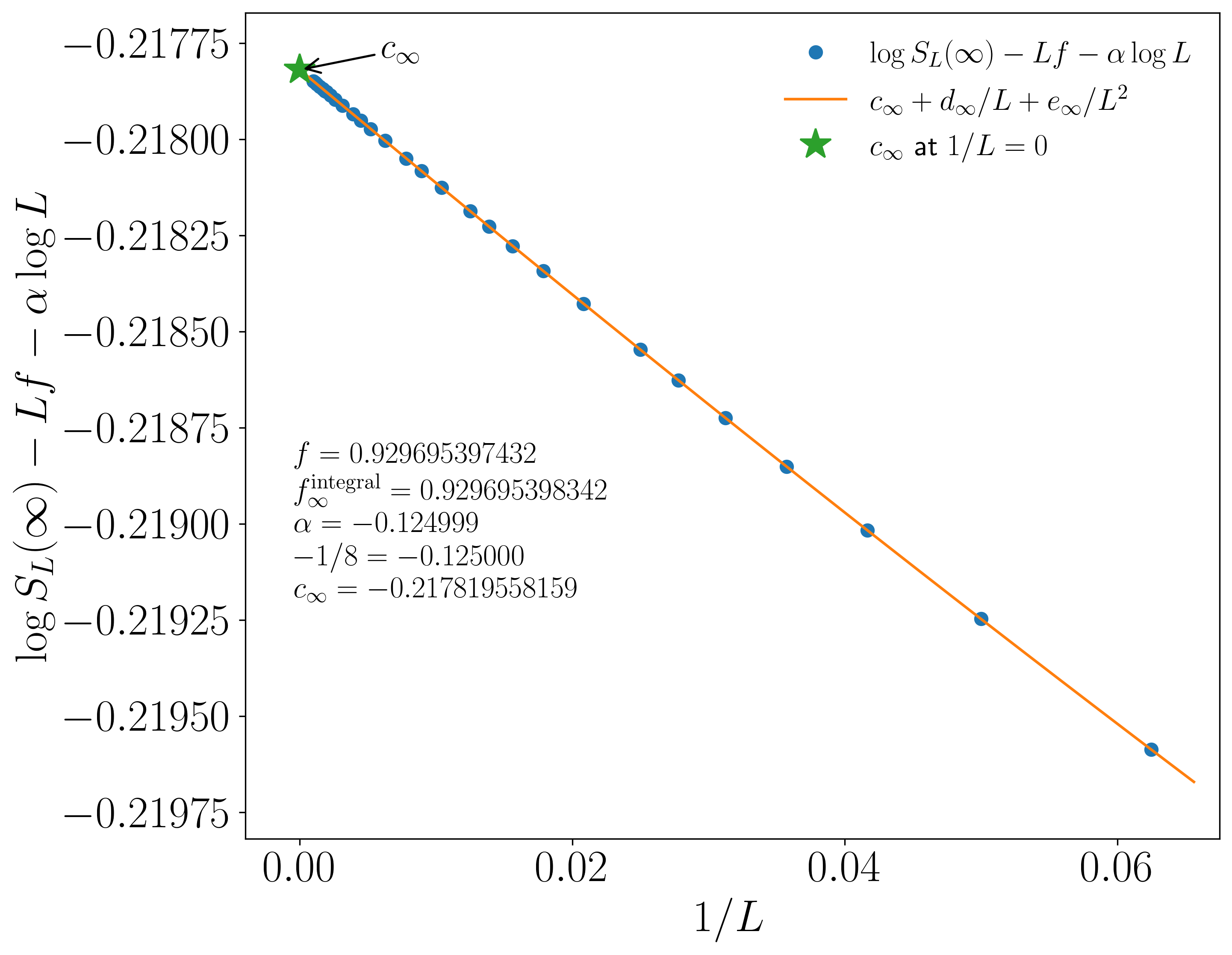}

  \caption{
Numerical confirmation of the Regime-III saturated finite-size scaling form for
the staggered-gauge Pfaffian numerator.  The data are computed from the Schur
complement determinant at \(\beta=\infty\).  Left: scaling of
\(\log S_L(\infty)\) with \(L\), fitted with the logarithmic correction included.
The fitted bulk coefficient agrees with the saturated symbol integral, and the
logarithmic coefficient is consistent with \(-1/8\).  Right: residual after
subtracting the fitted bulk and logarithmic terms, plotted against \(1/L\).  The
intercept gives the saturated constant \(c_\infty\), while the curvature is
captured by the \(1/L\) corrections.
}
    \label{fig:regimeIII_scaling_checks}
\end{figure}
Its determinant is
\begin{equation}
\Delta_\infty(\theta)
:=
\det a_\infty(e^{i\theta})
=
2+\frac{2}{\sin(\theta/2)}.
\label{eq:Delta_regimeIII_infty}
\end{equation}
Equivalently, with \(w=e^{i\theta/2}\),
\begin{equation}
\det a_\infty(w)
=
\frac{2(w+i)^2}{(w-1)(w+1)}.
\label{eq:det_a_regimeIII_w}
\end{equation}
On the unit circle this is the same positive function
\(\Delta_\infty(\theta)\). Near \(\theta=0\),
\begin{equation}
\Delta_\infty(\theta)
\sim
\frac{4}{\theta}.
\label{eq:Delta_regimeIII_endpoint}
\end{equation}
Thus the saturated symbol has an endpoint Fisher--Hartwig singularity.  The
bulk coefficient is nevertheless finite:
\begin{equation}
f_\infty
=
\frac{1}{2\pi}
\int_0^\pi
\log\Delta_\infty(\theta)\,d\theta=
\frac{1}{\pi}
\int_0^{\pi/2}
\log\left[
2+\frac{2}{\sin x}
\right]dx.
\label{eq:f_regimeIII_infty}
\end{equation}
\subsection{Saturated finite-size scaling}

Because of the endpoint singularity \eqref{eq:Delta_regimeIII_endpoint}, the
saturated determinant has a Fisher--Hartwig-type logarithmic correction.  The
numerically supported saturated asymptotic is
\begin{equation}
\log S_L(\infty)
=
L f_\infty
-\frac18\log L
+
c_\infty
+
\frac{d_\infty}{L}
+
o(L^{-1}).
\label{eq:S_regimeIII_asymptotic}
\end{equation}
Equivalently, at determinant level,
\begin{equation}
\log\det\mathcal D_L(\infty)
=
2L f_\infty
-\frac14\log L
+
2c_\infty
+
\frac{2d_\infty}{L}
+
o(L^{-1}).
\label{eq:det_regimeIII_asymptotic}
\end{equation}
The logarithmic coefficient is therefore $\alpha_\infty=-\frac18$ for the Pfaffian numerator. The constant is defined by the finite-section limit
\begin{equation}
c_\infty
=
\lim_{L\to\infty}
\left[
\log S_L(\infty)
-
L f_\infty
+
\frac18\log L
\right],
\label{eq:c_regimeIII_def}
\end{equation}
provided the limit exists.  Equivalently,
\begin{equation}
2c_\infty
=
\lim_{L\to\infty}
\left[
\log\det\mathcal D_L(\infty)
-
2L f_\infty
+
\frac14\log L
\right].
\label{eq:2c_regimeIII_def}
\end{equation}
This constant is a genuine block Toeplitz--Hankel Fisher--Hartwig constant.
The scalar determinant \(\Delta_\infty\) determines the bulk term and the
location of the singularity, but it does not determine \(c_\infty\) by itself.

The scaling form \eqref{eq:S_regimeIII_asymptotic} is supported numerically in
Fig.~\ref{fig:regimeIII_scaling_checks}.  The direct fit of
\(\log S_L(\infty)\) confirms the bulk coefficient \(f_\infty\) obtained from
the saturated symbol and gives a logarithmic coefficient consistent with
\(-1/8\).  After subtracting the fitted extensive and logarithmic terms, the
remaining finite-size data are well described by the expected expansion in
powers of \(1/L\), with intercept \(c_\infty\).

Combining the fixed-\(L\) thermal expansion with the saturated large-\(L\)
asymptotic, Regime III is summarized by
\begin{equation}
\log S_L(\beta)
=
L f_\infty
-\frac18\log L
+
c_\infty
+
\frac{d_\infty}{L}
+
o(L^{-1})
+
\text{thermal corrections},
\label{eq:regimeIII_final}
\end{equation}
where the thermal corrections vanish when \(\beta/L\to\infty\), with leading
scale \(e^{-\pi J\beta/L}\).

\subsection{Relation to the normalized mixed-state quantity}

At saturation, \(\Pi_L(\infty)=L\log2\). Therefore the normalized quantity has the zero-temperature asymptotic
\begin{equation}
M_{1/2,L}(\infty)
=
2\log S_L(\infty)
-
2L\log2.
\end{equation}
Using \eqref{eq:S_regimeIII_asymptotic}, this gives
\begin{equation}
M_{1/2,L}(\infty)
=
L(2f_\infty-2\log2)
-\frac14\log L
+
2c_\infty
+
\frac{2d_\infty}{L}
+
o(L^{-1}).
\label{eq:M_regimeIII_infty}
\end{equation}

\subsection{Remarks on the saturated constant}

The saturated constant \(c_\infty\) is not expected to reduce to a scalar
Fisher--Hartwig constant built only from \(\Delta_\infty\).  The reason is the
same as in Regime I: the full object is a block Toeplitz--Hankel determinant,
and the constant depends on the matrix factorization and on the reflected
Hankel part, not only on the scalar determinant of the Toeplitz symbol.

One may define a determinant-level constant
\begin{equation}
E_\infty
=
\lim_{L\to\infty}
L^{1/4}e^{-2Lf_\infty}
\det\mathcal D_L(\infty),
\label{eq:E_regimeIII_def}
\end{equation}
so that
\begin{equation}
c_\infty=\frac12\log E_\infty.
\end{equation}
This is the most compact exact definition of the saturated constant.  A closed
scalar expression for \(E_\infty\) is not known from the present reduction.

\section{Asymptotic scaling regimes for the Pfaffian formula: Regime IV( the crossover scaling $\beta=L\tau$)}\label{Sec7}

In this section we study the crossover regime between the fixed-temperature
limit and the saturated low-temperature limit.  The scaling window is
\begin{equation}
\beta=\tau L,
\qquad
0<\tau<\infty,
\qquad
L\to\infty .
\label{eq:regimeIV_scaling}
\end{equation}
The parameter \(\tau\) is kept fixed and measures the inverse temperature in
units of the system size.  Throughout this section we use the same staggered
Jordan--Wigner gauge as in the previous sections.  For simplicity we set
\(J=1\); the dependence on \(J\) is restored by replacing \(\tau\) by
\(J\tau\).

The key point is that this scaling keeps the lowest open-boundary modes at
finite effective temperature, while all bulk modes are already saturated.
Thus Regime IV inherits the bulk and logarithmic terms from Regime III, but it
also contains a nontrivial crossover function coming from the thermal edge
modes.

\subsection{Crossover scaling of the staggered kernel}

We start from the critical staggered-gauge correlation matrix
\eqref{eq:app_G_critical_stag}, with the mode factor \(t_p(\beta)\) defined in
Eq.~\eqref{eq:t_p_def_purity}. In the crossover scaling \(\beta=\tau L\), the low-lying modes have a
nontrivial limit.  For fixed \(p\),
\begin{equation}
2\tau L\sin\frac{k_p}{2}
\longrightarrow
\pi\tau\left(p-\frac12\right),
\qquad
L\to\infty .
\end{equation}
Hence
\begin{equation}
t_p(\tau L)
\longrightarrow
\tanh\!\left(\pi\tau\left(p-\frac12\right)\right).
\label{eq:tp_edge_limit_regimeIV}
\end{equation}
By contrast, if \(p\) is proportional to \(L\), then \(k_p=O(1)\), and
therefore $t_p(\tau L)\to1.$ Thus Regime IV is an edge scaling regime: the bulk modes are saturated, while
the lowest open-boundary modes remain thermal.

Let $Q=e^{-\pi\tau}$, then the limiting defect from saturation is
\begin{equation}
\delta_p(\tau)
:=
\tanh\!\left(\pi\tau\left(p-\frac12\right)\right)-1
=
-\frac{2Q^{2p-1}}{1+Q^{2p-1}}.
\label{eq:delta_p_regimeIV}
\end{equation}
Equivalently, if
\begin{equation}
n_p(\tau)
=
\frac{Q^{2p-1}}{1+Q^{2p-1}}
=
\frac{1}{e^{2\pi\tau(p-\frac12)}+1},
\label{eq:np_regimeIV}
\end{equation}
then
\begin{equation}
\delta_p(\tau)=-2n_p(\tau).
\end{equation}

\subsection{Mode decomposition and Toeplitz--Hankel form}

It is useful to separate the thermal factors from the geometric mode matrices.
Define
\begin{equation}
\left(B_p^{(L)}\right)_{jk}
=
(-1)^{j+k}
\frac{4}{2L+1}
\cos\!\left[\left(j-\frac12\right)k_p\right]
\sin(k k_p).
\label{eq:Bp_regimeIV}
\end{equation}
Then
\begin{equation}
G(\beta)
=
\sum_{p=1}^{L}t_p(\beta)B_p^{(L)}.
\label{eq:G_mode_decomp_regimeIV}
\end{equation}
The saturated matrix is
\begin{equation}
G_\infty
=
\sum_{p=1}^{L}B_p^{(L)}.
\label{eq:Ginf_mode_decomp_regimeIV}
\end{equation}
Therefore, in the crossover window,
\begin{equation}
G(\tau L)
=
G_\infty
+
\sum_{p=1}^{L}\delta_p^{(L)}(\tau)B_p^{(L)},
\label{eq:G_crossover_pert_regimeIV}
\end{equation}
where
\begin{equation}
\delta_p^{(L)}(\tau)
=
t_p(\tau L)-1
=
\tanh\!\left(2\tau L\sin\frac{k_p}{2}\right)-1.
\label{eq:delta_p_L_regimeIV}
\end{equation}
For fixed \(p\),
\[
\delta_p^{(L)}(\tau)\to\delta_p(\tau).
\]
This decomposition isolates the finite-temperature correction as a sum of
low-mode defects added to the saturated matrix.

The same kernel has the finite Toeplitz--Hankel form used in
Sec.~\ref{Sec3}.  Define
\begin{equation}
q_m^{(L,\beta)}
=
(-1)^m
\frac{2}{2L+1}
\sum_{p=1}^{L}
t_p(\beta)
\sin\!\left[\left(m+\frac12\right)k_p\right],
\qquad m\in\mathbb Z .
\label{eq:qm_regimeIV_stag}
\end{equation}
Then
\begin{equation}
G_{jk}(\beta)
=
q_{k-j}^{(L,\beta)}
-
q_{k+j-1}^{(L,\beta)}.
\label{eq:G_TH_regimeIV_stag}
\end{equation}
At saturation,
\begin{equation}
q_m^{(L,\infty)}
=
(-1)^m
\frac{2}{2L+1}
\sum_{p=1}^{L}
\sin\!\left[\left(m+\frac12\right)k_p\right].
\label{eq:qm_inf_regimeIV}
\end{equation}
Equivalently,
\begin{equation}
q_m^{(L,\infty)}
=
(-1)^m
\frac{2}{2L+1}
\frac{
\sin^2\!\left(\dfrac{L(m+\frac12)\pi}{2L+1}\right)
}{
\sin\!\left(\dfrac{(m+\frac12)\pi}{2L+1}\right)
}.
\label{eq:qm_inf_explicit_regimeIV}
\end{equation}
For fixed \(m\),
\begin{equation}
q_m^{(L,\infty)}
\longrightarrow
\frac{(-1)^m}{\pi(m+\frac12)}.
\label{eq:qm_inf_limit_regimeIV}
\end{equation}
The slow \(1/(m+1/2)\) decay is the endpoint singularity responsible for
the Fisher--Hartwig logarithm in the saturated regime.

\subsection{Finite-size crossover ratio}

The finite-\(L\) crossover ratio is defined by
\begin{equation}
\mathcal F_L(\tau)
:=
\frac{S_L(\tau L)}{S_L(\infty)}
=
\frac{\operatorname{Pf}K_L(\tau L)}{\operatorname{Pf}K_L(\infty)}.
\label{eq:FL_def_regimeIV}
\end{equation}
Here
\begin{equation}
K_L(\beta)
=
\begin{pmatrix}
\mathcal A(G(\beta)) & I_{2L}\\
-I_{2L} & -\mathcal J'_{2L}
\end{pmatrix},
\label{eq:K_regimeIV}
\end{equation}
with the same staggered selector \(\mathcal J'_{2L}\) as in
Sec.~\ref{SecPfaffian}.  The limiting crossover function is
\begin{equation}
\mathcal F(\tau)
:=
\lim_{L\to\infty}\mathcal F_L(\tau),
\label{eq:F_limit_regimeIV}
\end{equation}
whenever the limit exists.

Using the Schur complement, the same ratio may be written as
\begin{equation}
\mathcal F_L(\tau)
=
\left[
\frac{
\det\!\left(\mathcal A(G(\tau L))-(\mathcal J'_{2L})^{-1}\right)
}{
\det\!\left(\mathcal A(G_\infty)-(\mathcal J'_{2L})^{-1}\right)
}
\right]^{1/2}.
\label{eq:FL_schur_regimeIV}
\end{equation}
The positive square-root branch is chosen by continuity from the positive
absolute-minor sum.

Because each \(B_p^{(L)}\) is rank one, the ratio can be reduced to a finite
determinant in mode space.  Define
\begin{equation}
u_p^{(L)}(j)
=
\sqrt{\frac{4}{2L+1}}\,
(-1)^j
\cos\!\left[\left(j-\frac12\right)k_p\right],
\qquad
v_p^{(L)}(j)
=
\sqrt{\frac{4}{2L+1}}\,
(-1)^j
\sin(jk_p).
\label{eq:up_vp_regimeIV}
\end{equation}
Then
\begin{equation}
B_p^{(L)}
=
u_p^{(L)}(v_p^{(L)})^{T}.
\label{eq:Bp_rank_one_regimeIV}
\end{equation}
Embed these vectors into the interleaved \(2L\)-dimensional unbarred layer by
\[
a_p^{(L)}
=
\bigl(u_p^{(L)}(1),0,u_p^{(L)}(2),0,\ldots,u_p^{(L)}(L),0\bigr)^T,
\]
and
\[
b_p^{(L)}
=
\bigl(0,v_p^{(L)}(1),0,v_p^{(L)}(2),\ldots,0,v_p^{(L)}(L)\bigr)^T.
\]
Let
\[
W_p^{(L)}
=
\begin{pmatrix}
a_p^{(L)} & b_p^{(L)}
\end{pmatrix},
\qquad
J_2=
\begin{pmatrix}
0&1\\
-1&0
\end{pmatrix}.
\]
Then
\begin{equation}
\mathcal A(B_p^{(L)})
=
W_p^{(L)}J_2(W_p^{(L)})^T.
\label{eq:A_Bp_rank_two_regimeIV}
\end{equation}

Introduce the saturated Schur resolvent
\begin{equation}
R_L^\infty
:=
\left(
\mathcal A(G_\infty)-(\mathcal J'_{2L})^{-1}
\right)^{-1}.
\label{eq:Rinf_regimeIV}
\end{equation}
Then
\begin{equation}
\mathcal F_L(\tau)^2
=
\det\left[
I+
R_L^\infty
\sum_{p=1}^{L}
\delta_p^{(L)}(\tau)\mathcal A(B_p^{(L)})
\right].
\label{eq:FL_schur_pert_regimeIV}
\end{equation}
By Sylvester's determinant identity, this becomes a determinant on mode space:
\begin{equation}
\mathcal F_L(\tau)
=
\det{}^{1/2}_{1\le p,q\le L}
\left[
I_2\delta_{pq}
+
\delta_p^{(L)}(\tau)\,
J_2\mathcal M_L(p,q)
\right],
\label{eq:FL_edge_det_regimeIV}
\end{equation}
where
\begin{equation}
\mathcal M_L(p,q)
=
(W_p^{(L)})^T R_L^\infty W_q^{(L)}.
\label{eq:MLpq_regimeIV}
\end{equation}
This identity is exact for every finite \(L\).  It shows that the crossover is
controlled by the thermal occupation of the low-lying edge modes, dressed by
the saturated Pfaffian background.

The projected kernel is not diagonal in the raw open sine--cosine mode basis.
This is important: replacing it by a diagonal free-fermion kernel would give
only the ordinary free-free Majorana factor.  The actual Pfaffian ratio contains
additional Toeplitz--Hankel reflection data, which appear below as a
level-eight correction.

\subsection{Regime-IV asymptotic form}

Since all bulk modes are saturated in the scaling \(\beta=\tau L\), the
extensive coefficient is the saturated bulk coefficient.  With the staggered
normalization,
\begin{equation}
f_\infty
=
\frac{1}{\pi}
\int_0^{\pi/2}
\log\left(
2+\frac{2}{\sin x}
\right)\,dx.
\label{eq:finfty_regimeIV}
\end{equation}
The endpoint Fisher--Hartwig singularity is also the saturated one, so the
logarithmic coefficient is
\begin{equation}
\alpha_\infty=-\frac18.
\end{equation}
Thus the Regime-IV scaling form is
\begin{equation}
\log S_L(\tau L)
=
L f_\infty
-\frac18\log L
+
c_\infty
+
\log\mathcal F(\tau)
+
o(1).
\label{eq:regimeIV_asymptotic}
\end{equation}
Here \(c_\infty\) is the saturated constant defined in Regime III, and
\(\mathcal F(\tau)\) is the universal crossover factor.

As \(\tau\to\infty\), the edge modes also saturate and
\[
\mathcal F(\tau)\to1.
\]
As \(\tau\downarrow0\), the number of active edge modes grows like \(1/\tau\),
and the crossover factor develops an essential singularity.  These two limits
are described explicitly below.

\begin{figure}[t]
    \centering
    \includegraphics[width=0.4\textwidth]{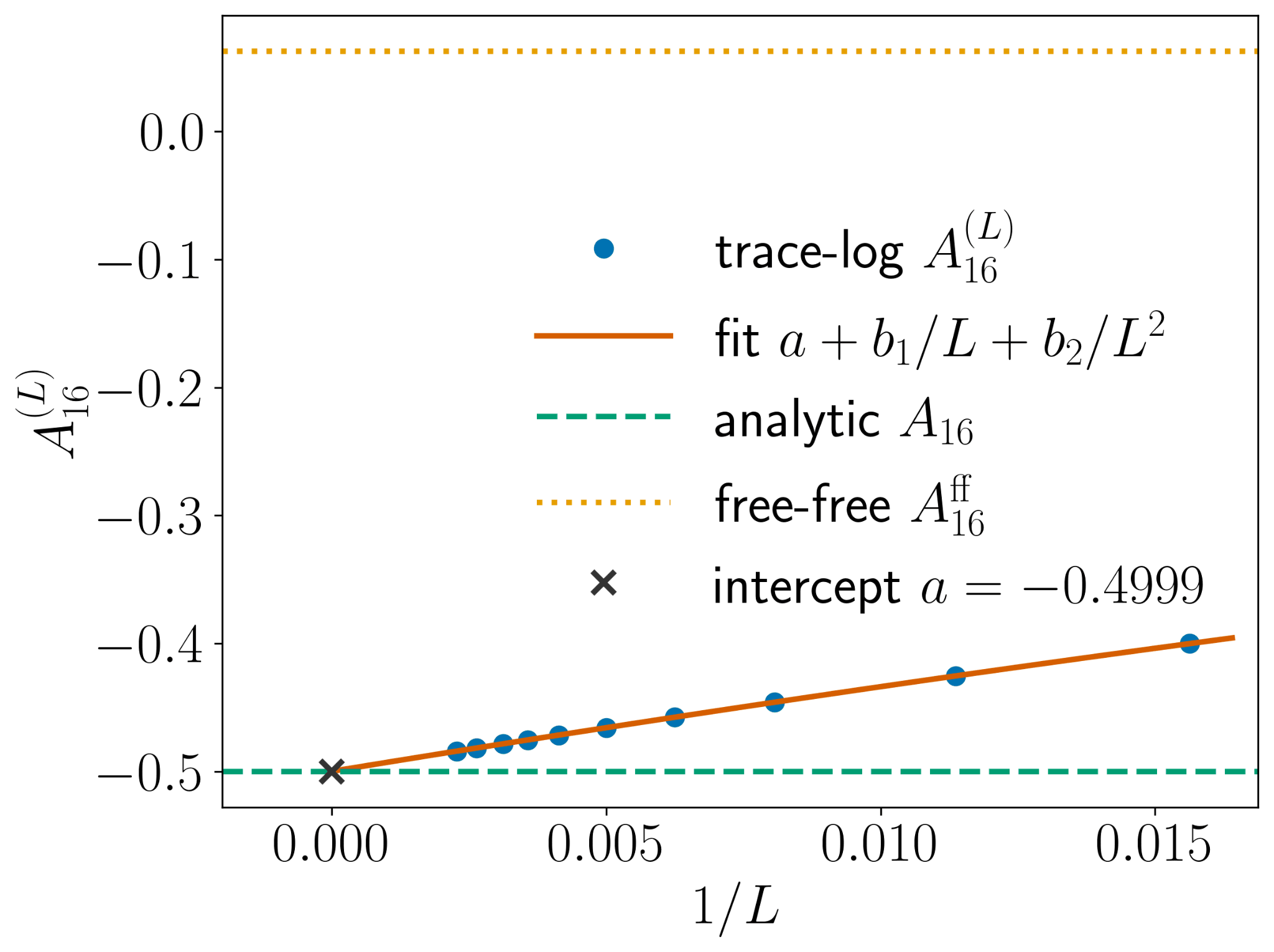}
    \includegraphics[width=0.4\textwidth]{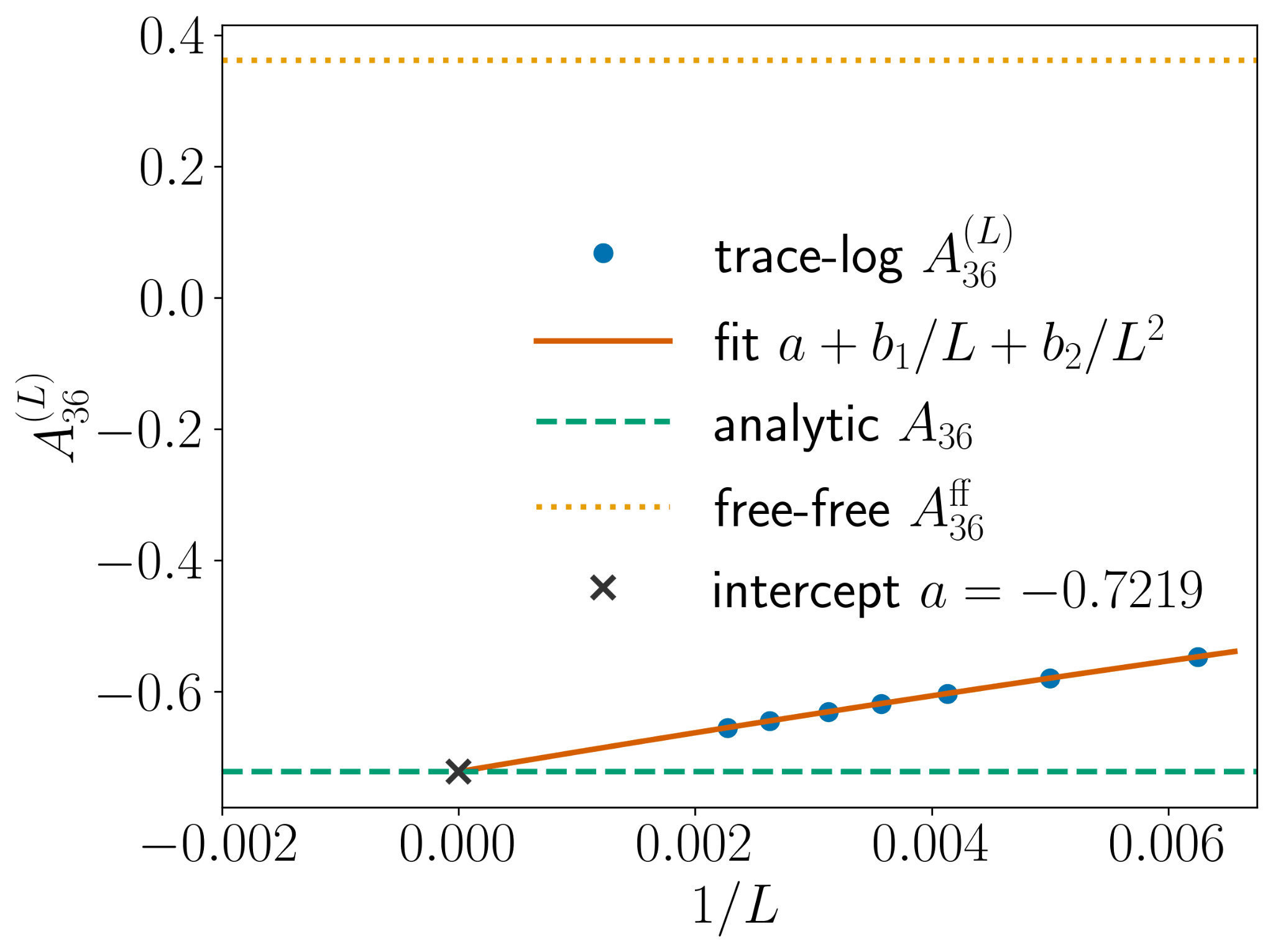}
    \caption{
Trace-log extraction of the Regime-IV coefficients \(A_j\).  The blue points
are the finite-section coefficients \(A_j^{(L)}\), and the orange curve is the
quadratic extrapolation \(A_j^{(L)}=a_j+b_1/L+b_2/L^2\).  The extrapolated
intercept agrees with the eta-quotient prediction \(A_j\) shown by the green
dashed line.  The orange dotted line is the free-free value \(A_j^{\rm ff}\).
The examples \(j=16\) and \(j=36\) are in the \(Q^{4m}\) sector, where the full
Regime-IV coefficient differs from the free-free Majorana benchmark.
}
\label{fig:regimeIV_trace_log_coeff_extrap}
\end{figure}

\subsection{Coefficient extraction and eta-quotient conjecture}

The exact ratio \eqref{eq:FL_edge_det_regimeIV} can be expanded at large
\(\tau\), or equivalently at small \(Q=e^{-\pi\tau}\).  For fixed \(p\),
\[
\delta_p(\tau)
=
-\frac{2Q^{2p-1}}{1+Q^{2p-1}}
=
-2Q^{2p-1}
+2Q^{2(2p-1)}
-2Q^{3(2p-1)}
+\cdots .
\]
Therefore, to determine the coefficient of \(Q^N\), only modes satisfying
\(2p-1\le N\) are needed.  The coefficient extraction is finite-dimensional at
every fixed order.

Writing
\begin{equation}
\log\mathcal F(\tau)
=
\sum_{j=1}^{\infty}A_jQ^j,
\label{eq:logF_Aj_regimeIV}
\end{equation}
the finite-\(L\) Pfaffian expansion suggests the following divisor-sum rule.
Let
\[
j=2^r n,
\qquad n\ \text{odd},
\]
and let
\[
\sigma(n)=\sum_{d\mid n}d.
\]
Then
\begin{equation}
A_j
=
\begin{cases}
-\dfrac{\sigma(n)}{n}, & r=0,\\[8pt]
+\dfrac{\sigma(n)}{2n}, & r=1,\\[8pt]
-\dfrac{\sigma(n)}{2n}, & r\ge2.
\end{cases}
\label{eq:Aj_rule_regimeIV}
\end{equation}
Equivalently,
\begin{align}
\log\mathcal F(\tau)
={}&
-Q+\frac12Q^2-\frac43Q^3-\frac12Q^4
-\frac65Q^5+\frac23Q^6-\frac87Q^7-\frac12Q^8
\nonumber\\
&-\frac{13}{9}Q^9+\frac35Q^{10}
-\frac{12}{11}Q^{11}
-\frac23Q^{12}
+\cdots .
\label{eq:logF_first_terms_regimeIV}
\end{align}

The ordinary free-free Majorana factor would be
\begin{equation}
\mathcal F_{ff}(\tau)
=
\prod_{p=1}^{\infty}(1+Q^{2p-1})^{-1}.
\label{eq:Fff_regimeIV}
\end{equation}
It gives the correct coefficients for powers not divisible by \(4\), but it
fails in the \(Q^{4m}\)-sector.  The first discrepancy occurs at \(Q^4\):
the free-free factor gives \(+\frac14 Q^4\) in \(\log\mathcal F_{ff}\), while
the Pfaffian extraction gives \(-\frac12 Q^4\).

This discrepancy can be checked directly at the level of the finite-section
trace-log expansion.  Starting from the Schur-complement ratio
\eqref{eq:FL_schur_pert_regimeIV}, define
\begin{equation}
X_p^{(L)}
:=
R_L^\infty \mathcal A(B_p^{(L)}),
\qquad
R_L^\infty
=
\left[
\mathcal A(G_\infty)-(\mathcal J'_{2L})^{-1}
\right]^{-1}.
\label{eq:Xp_trace_log_regimeIV}
\end{equation}
For coefficient extraction, we replace the low-mode defects by their
Regime-IV edge limits \(\delta_p(Q)\) and introduce the finite-section
generating function
\begin{equation}
\log \widehat{\mathcal F}_L(Q)
=
\frac12\operatorname{Tr}
\log\left[
I+\sum_{p=1}^{L}\delta_p(Q)X_p^{(L)}
\right].
\label{eq:trace_log_coeff_regimeIV}
\end{equation}
Expanding Eq.~\eqref{eq:trace_log_coeff_regimeIV} in powers of \(Q\) gives
finite-section coefficients
\begin{equation}
\log \widehat{\mathcal F}_L(Q)
=
\sum_{j\ge1} A_j^{(L)} Q^j .
\label{eq:AjL_def_regimeIV}
\end{equation}
For each fixed \(j\), only modes with \(2p-1\le j\) contribute, and we then
extrapolate
\begin{equation}
A_j^{(L)}
=
A_j
+
\frac{b_1(j)}{L}
+
\frac{b_2(j)}{L^2}
+\cdots .
\label{eq:Aj_finite_L_extrap_regimeIV}
\end{equation}
Representative examples in the anomalous \(Q^{4m}\)-sector are shown in
Fig.~\ref{fig:regimeIV_trace_log_coeff_extrap}.  The extrapolated intercepts
approach the coefficients predicted by the divisor-sum rule
\eqref{eq:Aj_rule_regimeIV}, rather than the ordinary free-free values.

The divisor-sum rule \eqref{eq:Aj_rule_regimeIV} is generated by the product
\begin{equation}
\mathcal F(\tau)
=
(Q;Q^2)_\infty
(Q^2;Q^4)_\infty^{-1}
(Q^4;Q^8)_\infty^{3/4}
(Q^8;Q^8)_\infty^{1/4},
\label{eq:F_product_regimeIV}
\end{equation}
where
\[
(a;q)_\infty=\prod_{m=0}^{\infty}(1-aq^m).
\]
Equivalently,
\begin{equation}
\mathcal F(\tau)
=
\frac{
(Q;Q)_\infty
(Q^4;Q^4)_\infty^{7/4}
}{
(Q^2;Q^2)_\infty^2
(Q^8;Q^8)_\infty^{1/2}
}.
\label{eq:F_product_alt_regimeIV}
\end{equation}
Using
\[
\eta\!\left(\frac{i a\tau}{2}\right)
=
Q^{a/24}(Q^a;Q^a)_\infty,
\]
the powers of \(Q\) cancel and one obtains
\begin{equation}
\mathcal F(\tau)
=
\frac{
\eta(i\tau/2)\eta(2i\tau)^{7/4}
}{
\eta(i\tau)^2\eta(4i\tau)^{1/2}
}.
\label{eq:F_eta_regimeIV}
\end{equation}
Equivalently, if \(z=i\tau/2\), then
\begin{equation}
\mathcal F(\tau)^4
=
\frac{\eta(z)^4\eta(4z)^7}
{\eta(2z)^8\eta(8z)^2}.
\label{eq:F_fourth_power_regimeIV}
\end{equation}
Thus the fourth power is an ordinary level-eight eta quotient.

It is useful to separate the ordinary Majorana factor from the correction:
\begin{equation}
\mathcal F(\tau)
=
\mathcal F_{ff}(\tau)\,
\mathcal F_{\rm Bell}(\tau),
\label{eq:F_factorized_regimeIV}
\end{equation}
where
\begin{equation}
\mathcal F_{\rm Bell}(\tau)
=
(Q^4;Q^8)_\infty^{3/4}
(Q^8;Q^8)_\infty^{1/4}.
\label{eq:FBell_regimeIV}
\end{equation}
The factor \(\mathcal F_{ff}\) accounts for the half-integer open-chain
Majorana modes.  The additional factor \(\mathcal F_{\rm Bell}\) is supported
only on the \(Q^{4m}\)-sector and encodes the stabilizer-Rényi, or
Bell-defect, correction.

\begin{figure}[t]
    \centering
        \includegraphics[width=0.48\textwidth]{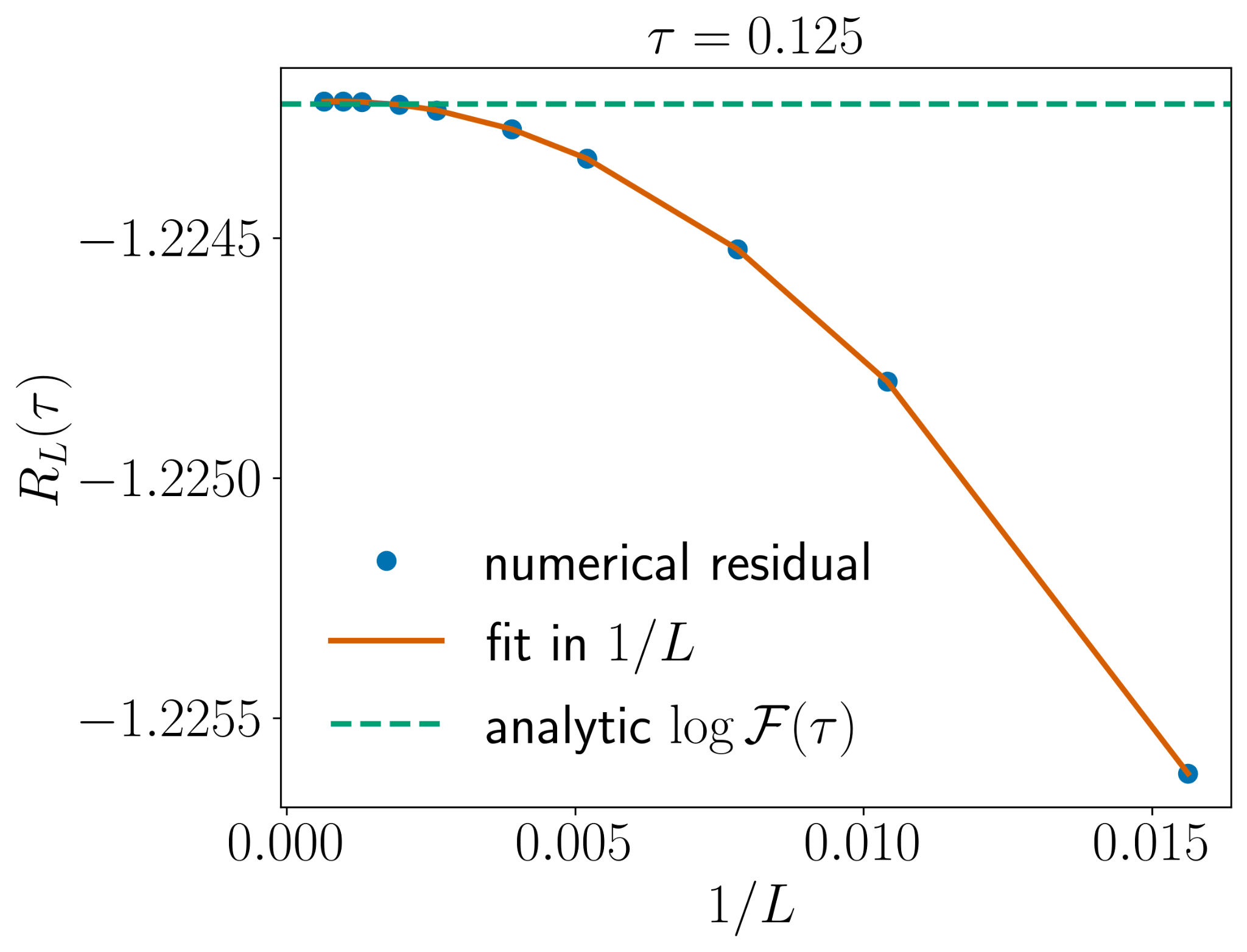}
        \includegraphics[width=0.48\textwidth]{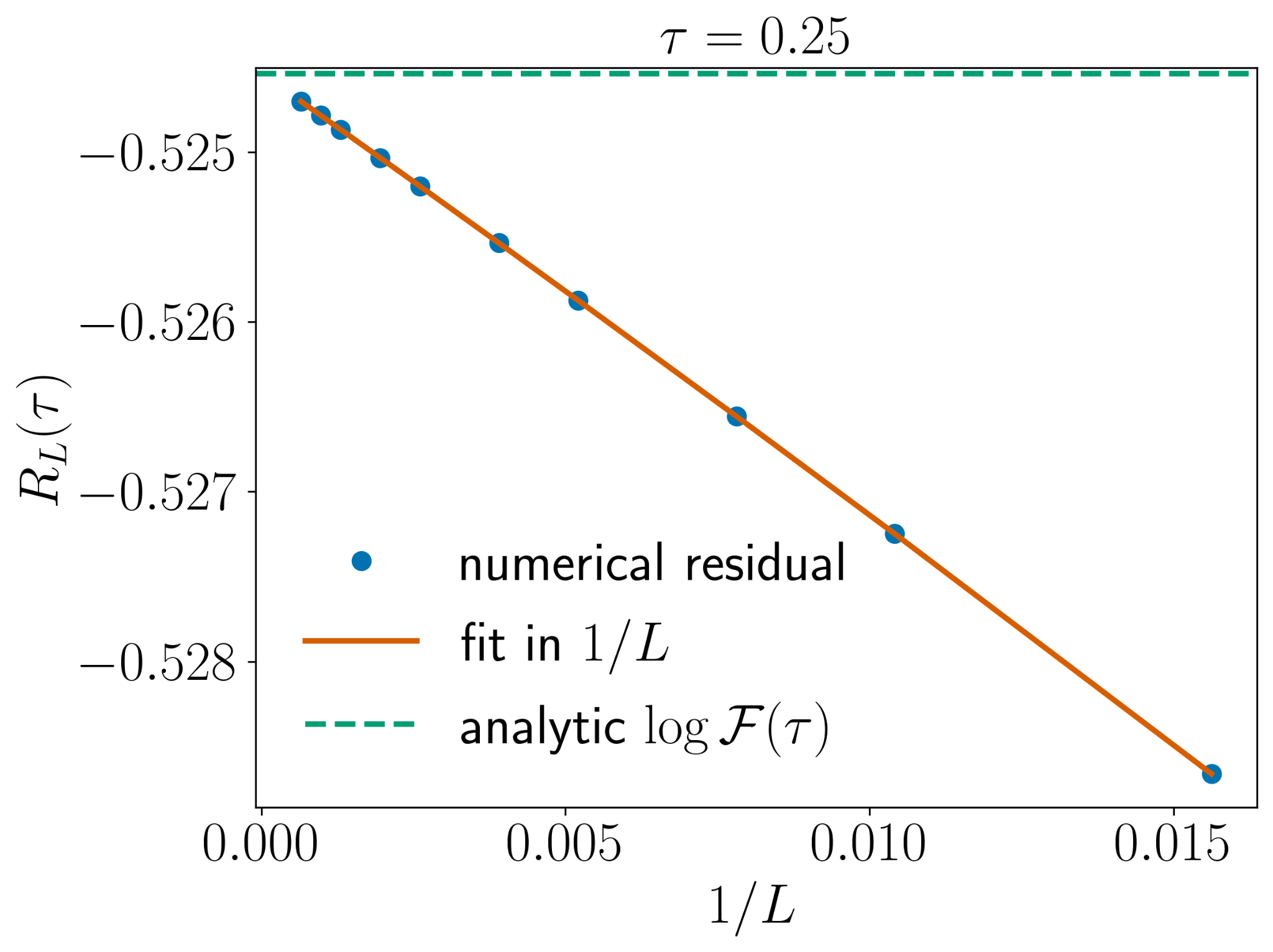}
    \caption{
    Finite-size approach of the Regime-IV residual \(R_L(\tau)\) defined in
    Eq.~\eqref{eq:RL_regimeIV_def}.  The blue points are the exact finite-\(L\)
    Pfaffian data after subtracting the saturated bulk contribution
    \(L f_\infty\), the Fisher--Hartwig logarithm \(-\frac18\log L\), and the
    saturated constant \(c_\infty\).  The orange curve is a fit in powers of
    \(1/L\), as in Eq.~\eqref{eq:RL_regimeIV_finite_size}.  The green dashed
    line is the eta-quotient prediction \(\log\mathcal F(\tau)\).  For
    \(\tau=0.25\), the leading \(1/L\) correction dominates over the displayed
    range and the data are nearly linear.  For \(\tau=0.125\), higher-order
    inverse-size corrections are more visible, producing curvature.  In both
    cases the extrapolation is consistent with the Regime-IV crossover
    function.
    }
    \label{fig:regimeIV_residuals}
\end{figure}

\subsection{Large- and small-\(\tau\) limits}

The large-\(\tau\) limit follows directly from the product
\eqref{eq:F_product_regimeIV}.  Since \(Q=e^{-\pi\tau}\),
\begin{align}
\log\mathcal F(\tau)
&=
-Q+\frac12Q^2-\frac43Q^3-\frac12Q^4
-\frac65Q^5+\frac23Q^6-\frac87Q^7-\frac12Q^8
+O(Q^9)
\nonumber\\
&=
-e^{-\pi\tau}
+\frac12e^{-2\pi\tau}
-\frac43e^{-3\pi\tau}
-\frac12e^{-4\pi\tau}
+O(e^{-5\pi\tau}).
\label{eq:F_large_tau_regimeIV}
\end{align}
The leading scale \(e^{-\pi\tau}\) matches the Regime-III thermal scale
\(e^{-\pi\beta/L}\) after setting \(\beta=\tau L\).  The first correction to
the ordinary free-free factor appears only at order \(e^{-4\pi\tau}\).

For \(\tau\downarrow0\), use the modular transformation
\[
\eta(ia\tau)
=
(a\tau)^{-1/2}\eta\!\left(\frac{i}{a\tau}\right).
\]
Applying this to \eqref{eq:F_eta_regimeIV} gives
\begin{equation}
\log\mathcal F(\tau)
=
-\frac{\pi}{16\tau}
-\frac18\log\tau
+\frac18\log2
+o(1),
\qquad
\tau\downarrow0.
\label{eq:F_small_tau_regimeIV}
\end{equation}
This is more singular than the ordinary free-free Majorana prediction
\(-\pi/(24\tau)\).  The stronger singularity is a finite-temperature signature
of the additional stabilizer-Rényi defect sector.

\subsection{Final Regime-IV statement}

The Regime-IV scaling window is
\[
\beta=\tau L,
\qquad
0<\tau<\infty.
\]
In this limit, the bulk is saturated but the lowest open-boundary modes remain
thermal.  The asymptotic form is
\begin{equation}
\log S_L(\tau L)
=
L f_\infty
-\frac18\log L
+
c_\infty
+
\log\mathcal F(\tau)
+
o(1),
\label{eq:regimeIV_final}
\end{equation}
where
\begin{equation}
f_\infty
=
\frac{1}{2\pi}
\int_0^\pi
\log\left(
2+\frac{2}{\sin x}
\right)\,dx,
\end{equation}
and \(c_\infty\) is the saturated constant from Regime III.

The crossover factor is conjectured to be
\begin{equation}
\mathcal F(\tau)
=
\frac{
\eta(i\tau/2)\eta(2i\tau)^{7/4}
}{
\eta(i\tau)^2\eta(4i\tau)^{1/2}
}.
\label{eq:F_regimeIV_final}
\end{equation}
Equivalently,
\begin{equation}
\mathcal F(\tau)
=
(Q;Q^2)_\infty
(Q^2;Q^4)_\infty^{-1}
(Q^4;Q^8)_\infty^{3/4}
(Q^8;Q^8)_\infty^{1/4},
\qquad
Q=e^{-\pi\tau}.
\end{equation}
The ordinary free-free Majorana factor is contained in this expression but is
not the full answer.  The additional level-eight factor is the
Toeplitz--Hankel, or stabilizer/Bell-defect, correction specific to the
\(\alpha=\frac12\) stabilizer-Rényi observable.

To test the final scaling form numerically, we subtract the saturated bulk,
logarithmic, and constant contributions and define
\begin{equation}
R_L(\tau)
:=
\log S_L(\tau L)
-
L f_\infty
+
\frac18\log L
-
c_\infty .
\label{eq:RL_regimeIV_def}
\end{equation}
Then the Regime-IV prediction \eqref{eq:regimeIV_final} is equivalent to
\begin{equation}
R_L(\tau)
\longrightarrow
\log\mathcal F(\tau),
\qquad
L\to\infty .
\label{eq:RL_regimeIV_limit}
\end{equation}
At finite size, the residual has an expansion in inverse powers of \(L\),
\begin{equation}
R_L(\tau)
=
\log\mathcal F(\tau)
+
\frac{a_1(\tau)}{L}
+
\frac{a_2(\tau)}{L^2}
+
\frac{a_3(\tau)}{L^3}
+\cdots .
\label{eq:RL_regimeIV_finite_size}
\end{equation}
Thus the intercept of \(R_L(\tau)\) plotted against \(1/L\) gives the
crossover value \(\log\mathcal F(\tau)\).  The higher-order terms in
Eq.~\eqref{eq:RL_regimeIV_finite_size} account for the curvature visible at
smaller \(\tau\), where more low-lying boundary modes remain thermally active. The numerical extrapolations of \(R_L(\tau)\) for representative values of
\(\tau\) are shown in Fig.~\ref{fig:regimeIV_residuals}. 

\section{Leading nontrivial minors of the correlation matrix}
\label{sec:supp-leading-minors}

In this section we describe how the dominant nontrivial minors of the correlation
matrix \(G(\beta)\) are identified.  For a chain of length \(L\), we consider
\begin{equation}
m_{\max}(L,\beta)
=
\max_{\substack{A,B\subseteq\{1,\ldots,L\}\\ |A|=|B|}}
\left|\det G_{A,B}(\beta)\right| .
\end{equation}
The empty minor gives the trivial value
\begin{equation}
\left|\det G_{\varnothing,\varnothing}(\beta)\right|=1,
\end{equation}
and is therefore excluded when discussing the leading nontrivial minors.

\begin{figure}[t]
    \centering
    \includegraphics[width=0.4\textwidth]{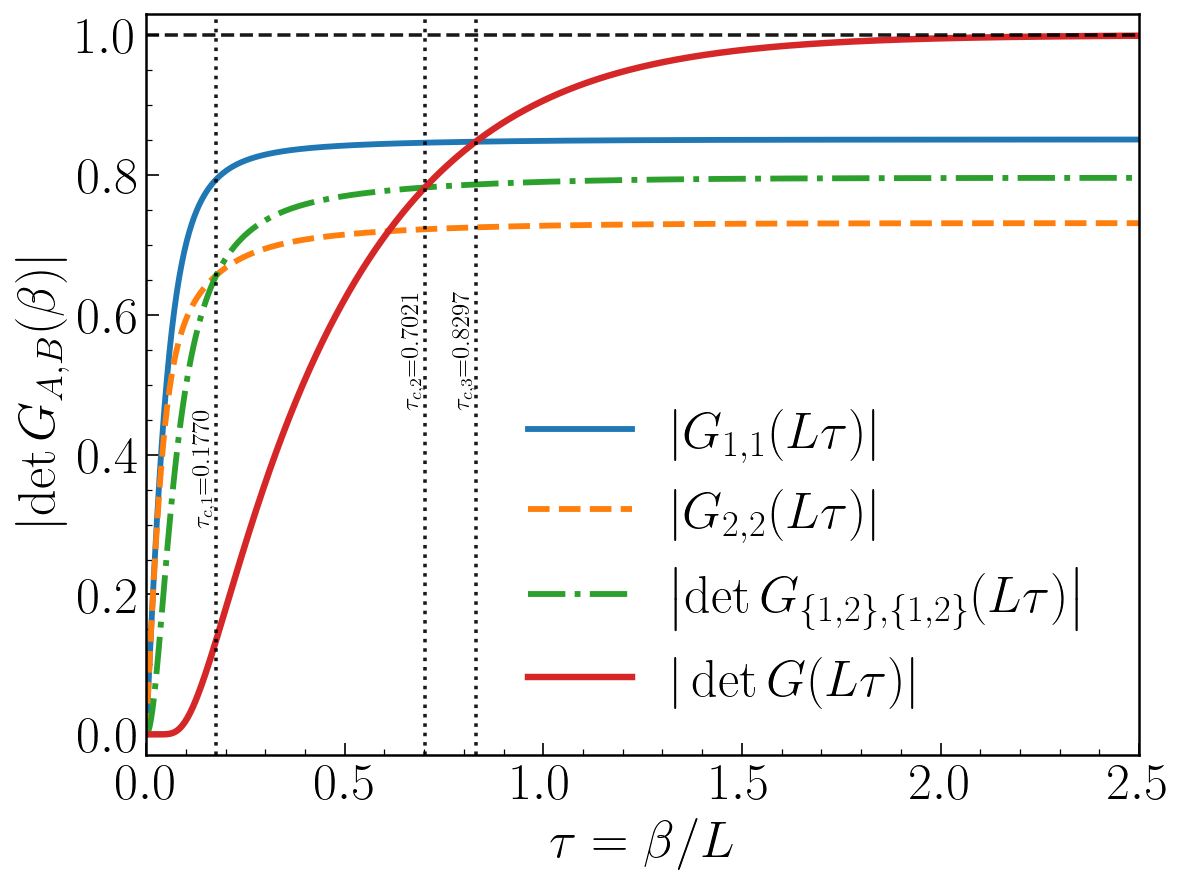}
    \caption{
    Leading nontrivial minor candidates as functions of
    \(\tau=\beta/L\) for \(L=12\).  The empty minor gives the trivial
    value \(1\) and is not included in the maximization over nontrivial
    minors.  The vertical dotted lines indicate the crossing values
    \(\tau_{c,1}\), \(\tau_{c,2}\), and \(\tau_{c,3}\), where the ordering of
    the leading nontrivial minors changes.
    }
    \label{fig:supp-leading-minors-L12}
\end{figure}

We parametrize the inverse temperature by the scaled variable $\tau=\beta/L$. For the range of system sizes studied here, the first two largest nontrivial
minor candidates are selected from the following four quantities:
\begin{equation}
\left|G_{1,1}(L\tau)\right|,
\qquad
\left|G_{2,2}(L\tau)\right|,
\qquad
\left|\det G_{\{1,2\},\{1,2\}}(L\tau)\right|,
\qquad
\left|\det G(L\tau)\right|.
\end{equation}
Figure~\ref{fig:supp-leading-minors-L12} shows these four quantities as
functions of \(\tau\) for \(L=12\).  The dashed horizontal line indicates the
trivial empty-minor value \(1\).  The vertical dotted lines mark the crossing
points at which the ordering of the leading nontrivial minors changes.

For \(L=12\), the first crossing occurs at
\(\tau_{c,1}=0.177\), where
\begin{equation}
\left|G_{2,2}\right|
=
\left|\det G_{\{1,2\},\{1,2\}}\right|.
\end{equation}
Thus, for small \(\tau\), the two leading nontrivial minors are
\(\left|G_{1,1}\right|\) and \(\left|G_{2,2}\right|\), while after this crossing
\(\left|\det G_{\{1,2\},\{1,2\}}\right|\) replaces
\(\left|G_{2,2}\right|\) as the second-largest nontrivial minor.

The second crossing occurs at \(\tau_{c,2}=0.702\), where
\begin{equation}
\left|\det G_{\{1,2\},\{1,2\}}\right|
=
\left|\det G\right| .
\end{equation}
Beyond this point, the full determinant becomes the second-largest nontrivial
minor.  Finally, at \(\tau_{c,3}=0.830\), one finds
\begin{equation}
\left|G_{1,1}\right|
=
\left|\det G\right| .
\end{equation}
For \(\tau>\tau_{c,3}\), the full determinant is the largest nontrivial minor,
while \(\left|G_{1,1}\right|\) becomes the second largest.

\begin{figure}[t]
    \centering
        \includegraphics[width=0.3\textwidth]{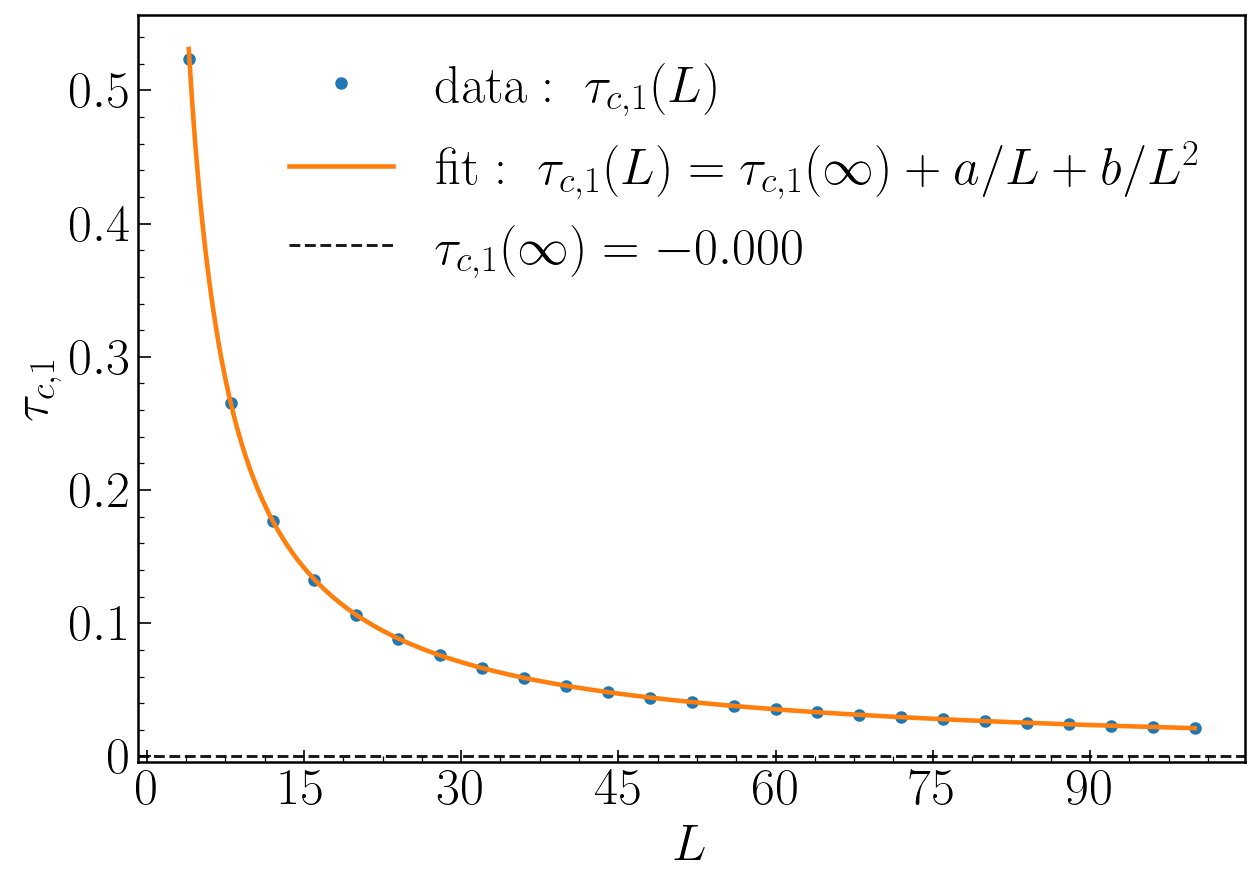}       \includegraphics[width=0.3\textwidth]{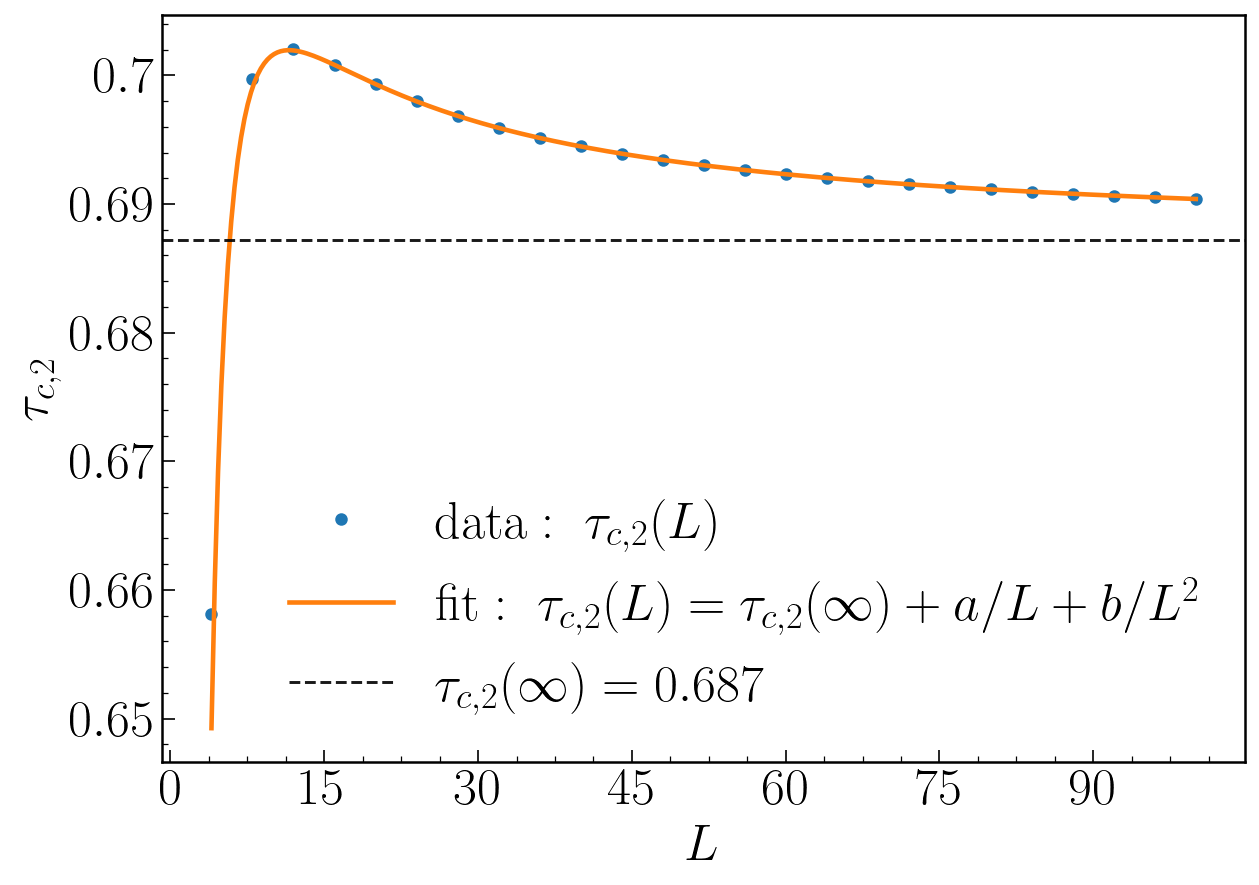}       \includegraphics[width=0.3\textwidth]{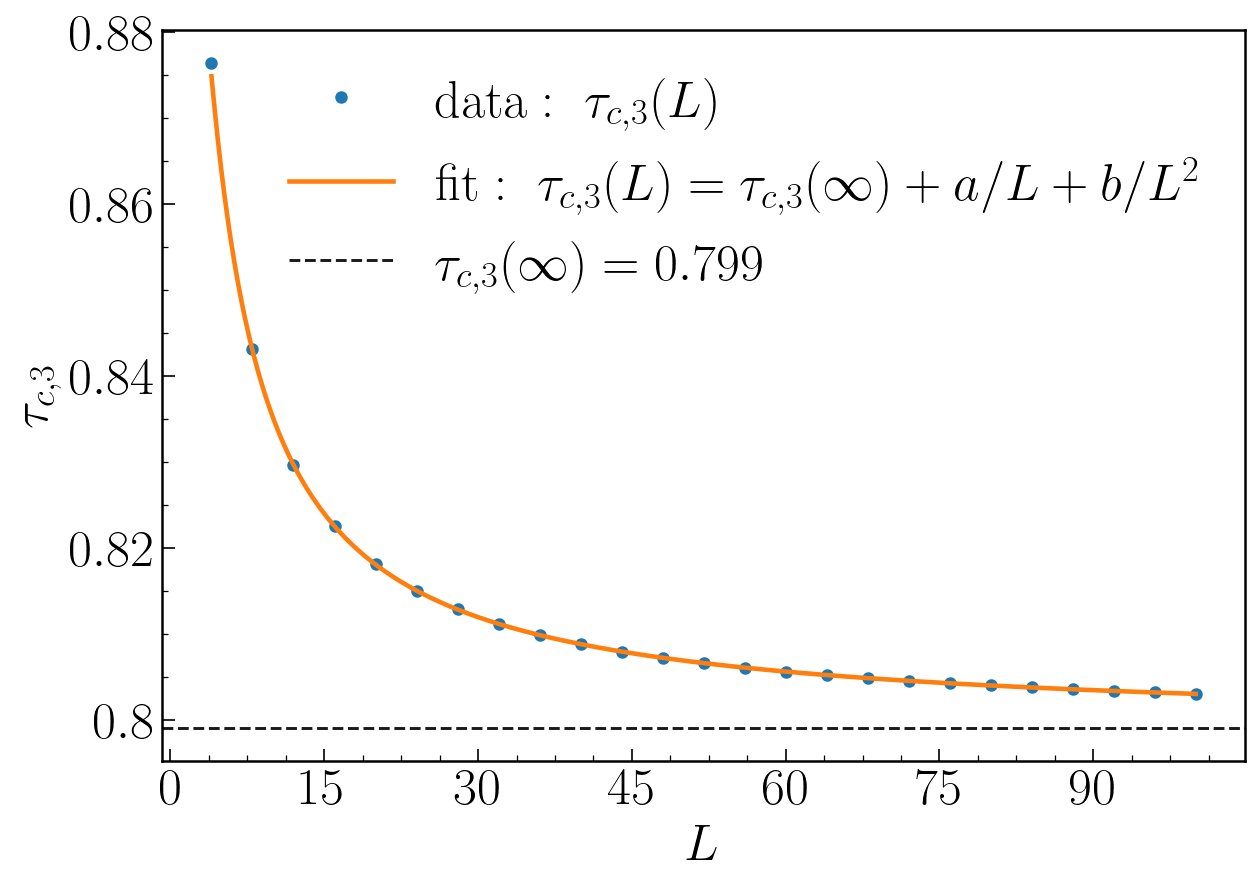}
    \caption{
    Finite-size extrapolation of the crossing points
    \(\tau_{c,i}(L)=\beta_{c,i}(L)/L\).  The data are fitted using
    \(\tau_{c,i}(L)=\tau_{c,i}(\infty)+a_i/L+b_i/L^2\).
    The extrapolated values are
    \(\tau_{c,1}(\infty)\simeq 0.000\),
    \(\tau_{c,2}(\infty)\simeq 0.687\), and
    \(\tau_{c,3}(\infty)\simeq 0.799\).
    }
    \label{fig:supp-critical-taus}
\end{figure}

To study the thermodynamic behavior of these crossings, we compute the critical
values \(\tau_{c,i}(L)=\beta_{c,i}(L)/L\) for several system sizes and fit them
with the finite-size form
\begin{equation}
\tau_{c,i}(L)
=
\tau_{c,i}(\infty)
+
\frac{a_i}{L}
+
\frac{b_i}{L^2}.
\end{equation}
The resulting extrapolations are shown in
Fig.~\ref{fig:supp-critical-taus}.  The first crossing extrapolates to zero,
\begin{equation}
\tau_{c,1}(\infty)\simeq 0.0,
\end{equation}
indicating that the region in which \(\left|G_{2,2}\right|\) is among the two
largest nontrivial minors shrinks to the origin in the large-\(L\) limit.
The other two crossings remain finite:
\begin{equation}
\tau_{c,2}(\infty)\simeq 0.687,
\qquad
\tau_{c,3}(\infty)\simeq 0.799 .
\end{equation}
Therefore, in the scaling limit with fixed positive \(\tau\), the relevant
changes in the ordering of the leading nontrivial minors occur near
\(\tau\simeq 0.687\) and \(\tau\simeq 0.799\).

\subsection{Approximation of the stabilizer R\'enyi entropy from leading minors}
\label{sec:supp-minor-approx-stabilizer-entropy}

We now use the hierarchy of minors to approximate the mixed-state stabilizer
R\'enyi entropy.  Starting from
\begin{equation}
\mathcal Z_\alpha(\rho)
=
2^{-L}
\sum_{P\in\mathcal P_L}
\left|\operatorname{Tr}(\rho P)\right|^{2\alpha},
\end{equation}
the stabilizer R\'enyi entropy is
\begin{equation}
M_{\alpha,L}(\beta)
=
\frac{1}{1-\alpha}
\log
\frac{\mathcal Z_\alpha(\rho_\beta)}
{\mathcal Z_1(\rho_\beta)} .
\end{equation}
Using the minor representation, this becomes
\begin{equation}
M_{\alpha,L}(\beta)
=
\frac{1}{1-\alpha}
\left[
\log
\sum_{\substack{A,B\subseteq\{1,\ldots,L\}\\ |A|=|B|}}
\left|\det G_{A,B}(\beta)\right|^{2\alpha}
-
\Pi_L(\beta)
\right],
\label{eq:supp-Malpha-minors}
\end{equation}
where
\begin{equation}
\Pi_L(\beta)
=
\log
\sum_{\substack{A,B\subseteq\{1,\ldots,L\}\\ |A|=|B|}}
\left|\det G_{A,B}(\beta)\right|^2
=
\log\!\left(2^L \operatorname{Tr}\rho_\beta^2\right).
\end{equation}
The empty minor is included in the sum and has value
\begin{equation}
\left|\det G_{\varnothing,\varnothing}(\beta)\right|=1 .
\end{equation}

For large \(\alpha\), the sum in Eq.~\eqref{eq:supp-Malpha-minors} is dominated
by the largest minors.  Since the empty minor is always equal to one, it gives
the leading contribution in the limit \(\alpha\to\infty\).  Therefore, with the
normalization used here,
\begin{equation}
M_{\infty,L}(\beta)=0 .
\end{equation}
The leading corrections at large but finite \(\alpha\) are controlled by the
largest non-empty minors. 

\begin{figure}[t]
    \centering
    \includegraphics[width=0.32\textwidth]{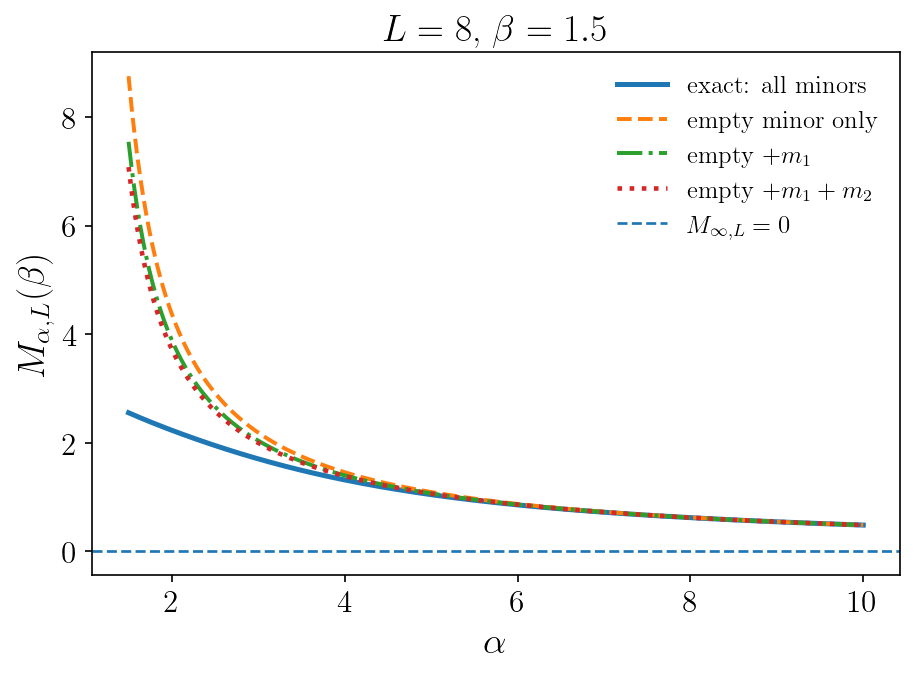}
    \includegraphics[width=0.32\textwidth]{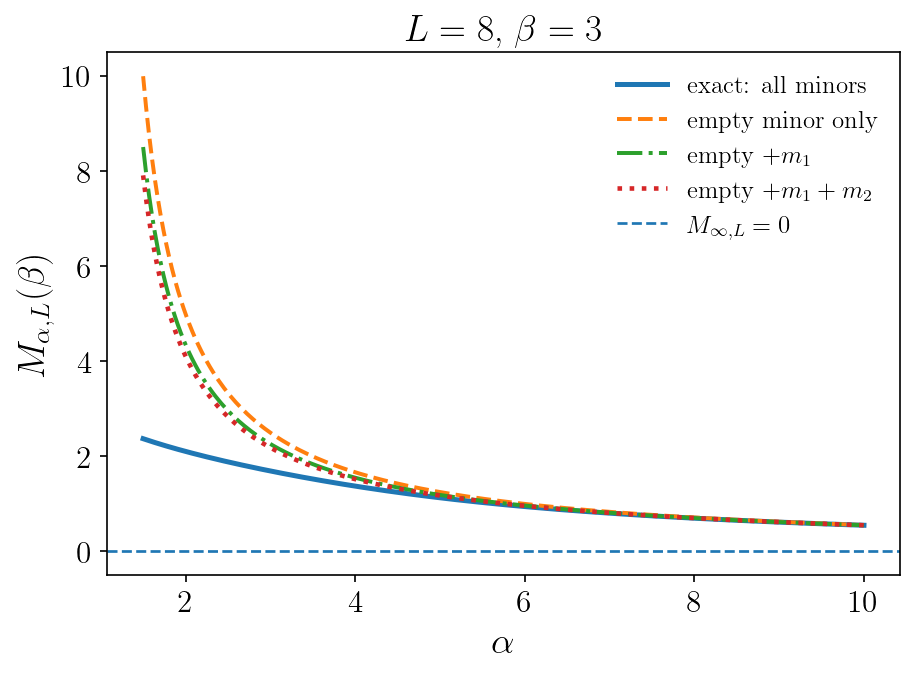}
    \hfill
    \includegraphics[width=0.32\textwidth]{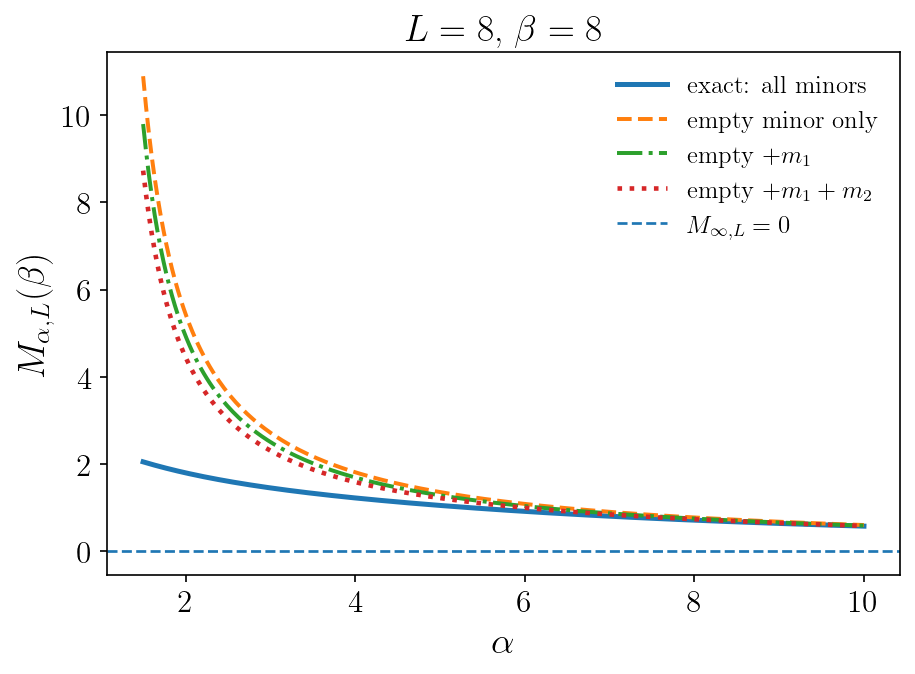}

   \caption{
    Stabilizer R\'enyi entropy \(M_{\alpha,L}(\beta)\) for \(L=8\), comparing
    the exact result from all minors with truncated approximations based on the
    empty minor and the leading non-empty minor groups. From left to right,
    the plots correspond to \(\beta=1.5\), \(\beta=3\), and \(\beta=8\).
    The horizontal dashed line indicates \(M_{\infty,L}=0\), which follows from
    the dominance of the empty minor in the \(\alpha\to\infty\) limit.
    Including the first and second largest non-empty minor groups improves the
    approximation at finite \(\alpha\), especially in the large-\(\alpha\)
    regime.
    }
    \label{fig:supp-stabilizer-leading-minors-L8}
\end{figure}

To illustrate how the leading minors control the large-\(\alpha\) behavior,
Fig.~\ref{fig:supp-stabilizer-leading-minors-L8} compares the exact stabilizer
R\'enyi entropy with truncated sums for \(L=8\) at three representative inverse
temperatures.  The blue curve is the exact result from all minors, while the
orange, green, and red curves retain respectively only the empty minor, the
empty minor plus the largest non-empty minor group, and the empty minor plus the
two largest non-empty minor groups.  As \(\alpha\) increases, the truncated
curves approach the exact result, which tends to \(M_{\infty,L}=0\).  This shows
that the large-\(\alpha\) behavior is controlled by the empty minor and the
first few largest non-empty minor groups.

\section{Exact XX--Ising correspondence for the stabilizer entropy}
\label{SecXXTFI}

In this section we record an exact finite-size relation between the
finite-temperature stabilizer entropy of the open XX chain and that of the
critical open transverse-field Ising chain studied in the main text.  The
relation is useful because it shows that the open XX problem on \(2L\) sites
factorizes, at the level of the Pauli-amplitude distribution, into two copies
of the open critical Ising problem on \(L\) sites.

Consider the open XX Hamiltonian on \(2L\) sites,
\begin{equation}
H_{\rm XX}^{\rm OBC}
=
\frac{J_{\rm XX}}{2}
\sum_{r=1}^{2L-1}
\left(
\sigma^x_r\sigma^x_{r+1}
+
\sigma^y_r\sigma^y_{r+1}
\right).
\label{eq:supp_H_XX}
\end{equation}
With this convention, the coupling which matches the critical Ising chain
Hamiltonian
\[
H_{\rm TFI}^{\rm OBC}
=
-J\sum_{j=1}^{L-1}\sigma^x_j\sigma^x_{j+1}
-J\sum_{j=1}^{L}\sigma^z_j
\]
is $J_{\rm XX}=2J$. After the Jordan--Wigner transformation, the open XX chain is a free hopping
Hamiltonian,
\begin{equation}
H_{\rm XX}^{\rm OBC}
=
J_{\rm XX}
\sum_{r=1}^{2L-1}
\left(
c_r^\dagger c_{r+1}
+
c_{r+1}^\dagger c_r
\right).
\label{eq:supp_HXX_fermion}
\end{equation}
The single-particle eigenfunctions and energies are
\begin{equation}
\phi_m(r)
=
\sqrt{\frac{2}{2L+1}}\sin(q_m r),
\qquad
q_m=\frac{\pi m}{2L+1},
\qquad
m=1,\ldots,2L,
\label{eq:supp_XX_modes}
\end{equation}
and
\begin{equation}
\epsilon_m=2J_{\rm XX}\cos q_m .
\label{eq:supp_XX_energies}
\end{equation}
In a compatible Majorana convention, the finite-temperature mixed correlation
matrix of the XX chain is
\begin{equation}
G_{rs}^{\rm XX}(\beta)
=
\frac{2}{2L+1}
\sum_{m=1}^{2L}
\sin(rq_m)\sin(sq_m)
\tanh\!\left(\beta J_{\rm XX}\cos q_m\right).
\label{eq:supp_GXX_full}
\end{equation}

The momenta occur in particle-hole pairs,
\[
q_{2L+1-m}=\pi-q_m .
\]
considering that one obtains
\begin{equation}
G_{rs}^{\rm XX}(\beta)
=
\frac{2}{2L+1}
\left[1-(-1)^{r+s}\right]
\sum_{m=1}^{L}
\sin(rq_m)\sin(sq_m)
\tanh\!\left(\beta J_{\rm XX}\cos q_m\right).
\label{eq:supp_GXX_pairing}
\end{equation}
Thus \(G_{rs}^{\rm XX}=0\) whenever \(r+s\) is even.  Ordering the XX sites as
odd sites followed by even sites,
\[
1,3,\ldots,2L-1 \;|\; 2,4,\ldots,2L ,
\]
the matrix has the off-diagonal block form
\begin{equation}
G_{\rm XX}^{\rm OBC}(\beta)
=
\begin{pmatrix}
0 & A(\beta)\\
A(\beta)^T & 0
\end{pmatrix},
\label{eq:supp_GXX_block}
\end{equation}
where
\begin{equation}
A_{jk}(\beta)
=
G_{2j-1,2k}^{\rm XX}(\beta).
\label{eq:supp_A_def}
\end{equation}

We now compare this block with the staggered-gauge Ising correlator derived in
Sec.~\ref{Sec1}.  Relabel
\[
m=L-p+1,
\qquad
p=1,\ldots,L .
\]
Then
\begin{equation}
q_m
=
\frac{\pi}{2}
-
\frac{k_p}{2},
\qquad
k_p=\frac{(2p-1)\pi}{2L+1}.
\label{eq:supp_q_k_relation}
\end{equation}
Therefore $\cos q_m=\sin\frac{k_p}{2}$. Together with \(J_{\rm XX}=2J\), this gives the matching of thermal factors,
\begin{equation}
\tanh\!\left(\beta J_{\rm XX}\cos q_m\right)
=
\tanh\!\left(2\beta J\sin\frac{k_p}{2}\right).
\label{eq:supp_thermal_match}
\end{equation}
The trigonometric factors satisfy
\begin{align}
\sin[(2j-1)q_m]
&=
(-1)^{j-1}
\cos\!\left[\left(j-\frac12\right)k_p\right],
\\
\sin(2kq_m)
&=
(-1)^{k+1}
\sin(kk_p).
\end{align}
Substituting these identities into Eq.~\eqref{eq:supp_A_def} gives
\begin{equation}
A_{jk}(\beta)
=
(-1)^{j+k}
\frac{4}{2L+1}
\sum_{p=1}^{L}
\tanh\!\left(2\beta J\sin\frac{k_p}{2}\right)
\cos\!\left[\left(j-\frac12\right)k_p\right]
\sin(kk_p).
\label{eq:supp_A_TFI_match}
\end{equation}
This is exactly the staggered-gauge Ising matrix
\(G_{\rm TFI}^{\rm OBC}(\beta)\) used throughout this Supplemental Material.
Equivalently,
\begin{equation}
A(\beta)=G_{\rm TFI}^{\rm OBC}(\beta).
\label{eq:supp_A_equals_GTFI_stag}
\end{equation}
If one uses the unstaggered Ising convention instead, the same statement is
written as \(A=D G_{\rm TFI}D\), with
\(D=\operatorname{diag}((-1)^1,\ldots,(-1)^L)\).  This difference is only a
diagonal sign convention and does not affect absolute minors or the purity.

We now define the absolute-minor sum of an \(N\times N\) matrix \(G\) by
\begin{equation}
\mathcal S_N(G)
=
\sum_{\substack{R,C\subseteq\{1,\ldots,N\}\\ |R|=|C|}}
\left|\det G_{R,C}\right|,
\label{eq:supp_general_minor_sum}
\end{equation}
including the empty minor.  For a block off-diagonal matrix
\[
M=
\begin{pmatrix}
0&A\\
A^T&0
\end{pmatrix},
\]
each nonzero square minor factorizes into one minor of \(A\) and one minor of
\(A^T\).  Hence the full absolute-minor sum factorizes as
\begin{equation}
\mathcal S_{2L}(M)
=
\mathcal S_L(A)^2 .
\label{eq:supp_block_minor_square}
\end{equation}
Applying this identity to Eq.~\eqref{eq:supp_GXX_block}, and using
Eq.~\eqref{eq:supp_A_equals_GTFI_stag}, gives
\begin{equation}
S_{2L}^{\rm XX}(\beta)
=
\left[
S_L^{\rm TFI}(\beta)
\right]^2 .
\label{eq:supp_S_XX_square}
\end{equation}

The purity normalization factor also squares.  For a Gaussian state with
correlation matrix \(G\),
\[
2^N\operatorname{Tr}\rho^2=\det(I_N+GG^T).
\]
Using Eq.~\eqref{eq:supp_GXX_block},
\begin{equation}
G_{\rm XX}G_{\rm XX}^T
=
\begin{pmatrix}
AA^T&0\\
0&A^TA
\end{pmatrix}.
\end{equation}
Therefore
\begin{equation}
2^{2L}\operatorname{Tr}\left(\rho_{\rm XX}^2\right)
=
\det(I_L+AA^T)\det(I_L+A^TA)
=
\det(I_L+AA^T)^2 .
\end{equation}
Since \(A\) is equal to the Ising correlator up to harmless diagonal signs,
\[
\det(I_L+AA^T)
=
\det(I_L+G_{\rm TFI}G_{\rm TFI}^T).
\]
Thus $ 2^{2L}\operatorname{Tr}\left(\rho_{\rm XX}^2\right)
=\left[
2^L\operatorname{Tr}\left(\rho_{\rm TFI}^2\right)
\right]^2$.
In terms of the logarithmic purity contribution,
\begin{equation}
\Pi_{2L}^{\rm XX}(\beta)
=
2\Pi_L^{\rm TFI}(\beta).
\label{eq:supp_Pi_XX_square}
\end{equation}

Combining Eqs.~\eqref{eq:supp_S_XX_square} and
\eqref{eq:supp_Pi_XX_square} with
\[
M_{1/2}(\beta)
=
2\log S(\beta)-2\Pi(\beta),
\]
we obtain the exact finite-size identity
\begin{equation}
M_{1/2,2L}^{\rm XX,OBC}(\beta;J_{\rm XX}=2J)
=
2M_{1/2,L}^{\rm TFI,OBC}(\beta;J).
\label{eq:supp_M_XX_2TFI}
\end{equation}
Equivalently, the XX numerator, purity factor, and normalized stabilizer
entropy are all obtained by doubling the corresponding logarithmic Ising
quantities:
\begin{equation}
\log S_{2L}^{\rm XX}(\beta)
=
2\log S_L^{\rm TFI}(\beta),
\qquad
\Pi_{2L}^{\rm XX}(\beta)
=
2\Pi_L^{\rm TFI}(\beta),
\qquad
M_{1/2,2L}^{\rm XX}(\beta)
=
2M_{1/2,L}^{\rm TFI}(\beta).
\label{eq:supp_all_XX_doubling}
\end{equation}

This identity is special to open boundary conditions.  The proof uses the
standing-wave momenta and the simple particle-hole pairing
\(q\leftrightarrow \pi-q\), which make the XX correlation matrix exactly
off-diagonal in the odd/even sublattice decomposition.  For periodic spin
chains, the Jordan--Wigner transformation introduces a boundary term depending
on the global fermion parity, and the spin Gibbs state involves a sum over
fermionic spin structures.  As a result, the periodic XX chain does not reduce
to a single square of the periodic Ising stabilizer entropy.


\section{Direct BCFT identification of the crossover function}

In the finite-size thermal scaling regime $\beta=\tau L$, the stabilizer
numerator obeys
\begin{equation}
\log S_L(\tau L)
=
\log S_L(\infty)+\log\cF(\tau)+o(1),
\end{equation}
with
\begin{equation}
\cF(\tau)
=
\frac{
\eta(i\tau/2)\eta(2i\tau)^{7/4}
}{
\eta(i\tau)^2\eta(4i\tau)^{1/2}
}.
\label{eq:F_eta_route2}
\end{equation}
Here we show that the complete scaling function in
Eq.~\eqref{eq:F_eta_route2} coincides with a known Ising boundary-CFT
amplitude.

For the critical Ising model with free boundaries, the one-point function of
the bulk disorder field $\mu$ on a cylinder is known exactly
\cite{Watts2000Supplement}.  Denoting by $C_\mu(x;\tau)$ the corresponding
free-boundary cylinder amplitude, and placing the bulk disorder field at the
reflection-symmetric midpoint $x=L/2$, we define the normalized quantity
\begin{equation}
\widehat C_\mu(\tau)
\equiv
\frac{C_\mu(L/2;\tau)}
     {C_\mu(L/2;\infty)}.
\label{eq:Cmu_norm_route2}
\end{equation}
In the conventions appropriate to the open Ising chain,
\begin{equation}
\widehat C_\mu(\tau)
=
\left[
\frac{\vartheta_3(1/4\,|\,i\tau)}
     {\vartheta_3(0\,|\,i\tau)}
\right]^{1/2}
\left[
\frac{\vartheta_1'(0\,|\,i\tau)}
     {\pi\,\vartheta_2(0\,|\,i\tau)}
\right]^{1/8}.
\label{eq:Cmu_theta_route2}
\end{equation}
Using the standard theta-function identities
\begin{align}
\vartheta_1'(0|z)
&=
\pi\vartheta_2(0|z)\vartheta_3(0|z)\vartheta_4(0|z),\\
\vartheta_3(1/4|z)^2
&=
\vartheta_3(0|2z)\vartheta_4(0|2z),
\end{align}
together with
\begin{equation}
\vartheta_4(0|z)
=
\frac{\eta(z/2)^2}{\eta(z)},
\qquad
\vartheta_3(0|z)
=
\frac{\eta(z)^5}
{\eta(z/2)^2\eta(2z)^2},
\end{equation}
one obtains
\begin{equation}
\widehat C_\mu(\tau)
=
\frac{
\eta(i\tau/2)\eta(2i\tau)^{7/4}
}{
\eta(i\tau)^2\eta(4i\tau)^{1/2}
}
=
\cF(\tau).
\label{eq:exact_match_route2}
\end{equation}

\begin{figure}[t]
\centering
\includegraphics[width=0.6\linewidth]
{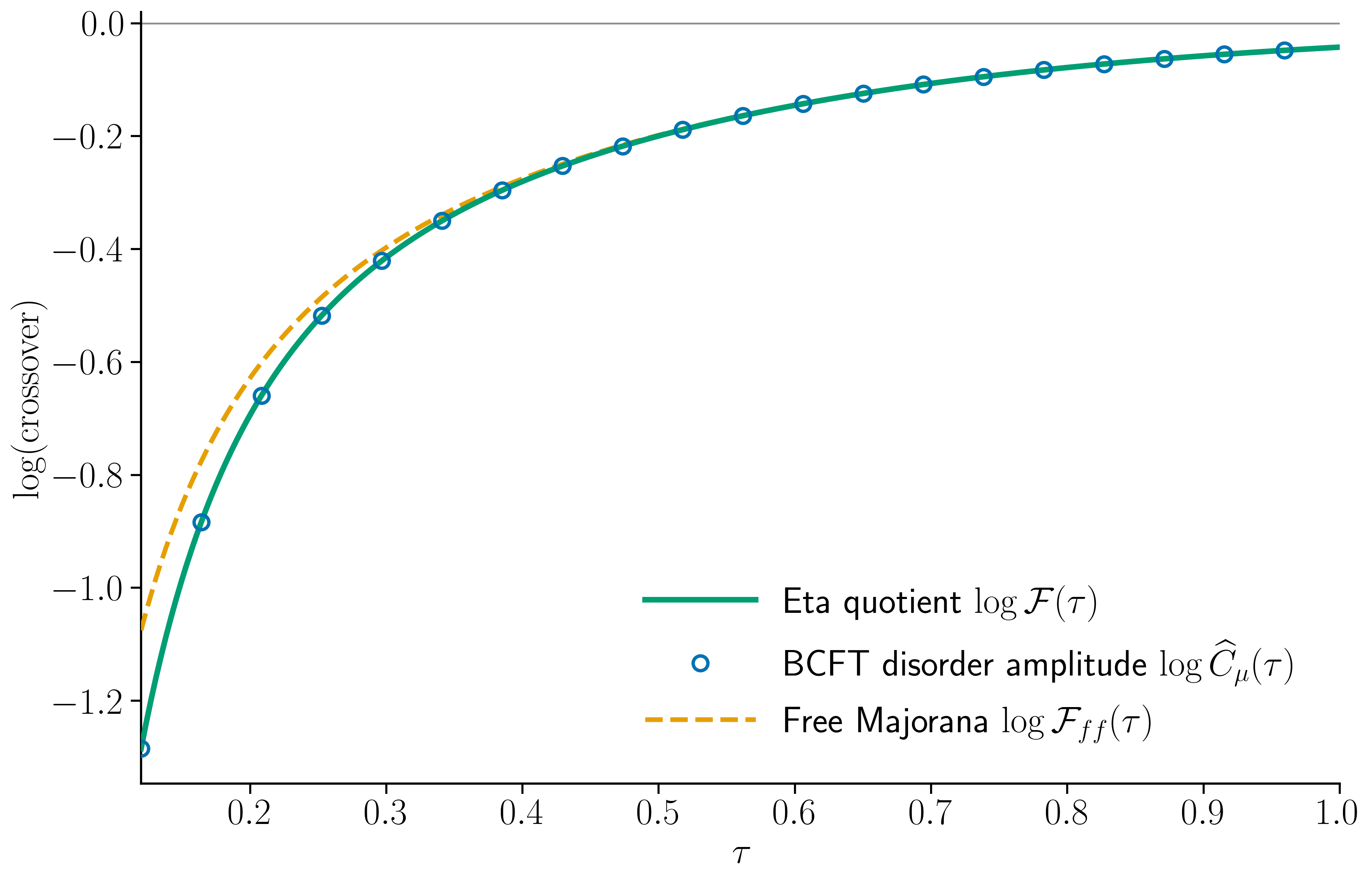}
\caption{
Direct comparison of the stabilizer crossover obtained from the eta quotient
(solid green line) with the independently evaluated midpoint Ising disorder
amplitude $\widehat C_\mu(\tau)$ (open blue symbols).  The two coincide over
the full range shown.  The dashed curve is the ordinary free-boundary Majorana
benchmark $\cFff(\tau)$, demonstrating that the disorder-decorated amplitude
contains universal information beyond the undecorated thermal factor.
}
\label{fig:route2_match}
\end{figure}

Equation~\eqref{eq:exact_match_route2} identifies the \emph{entire} universal
crossover, rather than only its leading small-$\tau$ coefficient, with the
normalized midpoint disorder-field amplitude of the free-boundary Ising CFT.
The equality is analytic; Fig.~\ref{fig:route2_match} provides a direct visual
comparison between the two independent representations.

The distinction from ordinary thermodynamics is therefore precise.  The
ordinary free-boundary Majorana factor,
\begin{equation}
\cFff(\tau)
=
\prod_{p=1}^{\infty}
\left(1+e^{-\pi\tau(2p-1)}\right)^{-1},
\end{equation}
describes the undecorated free-boundary thermal spectrum, whereas the
stabilizer numerator is governed by the same critical Ising theory with a
disorder-field insertion.

The modular asymptotic follows immediately from
Eq.~\eqref{eq:exact_match_route2},
\begin{equation}
\log\cF(\tau)
=
-\frac{\pi}{16\tau}
-\frac18\log\tau
+\frac18\log2
+o(1),
\qquad
\tau\to0^+.
\end{equation}
The coefficients $1/16$ and $1/8$ are precisely the chiral weight and scaling
dimension of the Ising disorder primary,
\begin{equation}
h_\mu=\bar h_\mu=\frac1{16},
\qquad
x_\mu=h_\mu+\bar h_\mu=\frac18.
\end{equation}
Thus the previously unusual modular coefficients are standard Ising conformal
data associated with the disorder sector, rather than a new bulk central
charge.

We emphasize that this identification concerns the CFT amplitude selected by
the stabilizer observable.  Neither the microscopic Hamiltonian nor the
thermal state contains a physical defect.  The disorder insertion appears in
the Euclidean representation of the nonlinear Pauli-weight observable.

\makeatletter
\let\arxivOriginalLabel\label
\renewcommand{\label}[1]{%
  \def\arxivLabelKey{#1}%
  \def\arxivLastBibKey{LastBibItem}%
  \ifx\arxivLabelKey\arxivLastBibKey
    \arxivOriginalLabel{LastSupplementBibItem}%
  \else
    \arxivOriginalLabel{#1}%
  \fi
}
\makeatother

\let\label\arxivOriginalLabel

\end{document}